\documentclass[fleqn,usenatbib]{mnras}

\usepackage[T1]{fontenc}

\DeclareRobustCommand{\VAN}[3]{#2}
\let\VANthebibliography\thebibliography
\def\thebibliography{\DeclareRobustCommand{\VAN}[3]{##3}\VANthebibliography}

\usepackage{graphicx}	
\usepackage{amsmath}	
\usepackage{colortbl} 
\usepackage[table,xcdraw]{xcolor} 
\usepackage{subcaption}
\usepackage{caption}

\def\sigu{\sigma_{\Delta u}}

\usepackage{pdfpages}
\usepackage{amssymb}	
\usepackage{amstext}
\usepackage[all]{hypcap}
\usepackage{afterpage}
\usepackage{gensymb}
\usepackage{color}
\usepackage{dblfloatfix}
\usepackage{enumitem}
\usepackage{soul}
\usepackage[flushleft]{threeparttable}
\usepackage{longtable}
\usepackage{pdflscape}
\usepackage{newtxtext,newtxmath}

\title[A new way to find symbiotic stars]{A new way to find symbiotic stars: accretion disc detection with continuum survey photometry}

\author[A. B. Lucy et al.]{A. B. Lucy,$^{1}$\thanks{E-mail: alucy@stsci.edu (ABL)}
J. L. Sokoloski,$^{2}$
G. J. M. Luna,$^{3,4}$
K. Mukai,$^{5,6}$
N. E. Nu\~nez,$^{7}$
D. A. H. Buckley,$^{8,9,10,11}$
\newauthor
H. Breytenbach,$^{8,9}$
B. Paul,$^{8,12}$
S. B. Potter,$^{8,12}$
R. Manick,$^{8,13}$
D. A. Howell,$^{14,15}$
C. Wolf,$^{16,17}$
\newauthor
and C. A. Onken$^{16,17}$
\\
$^{1}$Space Telescope Science Institute, 3700 San Martin Drive, Baltimore, MD 21218, USA\\
$^{2}$Columbia University, Dept. of Astronomy, 550 West 120th Street, New York, NY 10027, USA\\
$^{3}$Universidad Nacional de Hurlingham (UNAHUR). Laboratorio de Investigación y Desarrollo Experimental en Computación, Av.
Gdor. Vergara 2222,\\Villa Tesei, Buenos Aires, Argentina\\
$^{4}$Consejo Nacional de Investigaciones Científicas y Técnicas (CONICET)\\
$^{5}$CRESST II and X-ray Astrophysics Laboratory, NASA/GSFC, Greenbelt, MD 20771, USA\\
$^{6}$Department of Physics, University of Maryland, Baltimore County, 1000 Hilltop Circle, Baltimore, MD 21250, USA\\
$^{7}$Facultad de Ciencias Exactas, Físicas y Naturales, Universidad Nacional de San Juan, Av. Ignacio de la Roza 590 (O), Complejo Universitario ``Islas\\ Malvinas'', Rivadavia, J5402DCS, San Juan, Argentina\\
$^{8}$South African Astronomical Observatory, PO Box 9, Observatory 7935, Cape Town, South Africa\\
$^{9}$Department of Astronomy, University of Cape Town, Private Bag, Rondebosch 7701, Cape Town, South Africa\\
$^{10}$Southern African Large Telescope, PO Box 9, Observatory 7935, Cape Town, South Africa\\
$^{11}$Department of Physics, University of the Free State, PO Box 339, Bloemfontein 9300, South Africa\\
$^{12}$Department of Physics, University of Johannesburg, PO Box 524, Auckland Park 2006, South Africa\\
$^{13}$Univ. Grenoble Alpes, CNRS, IPAG, F-38000 Grenoble, France\\
$^{14}$Las Cumbres Observatory, 6740 Cortona Drive, Suite 102, Goleta,
CA 93117-5575, USA\\
$^{15}$Department of Physics, University of California, Santa Barbara, CA
93106-9530, USA\\
$^{16}$Research School of Astronomy and Astrophysics, Australian National University, Canberra, ACT 2611, Australia\\
$^{17}$Centre for Gravitational Astrophysics, Research Schools of Physics, and Astronomy and Astrophysics, Australian National University, Canberra, ACT 2611,\\Australia
}

\pubyear{2025}

\begin{document}
\label{firstpage}
\pagerange{\pageref{firstpage}--\pageref{lastpage}}
\maketitle

\begin{abstract}
Symbiotic stars are binaries in which a cool and evolved star of luminosity class I--III accretes onto a smaller companion. However, direct accretion signatures like disc flickering and boundary layer X-rays are typically outshone or suppressed by the luminous giant, shell burning on the accreting white dwarf, and the illuminated wind nebula. We present a new way to find symbiotics that is less biased against directly-detectable accretion discs than methods based on narrow-band H$\alpha$ photometry or objective prism plate surveys. We identified outliers in SkyMapper survey photometry, using reconstructed uvg snapshot colours and rapid variability among the three exposures of each 20-minute SkyMapper Main Survey filter sequence, from a sample of 366,721 luminous red objects. We found that SkyMapper catalog colours of large-amplitude pulsating giants must be corrected for variability, and that flickering is detectable with only three data points. Our methods probed a different region of parameter space than a recent search for accreting-only symbiotics in the GALAH survey, while being surprisingly concordant with another survey's infrared detection algorithm. We discovered 12 new symbiotics, including four with optical accretion disc flickering. Two of the optical flickerers exhibited boundary-layer hard X-rays. We also identified 10 symbiotic candidates, and discovered likely optical flickering in the known symbiotic V1044 Cen (CD-36 8436). We conclude that at least 20\% of the true population of symbiotics exhibit detectable optical flickering from the inner accretion disc, the majority of which do not meet the H$\alpha$ detection thresholds used to find symbiotics in typical narrow-band surveys.
\end{abstract}




\clearpage

\section{Introduction}\label{Introduction}

Symbiotic stars are binaries in which a cool evolved \mbox{G--M},~S,~or~C giant with luminosity class \mbox{I--III} accretes onto a smaller 
companion. The accreting companion is a white dwarf (WD) in most known symbiotic stars, evidenced by high 
luminosity ($\sim10^3\,L_\odot$) and high temperature ($\sim10^5$~K) in the shell-burning state and by nova-like behaviour \citep{Mikolajewska2003}, the short time-scale of accretion disc flickering in the accreting-only state \citep{Sokoloski2010,Luna2013}, and hard X-ray emission in the accreting-only state from the boundary layer between the accretion disc and the WD \citep{Luna2013}. In rare cases, the accretor is instead a neutron star \citep{Chakrabarty1997}.\footnote{Main-sequence accretors \citep{Sahai2015} and black hole accretors \citep{Lopez2017} may also be included. Their different contexts in stellar binary evolution have recently led some researchers to exclude them from the definition of symbiotics (e.g., \citealt{Merc2025}), but main sequence accretors can at times be difficult to rigorously distinguish from WD accretors \citep{Sahai2015}, and black hole accretors may share physical processes like wind accretion \citep{Wiktorowicz2021}.}

Judging by the high luminosities and high temperatures of the accreting WD \citep{Mikolajewska2003}, most known, confirmed symbiotic stars exhibit hydrogen shell burning on the WD, either due to stable/quasi-stable burning of accreted matter, or due to residual burning from a nova. WDs with shell burning are generally much more luminous than accreting-only WDs, because more than an order of magnitude more energy per nucleon is released from nuclear burning than from accretion onto a WD \citep{Sokoloski2001}. Accretion-fed shell burning is unstable outside of a specific region of accretion rate / WD mass parameter space \citep[e.g.,][]{Wolf2013}, and not many symbiotic recurrent novae are known, so it may be that there are symbiotic stars with lower accretion rates, very high WD masses, or very low WD masses that have been missed by existing surveys. Similarly, the duration of residual burning is inversely related to the WD mass, so it may be that existing samples of residually-burning WDs picked up by symbiotic surveys are primarily low-mass WDs. 

If these putative survey biases are real \citep{Mukai2016,Munari2021}, they likely originate in selection criteria based on optical emission lines developed throughout the history of symbiotic star surveys.

\subsection{The history of symbiotic star detection}\label{History}

At the same time as W. P. Fleming and A. J. Cannon were discovering a class of unusual nova-like stars that would later come to be known as symbiotic binaries \citep{Cannon1911,Cannon1923}, Fleming was also working on the taxonomy of ``Md class'' stars: M stars with Balmer emission, which she linked to their long-period variability \citep{flemingobit}. These pulsating red giants are these days known mostly as Mira variables and long period variables (LPVs), which according to our modern understanding produce Balmer emission through shocks in their atmospheres (\citealt{Fadeyev2004}, and references therein). Fleming showed that there was a continuum of Md stars, ranging from Md 1 in which H$\beta$ and H$\gamma$ were much stronger than H$\delta$, to Md 10 in which H$\delta$ was stronger by far \citep{Fleming1912}. Today, the holotype of Fleming's Md 10 stars is called an oxygen-rich Mira, in which TiO bands may absorb H$\alpha$ and H$\beta$ emission before it can escape the stellar atmosphere---and the holotype of her Md 1 stars is called an S Mira, in which spectral features from s-process products partially or fully replace the TiO molecular banding \citep{Castelaz2000}. 

But several objects that would later become known as symbiotic stars were also included in Fleming's Md sequence, including Y CrA, RW Hya, R Aqr, and Mira (Table IX and Plate II in \citealt{Fleming1912}). Here is where the fundamental tension of our subject began: when are emission lines on a luminous M spectrum emitted by the cool giant itself, and when are they emitted elsewhere?

Until recently, a simple resolution to this tension was sufficient. Over time, an understanding developed that high-ionization lines on cool giant spectra could only be produced by a hot component that accretes from the cool giant and photo-ionizes the giant's wind nebula \citep{Merrill1919, Plaskett1931,Merrill1932}. Merrill coined the term ``symbiotic star'' in 1941 \citep{McLaughlin1941}, referring to the sustained coexistence of the cool and hot components. The operational observational criteria eventually came to require a cool evolved star that has either {\it (a)} emission from an ion that takes at least 20 or 35 eV to create, typically [\ion{O}{iii}] (35 eV) or higher, or {\it (b)} an A/F star continuum with additional absorption lines signifying an outburst state in a companion to the evolved star \citep{Kenyon1986}. As discussed by \citet{Kenyon1986}, these observational criteria have always been porous, with a variety of exceptions made for special cases.

The fundamental reason for requiring a high-ionization emission line in quiescent symbiotics was that \citet{Fleming1912} had found many isolated M stars producing Balmer emission on their own in the course of their pulsations. Requiring [\ion{O}{iii}], \ion{He}{ii}, or the like was a practical and necessary observational cutoff to ensure that every symbiotic star had a hot component distinct from the cool giant, and to ensure that symbiotic star catalogs would not include the isolated giants that made up the majority of Fleming's Md class. This observational cutoff made the boundary between symbiotic stars and Md stars unambiguous in terms of observational properties, at the cost of possibly leaving some genuine accreting binaries mixed into the Md class.

Driven first by objective prism plate surveys, and then by narrow-band H$\alpha$ photometry coupled with follow-up spectroscopy to check for higher-ionization emission lines (e.g., \citealt{Miszalski2014}, \citealt{Mikolajewska2014}), the number of known and candidate symbiotics rose to 100 \citep{Allen1984}, then to between 188 and 218 \citep{Belczynski2000}, then to between 323 and 564 as of 2019 \citep{Akras2019b,Merc2019}. Ionization state aside, any weak emission lines can be missed by both objective prism plate surveys and narrow-band photometry. For the latter instruments, we can quantify this detection threshold; using the IPHAS narrow-band H$\alpha$ filter, \citet{Corradi2008} adopted an H$\alpha$ pseudo-equivalent width lower-limit of about 50\,\AA\, for their symbiotic candidate search. IPHAS is in principle sensitive to H$\alpha$ equivalent widths of a few Angstroms, but the threshold for detectability above the stellar locus reaches as high as 50\,\AA\, for red or reddened sources \citep{Drew2005}, further complicated by the wide range of molecular absorption band spectra in both symbiotic stars and their mimics.

\subsection{Accretion disc flickering}\label{introflickering}

If other signatures of accretion besides optical emission lines can be reliably detected in archival survey data, we can avoid reliance on large H$\alpha$ equivalent widths as an initial candidate selection cut, and it will not be necessary to require high-ionization emission from the giant's wind nebula to confirm an object's symbiotic nature. One such direct accretion signature is ``flickering''\footnote{``Flickering'' is not precisely synonymous with ``flicker noise,'' also known as pink noise, although the two terms are related. Accretion disc flickering can have both pink and red noise components \citep{Middleton2021}, or power spectrum slopes in between the two.} on time-scales of minutes, a superposition of multiplicatively-coupled fluctuations in the flux \citep{Uttley2005}. The exact physical cause of flickering is unknown, but models can produce it with variations in the mass-transfer rate propagating inwards through the disc, faster and faster variability building up at decreasing radii with shortening viscous time-scales \citep{Scaringi2014}. Flickering appears to be a universal phenomena of accretion, with similar phenomenology in the accretion discs of young stellar objects (YSOs), WDs, neutron stars, and supermassive black holes, with longer time-scales for higher-mass accretors \citep{Uttley2001,Vaughan2011,Scaringi2011,Scaringi2015,Middleton2021}.

Virtually all cataclysmic variables (CVs: symbiotic stars' smaller cousins, with main-sequence-like donors) exhibit accretion disc flickering at optical through X-ray wavelengths \citep{Bruch1992,Bruch2015,Balman2020}, but only about 11 or so symbiotics were known to exhibit optical flickering prior to 2021 (RS Oph, T CrB, MWC 560, V2116 Oph, CH Cyg, RT Cru, Mira $\equiv$ {\it o} Cet, V407 Cyg, V648 Car, EF Aql, CN Cha, and perhaps Y Gem; \citealt{Zamanov2017}, \citealt{Snaid2018}, \citealt{Lancaster2020}, \citealt{Guerrero2025}).\footnote{Z And is sometimes included in this list, but its rapid pulsations are usually coherent \citep{Sokoloski1999} and do not meet the above definition of accretion flickering; however, more recently, \citet{Merc2024} found transitory optical flickering in Z And during the decline from an active phase, as well as long-term variability in the oscillation period. Marginal or candidate optical flickering was found by \citet{Sokoloski2001} in EG And, BX Mon, CM Aql and BF Cyg; flickering in BF Cyg has since been confirmed by \citet{Merc2024}.} \citet{Munari2021} attest that there is optical flickering in a further 12 symbiotics from their GALAH Symbiotic Star (GaSS) sample, discovered through a novel survey design developed at the same time as our own, and \citet{Merc2024} have very recently used TESS to discover flickering in a further 13. Up to another 10 or so symbiotics flicker in the NUV \citep{Luna2013,Mukai2016,Lucy2020}, though sparse sampling with cadences of an hour or more currently precludes in-depth analysis. The relative rarity of detectable flickering from symbiotics is probably because shell-burning, when present on the WD and reprocessed into the UV and optical by the ubiquitous wind nebulae that characterize symbiotic stars, is much more luminous than the accretion disc, and shell-burning light is not likely to vary faster
than thermal-timescale variability of at most about $100$\% change per day, equivalent to at most about $1$\% change per 10 minutes \citep{Sokoloski2001}. Furthermore, it is difficult to detect an accreting-only symbiotic amid the light of the evolved giant itself, and quasi-spherical wind accretion from the ambient cool giant wind nebula might sometimes not lead to the formation of a disc in the first place.

The serendipitous discovery mechanisms of the 11 or so optically-flickering symbiotics known prior to 2021 reinforce the view that there may be a survey bias against accreting-only symbiotics. RS Oph and T CrB were discovered via recurrent novae observed by the 19th century or earlier (\citealt{Cannon1905}; \citealt{Bruch1986}; \citealt{Schaefer2023}), and after fading from their transient thermonuclear state became accreting-only until the next eruption, with flickering later discovered by \citealt{Walker1957}. V407 Cyg was first noticed due to an accretion disc outburst in 1936 (see \citealt{Giroletti2020}). Extraordinarily, CH Cyg was treated as an M6 III {\it standard star} reference until its accretion rate increased (\citealt{Deutsch1964}, as discussed by \citealt{Burmeister2009}; flickering was discovered soon after in \citealt{Wallertstein1968} and \citealt{Cester1968}). Mira, a holotype of the Mira class to which it gave its name, is nearby and one of the brightest variable stars in the sky. MWC 560 has perhaps the highest accretion rate of any known accreting-only symbiotic, with broad Balmer absorption to make it particularly interesting for follow-up \citep{Bond1984,Lucy2020}. V2116 Oph (GX 1+4) is a neutron star accretor, detected first in X-rays by balloon observations and then found to be flickering \citep{Jablonski1997}. V648 Car\footnote{Along with perhaps RT Cru, whose exact selection mechanism was not fully explained in its discovery paper \citep{Cieslinski1994}.} had probably the most normal discovery mechanism of the flickering symbiotics through objective prism plates \citep{Henize1952}, with flickering later discovered \citep{Angeloni2012} after an X-ray detection \citep{Masetti2006}. Most interestingly from the point of view of the present work, two more recent flickerers were discovered to be symbiotics through excess UV flux: EF Aql was a contaminant in the UV-bright Quasar Survey (\citealt{Monroe2016}, \citealt{Margon2016}; unusual infrared colours also implicitly played a role), after which \citet{Zamanov2017} found flickering, and Y Gem was discovered in an intentional search for asymptotic giant branch (AGB) stars in binaries with UV excess and X-rays \citep{Sahai2015,Snaid2018}.

\subsection{Overview of survey design}\label{overview}

It is an intuitive notion, and indeed it has been known for a long time \citep{Arkhipova1985}, that symbiotics should differ from the main stellar locus in their optical continuum colours due to the hot component contribution. And it is an equally intuitive notion that minutes-timescale accretion disc flickering (\autoref{introflickering}) could be detected if a survey contained multiple exposures in the same band separated by mere minutes, although one might wonder if many data points would be required to detect flickering with any statistical significance. A survey employing either or both of these two principles as its first-pass search criteria might be less biased against accreting-only symbiotics than H$\alpha$ methods, and in the latter flickering case might preferentially select symbiotics with detectable accretion discs. Aiming to test the idea that there is a hidden population of accreting-only symbiotics, we employed these concepts in the design of a new method of symbiotic star detection intended to be less biased against optically-flickering symbiotic stars than methods based on narrow-band H$\alpha$ photometry and objective prism plate surveys.

The SkyMapper Southern Sky Survey \citep{skymapperdr1,skymapperdr2} seemed to us well-suited to the task. The {\it uvgriz} filter set \citep{Bessell2011} has some unusual advantages over SDSS when it comes to disentangling accretion disc light from the luminous giant. The u filter is shifted further to shorter wavelengths than the SDSS U filter, emulating Str{\"o}mgren u by having the vast majority of its throughput shortward of the Balmer jump, which should be advantageous for diminishing the contribution of the cool giant as much as possible in our shortest filter. The v filter, an intermediate-band violet filter just longward of the Balmer jump, is also non-standard. Finally, with respect to accretion disc flickering, each SkyMapper Main Survey visit contains a total of three u-band exposures in the space of about 20 minutes, comparable to a typical peak-to-peak variability time-scale of symbiotic star accretion disc flickering \citep{Lucy2020}.

We conducted our survey in multiple steps, progressively narrowing our sample from 366,721 sources in our preparatory analysis of archival SkyMapper data, to 234 targets in our optical spectroscopy survey, to 10--11 targets in our follow-up optical fast photometry observations and shallow X-ray observations, to 2 targets in our follow-up deep X-ray observations. We present the curation of our working sample of 366,721 luminous red objects in \autoref{lro}, the distribution of these objects in SkyMapper parameter space and their archival object classifications in \autoref{theskymapperparameterspace}, target selection criteria for the optical spectroscopy survey in \autoref{targeting}, an overview of the new observations we conducted in \autoref{Observations}, and the results of our observations including source classification in \autoref{Results}. We discuss the implications of our survey in \autoref{Discussion}, and present our conclusions in \autoref{Conclusions}.

In order to clearly explain the development of our survey, ``known'' symbiotics in this paper refer to symbiotic stars reported in the literature prior to our first major optical spectroscopy observing run starting 2019 June 19. More recent discoveries by other groups are discussed with different wording in \autoref{comparison}.

\section{Sample curation: the luminous red objects}\label{lro}

We needed to choose a sample from which to draw SkyMapper outliers. The first\footnote{Methodology and results from a previous test of our symbiotic search design are outlined in \citet{Lucy2018b}. One major difference between that work and this is that our first search design did not use {\it Gaia}, which had only just become available with its DR1, so the substrate sample of ``cool giants'' in \citet{Lucy2018b} required multiple IR colour cuts and contained many cool dwarfs. It was also based on SkyMapper DR1.1, which did not include the minutes-time-scale variability data from the Main Survey, so \citet{Lucy2018b} instead attempted outburst-detection with long-timescale u-band and z-band variability.} step was to start building a working catalog of cool giants. Not every object in the sample ended up being a cool giant, so we call our final catalog a sample of ``luminous red objects''. This sample of luminous red objects is the substrate from which we drew symbiotic star candidates and other outliers of interest.

The philosophy underlying our sample selection was to balance the competing demands of purity and completeness. We aimed for sample purity, so that targets with outlying SkyMapper properties would, despite being outliers, still be systems containing cool giants. But we also aimed to build a selection function with as few constraints as possible, to include both known symbiotics {\it and} as-yet undiscovered symbiotics that could differ in unpredictable ways from known symbiotics; in other words, we wanted our sample to be as complete as possible with respect to a population whose properties we could not know in advance. At each step, we were also especially careful to avoid excluding the known symbiotic EF Aql, as it was the only known optical flickerer \citep{Zamanov2017} in the SkyMapper footprint that could be captured with reasonable photometry quality cuts.

The order of the selection filters applied in this section is described in \autoref{criteria}. We will first focus on the major scientific criteria in \autoref{choosing}, then provide a brief overview of the detailed implementation in \autoref{qualitycuts}.

\begin{figure*}
	\includegraphics[width=\textwidth]{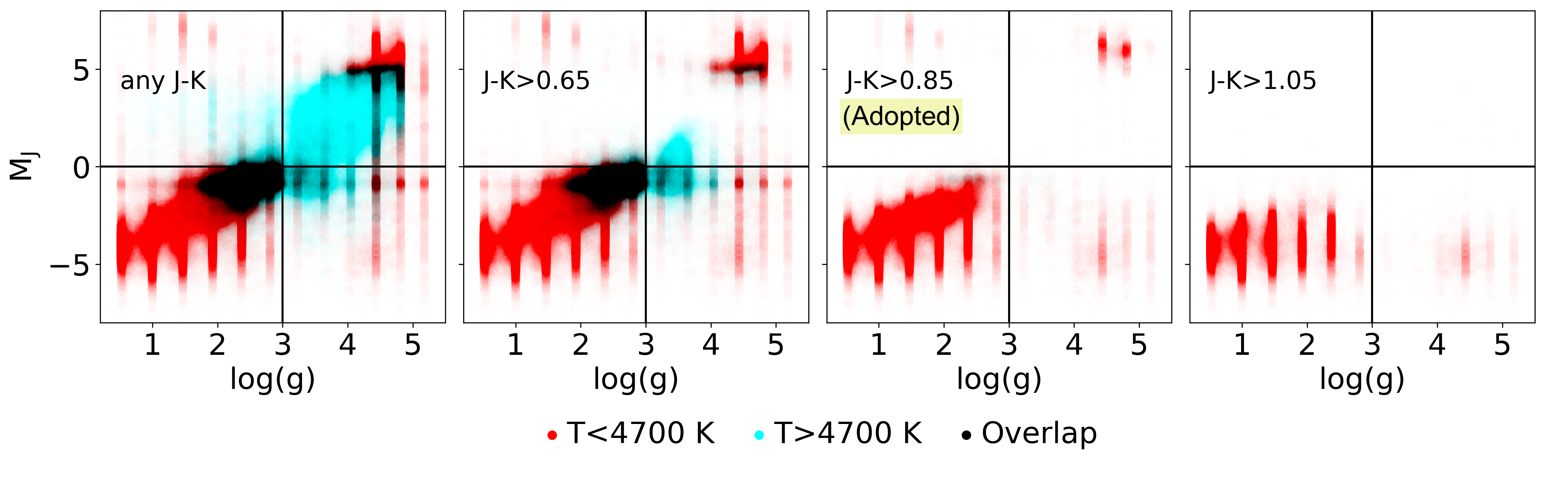}
    \caption{Scatter plot of RAVE stars, with absolute J-band magnitude from 2MASS and \citet{Bailer2018} {\it Gaia} DR2 distances, and calibrated surface gravity and temperature from RAVE DR5. The vertical streaks are artifacts of how surface gravity is binned in RAVE. Without a sufficiently strict \mbox{J-K$_{\rm s}$} colour cut from 2MASS, the region of cool giants in this parameter space blends into the region of warm subgiants, the blue region in the second panel. With an overly strict J-K$_{\rm s}$ colour cut, the completeness of the cool giant sample suffers, as seen in the fourth panel.}
    \label{2masscut}
\end{figure*}

\subsection{Choosing IR colour and luminosity criteria}\label{choosing}

Most papers published on cool giant selection methods are oriented towards using M giant samples to examine the structure of our Galaxy and its stellar streams \citep[e.g.,][]{Majewski2003,Bochanski2014,Li2016}. Selection functions have typically relied on a combination of several IR colour criteria (e.g., Equation 1 in \citealt{Li2016}). Relying on such strict criteria would pose a severe problem to us, because symbiotic stars can have different IR colours than isolated cool giants (e.g., \citealt{Akras2019}). But with the launch of the {\it Gaia} mission \citep{Gaia2016,Gaia2018}, parallax distance-informed luminosities can replace the more stringent colour criteria in the task of sorting giants from lower-luminosity objects. 

We adopted the Bayesian-inferred estimated distances that \citet{Bailer2018} derived from {\it Gaia} DR2 parallaxes.\footnote{Formally, as discussed by \citet{Bailer2018}, this solution is technically incorrect, because Bayesian inferences on calculated quantities like absolute magnitudes should be derived from first principles. Practically, our validation tests presented in this section suggest that the \citet{Bailer2018} distances are sufficient for our purposes, particularly in light of the other large uncertainties to which we are subject.} \citet{Bailer2018} distances use a Bayesian prior Galaxy model governed by a single length scale parameter, varying as a function of only Galactic longitude and latitude. We tested our final selection function both with the estimated distance and, more conservatively, with its lower uncertainty bound (which would be less likely to misclassify a nearby dwarf as a distant giant), and found barely any difference in the selected samples. Another concern is that orbital motion could cause deviant parallax measurements; this issue has the strongest potential effect for very wide binaries of unequal brightness and orbital periods close to a year (D. Pourbaix 2018, private communication; \citealt{Penoyre2020}), the approximate orbital period of {\it Gaia} around the sun. This could affect the parallax measurements of symbiotic binaries and pose an issue for completeness, but we are not aware of any classes of potential contaminant objects that could be introduced {\it into} the sample by this mechanism.

We chose a single colour cut to constrain the stellar temperature, paired with a luminosity cut. Among the various Two Micron All Sky Survey (2MASS; \citealt{2MASS}) and ALLWISE (\citealt{allwise}, based on the Wide-field Infrared Survey Explorer; \citealt{WISE}) colour cuts used to select red giants, J-K$_{\rm s}$ appears to be the best probe of a star's temperature (e.g., \citealt{Li2016}). To determine where to place our cutoffs, we matched the RAdial Velocity Experiment (RAVE) DR5 catalog \citep{Kunder2017} to \citet{Bailer2018} distances and 2MASS photometry using the pre-crossmatched source IDs {\it ravedr5\_best\_neighbour} and {\it tmass\_best\_neighbour} in the {\it Gaia} DR2 catalog, leaving about 456,000 RAVE stars of all types to test after initial quality cuts. In \autoref{2masscut}, we plot the absolute J magnitudes (from 2MASS J and \citealt{Bailer2018} distances) against the calibrated surface gravity from RAVE ({\it cloggk}), for RAVE stars with temperatures above and below {\it cteffk}=4700 K. 

 Without a sufficiently strict J-K$_{\rm s}$ cut, the placement of any M$_{\rm J}$ luminosity cut would arbitrarily carve into a continuous distribution. From left to right in \autoref{2masscut}, the application of an increasingly severe J-K$_{\rm s}$ cut is increasingly restricting the sample to the cool side of the Hertzsprung–Russell (HR) diagram, where the luminosities of evolved giants are increasingly well separated from the main sequence. With a relatively loose J-K$_{\rm s}$>0.65 cut, a continuous extension into warm subgiants evolving off the main sequence blurs the distinction between luminosity classes. With an excessively severe J-K$_{\rm s}$>1.05 cut, the extension into warm subgiants is gone, but the completeness of the cool giants sample suffers drastically. We instead adopted a compromise J-K$_{\rm s}$>0.85 cut, which might bias our sample against K giants. Finally, we adopted a M$_{\rm J}$<0 luminosity cut; the exact placement of this luminosity cut in the range of M$_{\rm J}$=0 to 3 remains quite arbitrary, but unimportant by design at the end of the HR diagram to which J-K$_{\rm s}$>0.85 restricted us. 

Next, we performed a validation test of our M$_{\rm J}$ and J-K$_{\rm s}$ cuts with the \citet{Zhong2015} catalogs of M giants and M to late-K dwarfs and subdwarfs from the LAMOST survey, where the discretely defined temperature and luminosity classes simplify our test.\footnote{We found similar results in RAVE; for example, less than 1\% of RAVE giants that remained after the J-K$_{\rm s}$>0.85 colour cut were excluded by taking M$_{\rm J}$<0.} After imposing a cut of J<14 to keep the uncertainties small and stable, the RAVE test sample was left with 7247 M giants and 58,525 cool dwarfs/subdwarfs. Selecting a test sample with M$_{\rm J}$<0 and J-K$_{\rm s}$>0.85 yielded a completeness for the M giants of 91\% (barely changed from 92\% without the J-K$_{\rm s}$ cut), and 99.9\% of Lamost cool dwarfs did not meet our cool giant selection criteria.  It may be that some of the remaining 0.1\% of LAMOST dwarfs are misclassified giants.

As a concession to the need for sample purity, and in the absence of sufficiently high-resolution 3D reddening maps, we de-reddened using the full \citet{Schlegel1998} Galactic interstellar extinction column, which is pre-matched in the SkyMapper {\it dr2.master} table, for our (J-K$_{\rm s}$)$_{0}$>0.85 colour cut.\footnote{We adopted R$_{\rm J}$=0.723 and and R$_{\rm K}$=0.310 from \citet{Schlafly2011}. The correction to de-redden J-K$_{\rm s}$ was -0.413 multiplied by the SkyMapper DR2 {\it ebmv\_sfd} column.} By the same logic, we did {\it not} de-extinct the luminosity cut, because a substantial overcorrection could make a low-luminosity source appear to be a high-luminosity giant. These decisions could have some impact on the spatial distribution of our sample, because cool giants on highly extincted lines of sight could be excluded from the sample by excessive colour de-reddening or an insufficient luminosity extinction correction if they are located only partway through the dust column on their sight-line.

\subsection{Implementation and quality cuts}\label{qualitycuts}

The full set of criteria used to create our luminous red objects sample is listed in \autoref{criteria}, and the procedure by which we implemented these cuts is described in detail in section 2.1.2 of \citet{Lucy2021}. The cuts were performed using the Tool for Operations on Catalogues and Tables (\textsc{TOPCAT}; \citealt{topcat}), pointed to the SkyMapper Data Release 2 (DR2) Table Access Protocol (TAP) endpoint and the ARI-GAIA TAP endpoint. The SkyMapper data release was pre-matched to 2MASS (which we used to perform the infrared colour cut) and Gaia DR2; SkyMapper reports the match distance to each in dedicated columns. \citet{Bailer2018} distances were obtained later by joining to their catalog on Gaia identifiers, to perform the luminosity cut. 

Various subtler quality cuts were also implemented. We required that the 2MASS J magnitude be brighter than 14 to ensure a high signal to noise ratio (SNR) in the infrared colour selection cuts.\footnote{For a typical M giant of say M$_{\rm J} \leq$ -2.5, ignoring reddening, J < 14 would not exclude any objects out to 20 kpc, outside of the Galaxy in most directions. For a less typical M$_{\rm J}$ = 0, the threshold is about 6 kpc. Extinction and other implicit selection effects from our various survey data are likely a bigger issue for 3-D spatial completeness, both intrinsically and due to our conservative approach of dereddening J-K$_{\rm s}$ by the full Galactic dust column in the direction of the object. On all these fronts, G and K stars will tend to be the first to fall out of the sample.} We also required that each selected source have at least one u, one v, and one g band measurement identified by the SkyMapper pipeline as good---and as explained in Appendix~\ref{Snapshots}, that there be sufficient SkyMapper data to compute at least one ``nightly'' u-g colour in which the u measurement(s) were conducted on the same night as the g measurements(s), and, separately, to likewise compute at least one ``nightly'' u-v colour.

SkyMapper PSF magnitudes can deviate by greater than 1\% if a neighbor of equal brightness is located within 5 arcsec of the source, and the 2MASS confusion limit is 6 arcsec, so we required that the selected sources must be isolated from their neighbors by at least 6 arcsec in SkyMapper using the SkyMapper {\it dr2.master} column ``{\it prox}''. This proximity limit removed about 5\% of the sample, and may spatially bias our sample against crowded fields. We selected only sources for which the SkyMapper {\it flags\_psf} flag did not indicate source contamination.

We required that the 2MASS {\it ext\_key} and SkyMapper {\it class\_star} indicate a point source, and that SkyMapper never resolved the source into multiple components in any observation. We checked a variety of symbiotic stars with known or hypothesized extended nebula or jets (the Southern Crab, Mira, R Aqr, HD 149427, MWC 560, BI Cru, Hen 2-147, AS 201, DT Ser, and RS Oph) and all of them met these requirements. Intriguingly, some of them did have ``extra'' sources in the wings of their 2MASS point spread function (PSF), which is why we implemented our proximity limit exclusively in SkyMapper and not in 2MASS. 

\subsection{Object type labels and sample completeness/purity}\label{crossmatch}

We retrieved object type labels for our working catalog of luminous red objects. We crossmatched to the Set of Identifications, Measurements, and Bibliography for Astronomical Data database (SIMBAD; \citealt{simbad}) with the CDS Upload X-Match in \textsc{TOPCAT} on a 3 arcsec matching radius, to retrieve names and SIMBAD ``main\_type'' object type labels from SIMBAD. We likewise crossmatched to the Naval Observatory Merged Astrometric Dataset (NOMAD; \citealt{nomad}) on a 2 arcsec matching radius, to obtain V band measurements used to estimate exposure times for follow-up observations. Finally, we crossmatched our working catalog to the list of Galactic and extragalactic symbiotics ``confirmed'' prior to our first observing run in the \citet{Merc2019} New Online Database of Symbiotic Variables.\footnote{We used the version released on May 16, 2019. We excluded Hen 3-1768, which was first discovered through an earlier test iteration of our search, reported in \citet{Lucy2018b} and subsequently incorporated into the Merc catalog.} We used a 10 arcsec matching radius, and we manually checked the results to ensure that no spurious matches were retrieved.

\begin{table}
\sffamily
\begin{tabular}{l}
\hline
\\
STEP 1\\
\bf{Distance to pre-matched 2MASS source:}\\ 
SkyMapper twomass\_dist < 2 arcsec 
\\
\bf{Distance to pre-matched Gaia DR2 source:}\\
SkyMapper gaia\_dr2\_dist1 < 2 arcsec \\
\bf{Distance to pre-matched ALLWISE source:}\\
SkyMapper allwise\_dist < 3 arcsec\\
\bf{SkyMapper DR2 source confusion:}\\ No other source within 6 arcsec (prox > 6)\\
\bf{SkyMapper DR2 data availability:}\\ u\_ngood > 0, v\_ngood > 0, g\_ngood > 0 \\
\bf{SkyMapper DR2 quality:}\\ class\_star > 0.9, flags\_psf = 0, nch\_max = 1\\
\bf{2MASS quality:}\\ ph\_qual='AAA', gal\_contam = 0, ext\_key = NULL, cc\_flg = '000'\\
J < 14.0 \\
\bf{2MASS colour cut:}\\
(J-K$_{\rm s}$)$_{0}$ > 0.85\\
\\
STEP 2\\
\bf{Get distances:}\\ Join to \citet{Bailer2018} on Gaia ID\\Check that converged distance exists\\
\bf{Luminosity cut:}\\
M$_{\rm J}$ < 0\\
\\
STEP 3\\
\bf{SkyMapper data availability for snapshot colours:}\\
\autoref{Snapshots}: sufficient data to reconstruct $\geq$1 nightly colour snapshot\\
{$\cdot$ in both u-g and u-v} \\
{$\cdot$ from individual measurements in SkyMapper dr2.photometry} \\
{$\cdot$ with used\_in\_clip = True or False, but not null.}
\\
\\
\hline
\end{tabular}
\caption{Criteria for inclusion in our luminous red objects sample, in the order they were implemented. The ALLWISE crossmatch was used for sample validation and can be discarded in future improvements on our technique.}\label{criteria}
\end{table}

\subsection{Initial assessment of completeness and purity}\label{agn_etc}

Our luminous red objects sample included a total of 58 known symbiotics, of which 56 were Galactic. To check how many Galactic symbiotics were lost through our cuts, we crossmatched the \citet{Merc2019} catalog to SkyMapper {\it dr2.master}; this yielded 200 Galactic symbiotics ``confirmed'' in \citet{Merc2019} prior to our first observing run, but only 110 with a u-band observation, only 73 with u and v, and only 71 with u, v, and g. That is, only 15 out of 71 known Galactic symbiotics with uvg photometry were removed by the cuts described above. Of these: one was demoted out of the ``confirmed'' status in the latest \citet{Merc2019} catalog as of February 2024; one other may also be misclassified or subject to source confusion\footnote{SWIFT J171951.7-300206, formerly considered an accreting-only symbiotic and associated with IGR J17197-3010, as discussed by \citet{Luna2013}. It was eliminated from our sample of luminous red objects on the basis of a {\it Gaia}-estimated absolute magnitude of M$_J$=4.5 (and a de-reddened (J-K$_{\rm s}$)$_{0}$ colour of 0.5), with a {\it Gaia}-estimated distance of 187 pc \citep{Bailer2018}, for the nearest source in 2MASS and {\it Gaia}.}; four had a SkyMapper neighbour within 6 arcsec; three were saturated in 2MASS; and two had minor data quality issues such as a possible diffraction spike in 2MASS H-band or merely lacking sufficient data to construct a nightly snapshot colour. Of the remaining five, all are currently labeled as shell-burning in the \citet{Merc2019} catalog, and all failed at our (J-K$_{\rm s}$)$_{0}$ > 0.85 criterion. The properties of these five confirmed our pre-existing inferences about the types of symbiotics that our J-K$_{\rm s}$ cut may exclude: three lie along highly extincted sight lines that we may have over-corrected when we conservatively used the entire Galactic extinction column, and three definitively have a G or K giant donor star. More interestingly, the fifth is a rare symbiotic slow nova, V4368 Sgr, in a burning state since 1994; its sightline is not highly extincted, but it might have a late-K giant donor inferred with IR photometry \citep{Akras2019b}, which has not been confirmed spectroscopically.

Between 83\% (without de-reddening) and 93\% (after dereddening by the full Galactic extinction column) of the luminous red objects sample met the 2MASS and ALLWISE colour criteria for the selection of M giants spelled out in Equation 1 of \citet{Li2016}, ignoring objects for which ALLWISE is saturated. In contrast, only 31\% (both with or without de-reddening) of the \citet{Merc2019} Galactic symbiotics meet the \citet{Li2016} criteria, supporting our decision to avoid the multiple strict IR colour cuts that would have given us a purer sample of cool giants. 

Some active galactic nuclei (AGN) remained in our luminous red objects sample, and we found that we could identify them with ALLWISE (AGN have W1-W2>0.5; see \citealt{Li2016}) and {\it Gaia} proper motions (AGN have {\it Gaia} $\sqrt{pmra^{2}+pmdec^{2}}<0.35$ mas year$^{-1}$). However, 38 out of 45 objects identified in this way were already known in SIMBAD to be AGN or galaxies, one was a known Mira, and furthermore the known flickering symbiotic EF Aql has an AGN-like W1-W2$\approx$0.9 colour\footnote{EF Aql was, in fact, discovered in part on the basis of a W1-W2 colour cut to select AGN in the UV-bright Quasar Survey of \citet{Monroe2016}---the survey which, as discussed in \autoref{introflickering}, yielded EF Aql's discovery as a symbiotic in \citet{Margon2016}.} and a borderline proper motion of 0.48 mas year$^{-1}$. So to avoid cutting out EF Aql-like symbiotics, we chose not to alter our selection function to exclude AGN, but rather to simply not observe any object already identified as an AGN by SIMBAD.

\begin{figure}
\centerline{\includegraphics[width=\columnwidth]{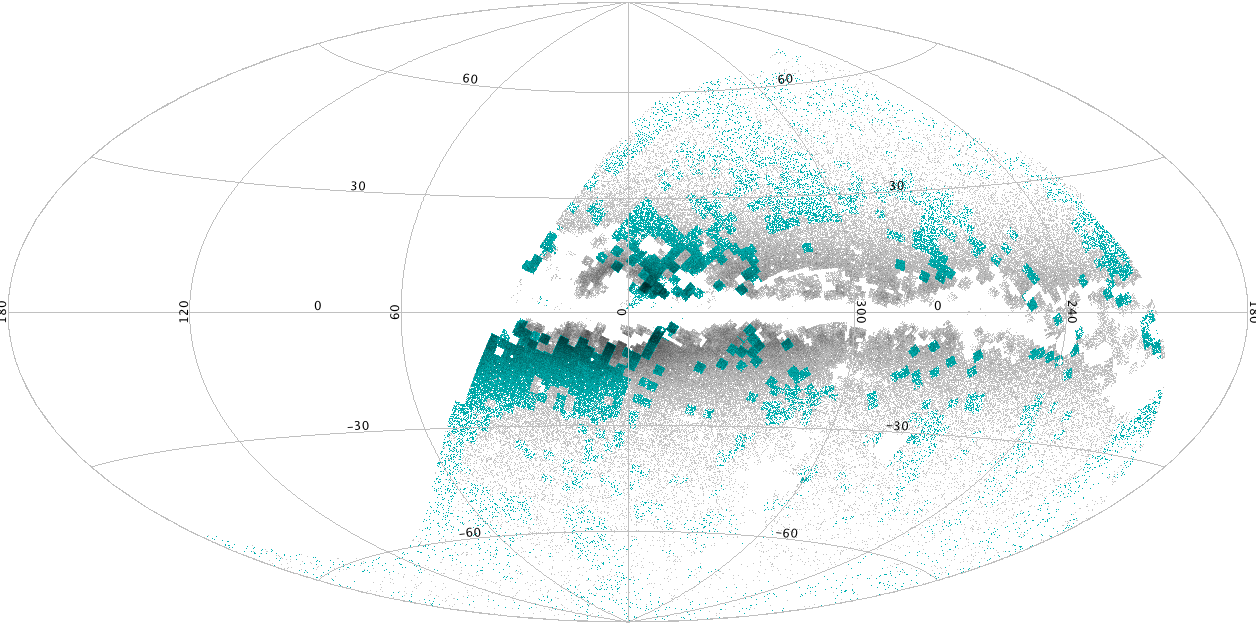}}
\caption{Density-shaded scatter plot of our 366,721 luminous red objects sample in Galactic coordinates (grey, corresponding to the objects in the right panel of \autoref{parameterspace}), with an overlay of the portion of the sample for which good short-timescale u-band variability information was available within at least one SkyMapper Main Survey filter sequence (blue, corresponding to the objects in the middle panel of \autoref{parameterspace}).}\label{map}
\end{figure}

Our selection criteria likely introduce some spatial biases, but as shown in \autoref{map}, the on-sky spatial distribution of our final luminous red objects sample is dominated by the sky coverage of the SkyMapper Southern Sky Survey. The most notable issue is that SkyMapper DR2 does not, for the most part, cover low Galactic latitudes in the u or v bands, leading to the exclusion of the majority of known symbiotics. The present work can be extended further into the Galactic plane with future SkyMapper data releases.

\section{SkyMapper parameter space}\label{theskymapperparameterspace}

Our analysis of our 366,721 luminous red objects in SkyMapper DR2 parameter space included generating a reconstructed colour-colour diagram (\autoref{color}) and calculating measurements of rapid variability (\autoref{rapid}). Even before conducting our spectroscopic survey and follow-up observations, the locations of known objects in this parameter space already made it clear that SkyMapper is a potent tool for identifying symbiotic stars, carbon giants, and s-process-enhanced giants. Below, we describe our development of a scheme to separate different types of objects in SkyMapper colour-colour and variability space. In this scheme, known symbiotic stars tend to be outliers in colour-colour and SkyMapper u-band rapid variability.

\subsection{Separation of object types in SkyMapper colour-colour space}\label{color}

We used the SkyMapper colours u-g and u-v to explore SkyMapper colour-colour parameter space. Based on the physics of symbiotic binaries, we expected that the symbiotic hot component would produce an excess in SkyMapper u, the shortest-wavelength filter, relative to redder bands. Of all the SkyMapper colours, u-g turned out to be the most effective at identifying symbiotic stars as colour outliers. After choosing u-g, the second-most effective colour at isolating symbiotic stars was u-v; it acted on our sample in a way that was, out of all the remaining colours, most orthogonal to u-g. A two-dimensional principal component analysis (PCA) of u-based colours confirmed these impressions, loading u-g, u-r, u-i, and u-z onto the same eigenvector with similar loading scores, with the other eigenvector dominated by u-v. We also aimed to use as few filters as possible, because requiring good data in additional filters would have led to the exclusion of more objects from the sample.

We did not correct our SkyMapper colours for extinction. An attempt to do so using a 3D dust map (from \citealt{Bovy2016}) had the counterproductive effect of narrowing the separation between known symbiotics and the rest of our luminous red objects. This abandoned effort, and a variety of possible explanations for its failure, are described in section 2.1.3.3 of \citet{Lucy2021}.

Finally, we found that it was necessary to re-compute SkyMapper colours using an average of ``nightly snapshot'' colours derived from the full table of individual photometry measurements. Magnitudes in the {\it dr2.master} catalog are strongly affected by variability in large-amplitude pulsating stars like Mira variables, inconsistently between the different filters, such that colours computed from the {\it dr2.master} catalog for such stars are often off by several magnitudes. Throughout this paper, ``u-g'' and ``u-v'' refer instead to colours reconstructed to ensure that both of the filters making up a colour were observed on the same nights as each other. The details of this problem, and our solution, are described in Appendix~\ref{Snapshots}.

\begin{figure*}
\centerline{\includegraphics[width=\textwidth]{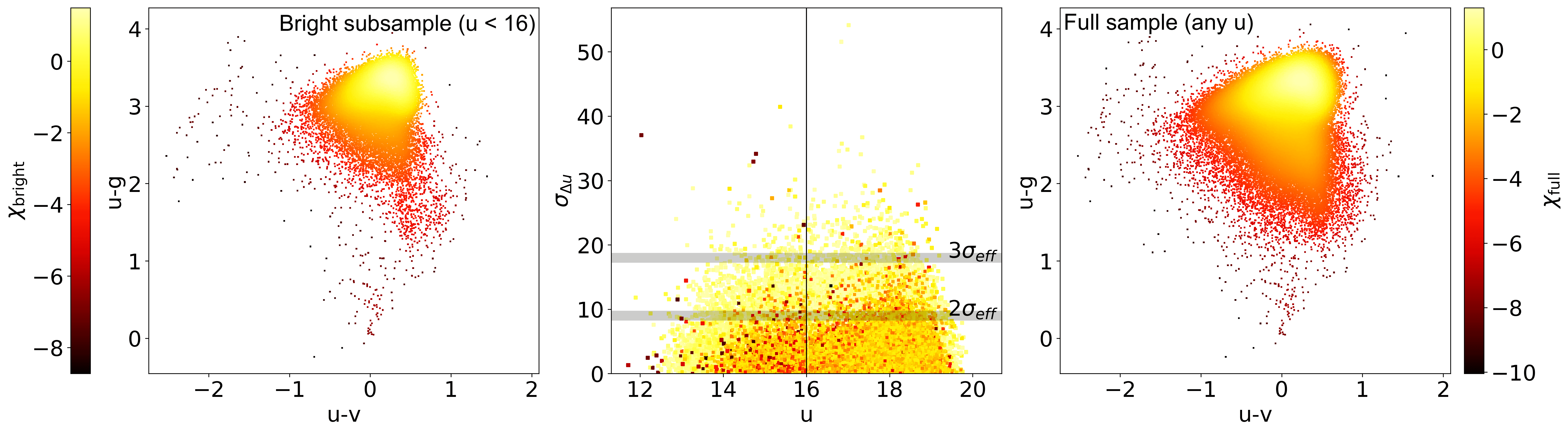}}
\caption{Scatter plots of our luminous red objects sample in SkyMapper parameter space, from which outlying targets were selected. \textbf{Left panel}: The u<16 bright subsample (97,957 objects) in our reconstructed SkyMapper u-g and u-v colour-colour diagram. \textbf{Right panel}: the same for the full sample (366,721 objects) with any u magnitude. \textbf{Middle panel}: The statistical significance of rapid u-band variability within a filter sequence for each object with a good filter sequence, translating to magnitude-independent effective significance thresholds of $\sigma_{\rm eff}$=2 at statistical $\sigu$=9, $\sigma_{\rm eff}$=3 at statistical $\sigu$=18, so that about 95\% of the sample lies below the 2$\sigma_{\rm eff}$ line and about 99.7\% of the sample lies below the 3$\sigma_{\rm eff}$ line in any given u magnitude bin. There are 25,619 objects in the left half of the middle panel, and 146,577 objects in the entire middle panel. \textbf{Colourbars}: Points in all three panels, including the middle one, are mapped to their colour-colour KDE score  $\chi$ delineated in the colourbars, which go from light in the densest region of u-g/u-v parameter space to black for u-g/u-v outliers. At the u=16 vertical line in the middle panel, the mapping is switched from the bright subsample's KDE score $\chi_{\rm bright}$ (left colourbar) to the full sample's KDE score $\chi_{\rm full}$ (right colourbar). In the middle panel, to ensure that colour-colour outliers are noticeable, objects with high-outlier $\chi$ are overplotted after first plotting the objects with low-outlier $\chi$.} \label{parameterspace}
\end{figure*}

The finalized SkyMapper u-g/u-v colour-colour diagram for our sample of luminous red objects is shown in the left (u<16 mag: bright subsample) and right (any u mag: full sample) panels of \autoref{parameterspace}. In the left panel, the concentration of sources into a ``tail'' feature in the lower-right region (a useful distribution pattern that will help to distinguish known symbiotics in the lower-density region to the tail's left) has been preserved by limiting the sample to u<16. In the right panel, the distinction between the tail and the lower-density region to its left vanishes. With some experimentation, we found that the transition over which the distinctiveness of the tail vanishes is in the u=16--17 range. Accordingly, in this plot and in all other plots based on it, we plot the bright subsample of our luminous red objects with SkyMapper u<16 mag in the left panel, and the full sample of our luminous red objects with any u magnitude (including the bright subsample) in the right panel. The median statistical uncertainty on each colour is about 0.02 mag, with tails out to 0.2 mag or so\footnote{There are about 1600 objects with uncertainties larger than 0.1 mag in the right panel, and about 100 objects with uncertainties larger than 0.1 mag in the left panel, for both u-g and u-v.}. For convenience, we calculated colour outlier scores ($\chi$) for each object, calculated with the Kernel Density Estimation (KDE) function {\it sklearn.neighbors.KernelDensity} \citep{sklearn}. Scores were calculated separately in the left ($\chi_{\rm bright}$) and right ($\chi_{\rm full}$) panels. We used a Gaussian kernel of bandwidth 0.1 mag, chosen to be just small enough to detect subtle features in the colour-colour distribution that could guide our symbiotic candidate targeting strategy.

\begin{figure*}
\centerline{\includegraphics[width=\textwidth]{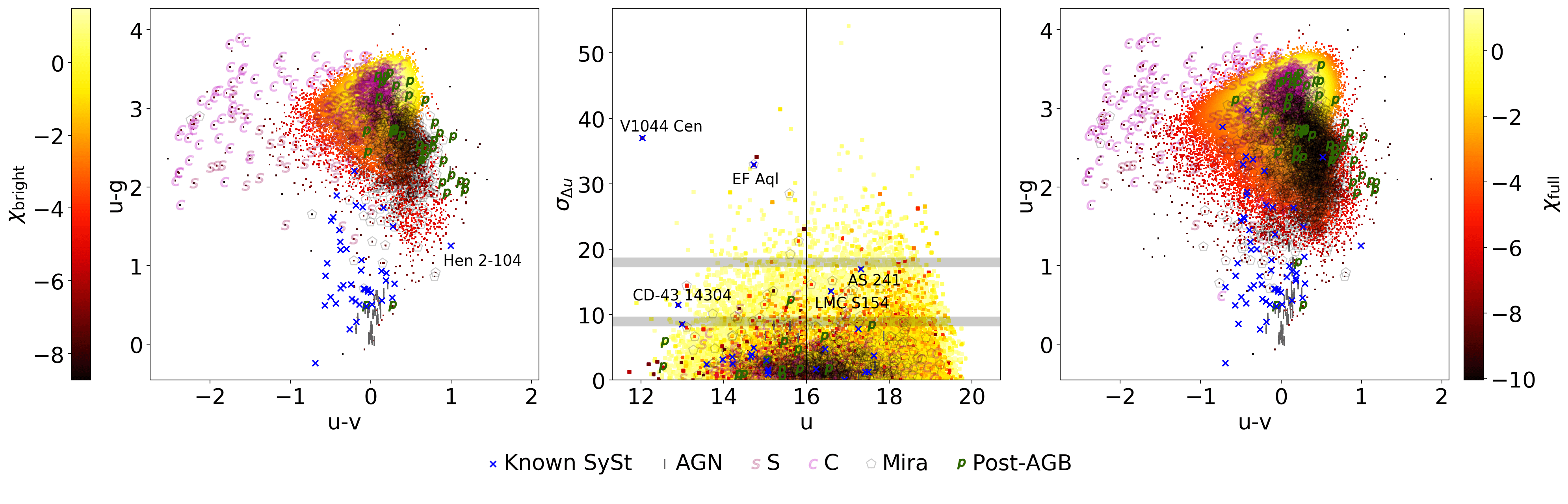}}
\caption{Our sample of luminous red objects crossmatched to external catalogs, including objects with ``main type'' SIMBAD \citep{simbad} classification as an S star (S symbols), carbon star (C symbols), Mira (open pentagons), or post-AGB (P symbols; including SIMBAD labels post-AGB*, PulsV*RVTau, and PulsV*WVir under the post-AGB umbrella), and symbiotics (SySt; blue crosses) from the New Online Database of Symbiotic Variables \citep{Merc2019}. We identified AGN with W1-W2>0.5 and {\it Gaia} proper motion < 0.35 mas year$^{-1}$ criteria, and the vast majority also have a variety of AGN or galaxy labels in SIMBAD. A few notable symbiotics are labelled individually. Here and in all future overlay plots on \autoref{parameterspace}, every object plotted is in our luminous red objects sample, only objects with u<16 are plotted in the left panel, and the point centered in a hollow symbol is the same object represented by the hollow symbol.} \label{simbadlabels}
\end{figure*}

In \autoref{simbadlabels}, we display the results of crossmatches (\autoref{crossmatch}) between our luminous red objects sample and external catalogs of known objects. Most importantly, we see that the vast majority of known symbiotics in our luminous red objects sample are very well isolated by u-g and u-v colours in the bright sub-sample, albeit somewhat less so in the full sample. In the leftward direction of u excess in u-v, we find a substantial fraction (but not all) of the carbon stars and s-process-enhanced stars (S stars) in the sample; in these giants, s-process and carbon products have been dredged up on the thermally-pulsing AGB (either in the current giant, or accreted into the current giant from a more evolved companion) leading to different molecular banding. Meanwhile, Mira variables are well localized, with a slight extension into the lower-right tail of the scatter plot. Most AGN are located just down and to the right of the concentration of symbiotics. A small number of post-AGB stars (including those labelled as post-AGBs, RV Tau, or W Vir) are slight outliers in a long line along the upper-right of the scatter plot. These are the largest SIMBAD categories by far in the luminous red objects sample, and we did not notice any other patterns of interest.

\subsection{SkyMapper u-band flickering}\label{rapid}

We used SkyMapper u to detect minutes-timescale flickering. Each of the SkyMapper Main Survey visits (of which there are typically one or two) to each field comprises a sequence of exposures patterned as {\it uvgruvizuv}, typically completed in 20 minutes, with 100-second exposures in each filter and a median 21-second overhead per exposure \citep{skymapperdr2}. There were thus three 100-second u-band exposures performed with a cadence of typically 8 minutes within a sequence. The same is true for v-band, but we focused on u because it inevitably had higher SNR (due to its broader throughput) at even shorter wavelengths, where the accretion disc was likely to have a greater contribution relative to the giant. Our goal was to see if variability between these three 8-minute-cadence exposures within a filter sequence could be sufficient to detect accretion disc flickering.

After testing a variety of metrics, we settled on a simple one that made the known optically-flickering symbiotic EF Aql \citep{Zamanov2017} stand out: within each Main Survey filter sequence, we calculated

\begin{equation}
\sigu
= {\rm max}_{i,j} \left(\frac{|{\rm u_{i} - u_{j}}|}{\sqrt{\rm (e\_mag\_psf)_{i}^{2} + (e\_mag\_psf)_{j}^{2}}}\right), \label{equation1}
\end{equation}

\noindent where ${\rm max}_{i,j}$ is defined as the maximum value for any i and j in the two or three good u-band PSF magnitude measurements within the Main Survey filter sequence, using the {\it e\_mag\_psf} column of the {\it dr2.photometry} table as the uncertainty for each measurement (without incorporating any systematic error floor, because incorporating a floor made EF Aql stand out less). To enforce a minimum data quality, we only used measurements with {\it used\_in\_clipped} = True or False, not null, equivalent to excluding measurements with source extractor flags$\geq$4 and {\it nimaflags}$\geq$5. We excluded Shallow Survey measurements. A few objects and nights had unusual sequence patterns with Main Survey u-band measurements separated by up to 4 hours, but we found that these cases had minimal effect on the statistical distribution of the sample and our selection thresholds, so we allowed them to be incorporated. Some Main Survey sequences only had two u-band measurements, but again we found that the statistical distribution was not significantly affected by incorporating sequences with either two or three u measurements indiscriminately.  

There were several hundred objects that had more than three Main Survey u-band measurements within the selected night (mostly with four measurements), and these tended to yield a much higher value from \autoref{equation1}, so we excluded them by only including objects that had only two or three Main Survey u measurements per night, in order to ensure that we were comparing apples to apples.\footnote{The extra measurement sometimes only added 8 minutes to the time spanned by a night's u-band observation, but other times the measurement occurred up to over 80 minutes later. This might be an opportunity for a variety of differently-timescaled variability, both astrophysical and instrumental/weather-related, to arise. As discussed later, systematic uncertainties overwhelm statistical uncertainties in these short-term variability data generally. We infer that the systematic uncertainties on longer timescales may be much different than those on 8 to 20 minute timescales.} Where two or more nights of Main Survey filter sequences were obtained (i.e., objects with two or more filter sequences), we assigned that object the night that gave it its maximum value of \autoref{equation1}, which we found did not significantly affect the statistical distribution of the sample. In future data releases, where eventually almost 100\% of objects will have two nights of main survey data, information from both nights should be incorporated more carefully.

The resultant value of $\sigu$ from \autoref{equation1} is plotted in the middle panels of \autoref{parameterspace} and \autoref{simbadlabels} against the {\it dr2.master} u magnitude for each of our luminous red objects, colour-coded by the KDE colour outlier score. There are 25,619 objects in the u<16 bright subsample with a $\sigu$ value, and 146,577 objects in the full sample with a $\sigu$ value. The spatial distribution of these objects is shown with an overlay in \autoref{map}, illustrating the sky regions over which usable Main Survey filter sequences were available.

Although $\sigu$ (defined in \autoref{equation1}) is nominally the statistical significance of the maximum variability between any two u exposures within a filter sequence, it was clearly far too high for far too many objects. We binned the data from the middle panel of \autoref{parameterspace} into 0.5 mag bins, and calculated the value of $\sigu$ below which 99.73\% and 95.45\% of the objects in each bin lay. We found that these percentile thresholds were very close to flat as a function of u magnitude, suggesting that the distribution of $\sigu$ values is dominated by systematic errors. Because these percentile thresholds were nearly flat as a function of brightness, we defined flat 3$\sigma_{\rm eff}$ and 2$\sigma_{\rm eff}$ thresholds at the approximate percentile levels of 99.73\% (3$\sigma_{\rm eff}$ at $\sigu$ = 18) and 95.45\% (2$\sigma_{\rm eff}$ at $\sigu$ = 9). These are the horizontal lines in the middle panels of our parameter space plots.

\section{Target selection for the optical spectroscopy survey}\label{targeting}

To confirm the presence of a cool giant and search for emission lines indicative of symbiotic binarity in SkyMapper outliers from the distribution of luminous red objects, we set out to obtain optical spectroscopy for objects within a set of regions in parameter space. Our two main goals were to (1) determine the boundaries within which symbiotic binaries of any type can be found in SkyMapper parameter space, and (2) develop techniques to find optically-flickering, accreting-only symbiotics. Optical spectroscopy constituted an intermediary step to the latter goal, because the length of fast-cadence optical light curves and the depth of X-ray exposures needed to detect accretion discs would be prohibitively resource-intensive without at least some independent confirmation of the presence of a cool giant and a hint of emission line properties not seen in isolated cool giants.

To those ends, and informed by the locations of known symbiotics in parameter space, we defined four categories of target selection for optical spectroscopy. The depth we explored into the sample distribution in each category was determined by both the amount of available observing time with good weather and on-the-fly re-prioritization between the different categories enumerated below over the course of the observing runs. Constrained by the timing of our observation runs, we targeted objects with right ascensions between 10--24 and 0--2 hours, towards the Galactic Center, excluding the 2--10 hour range and the Galactic anti-center where star densities are lower. In \autoref{targets}, we show the target selection category for each observed target in SkyMapper colour-colour parameter space. The target selection categories are as follows:

\begin{figure*}
\centerline{\includegraphics[width=\textwidth]{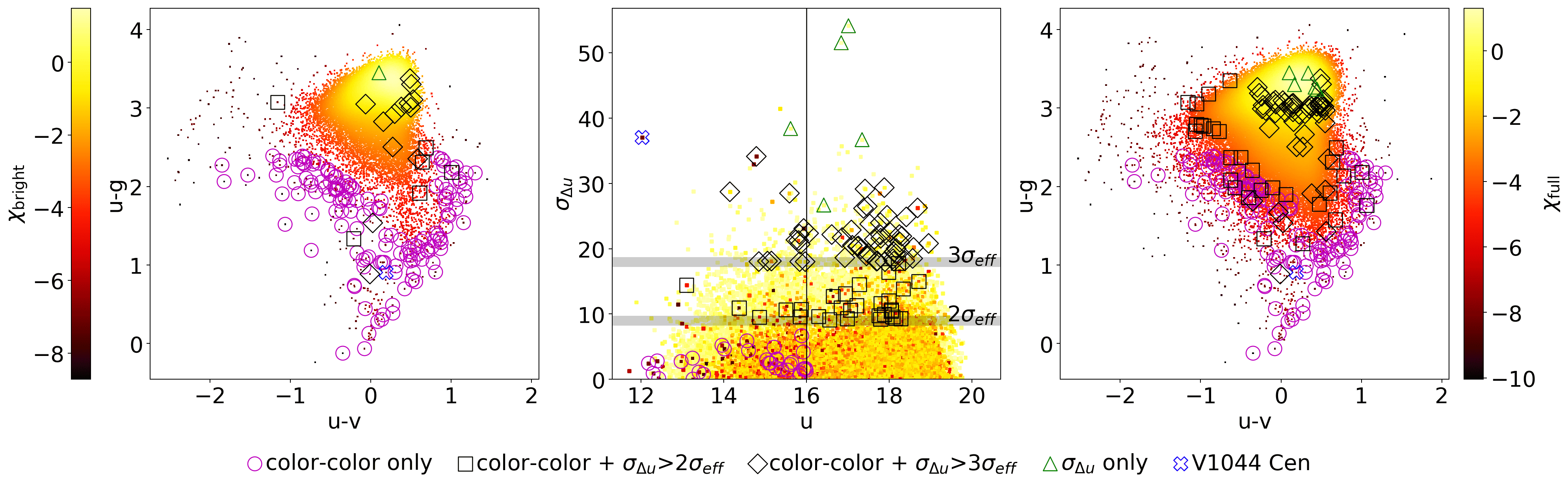}}
\caption{Selection criteria of targets for which we obtained optical spectroscopy, overlaid on \autoref{parameterspace}. The targeted zones, in order of how far they venture into the densest region of the colour-colour distribution, are: the outskirts of the bright sample with u-g $\le 2.4$ and colour score $\chi_{\rm bright} \le -5.3$ from the left colourbar (146 magenta circles), variability measure $\sigu \ge 2\sigma_{\rm eff}$ in the middle panel and colour score $\chi_{\rm full} \le -3.5$ from the right colourbar (29 black squares), variability of $ \sigu \ge 3\sigma_{\rm eff}$ in the middle panel and colour score $\chi_{\rm full} \le 0.2$ on the right colourbar (53 black diamonds), and variability  $\sigu \ge 26$ without colour criteria (5 green triangles). We also observed the known symbiotic V1044 Cen (blue hollow cross).}\label{targets}
\end{figure*}

\begin{enumerate}[align=left, leftmargin=*, rightmargin=0ex, labelsep=0ex]\itemsep0.3em
    \item We selected 146 targets (plus an additional three that also qualified under the categories below) by u-g/u-v colour-colour alone in the u<16 bright subsample, exploring the outskirts of the colour-colour distribution with u-g $\le$ 2.4 and a KDE colour score of $\chi_{\rm bright} \le -5.3$. Our intent here was not to maximize the number of symbiotics discovered. Rather, our intention was to fully explore the blue outskirts of the parameter space we had built, to determine where future searches should be focused. We had in mind that accreting-only flickering symbiotics, symbiotics with S star donors, or symbiotics with unusual nebulosity like the Southern Crab (Hen 2-104, labelled in the left panel of \autoref{simbadlabels} and notably separated from the concentration of symbiotics), might be located in different regions of parameter space from most known symbiotics. We also had in mind that it could be useful to have a more complete set of outlier object type labels for supervised machine learning in the future.
    \item We selected 53 targets on the basis of having both $\sigu >3\sigma_{\rm eff}$ and KDE colour score $\chi_{\rm full} \le 0.2$ in the full sample. Our intent here was to test the $\sigu$ metric as a flickering detector, while increasing the likelihood of successfully finding new symbiotics by prioritizing targets with at least some small blue excess relative to the mode of the distribution. Two of these targets would also have qualified with only the u<16 colour-colour criterion above.
    \item We selected 29 targets on the basis of having both $2\sigma_{\rm eff} < \sigu < 3\sigma_{\rm eff}$ and KDE colour score $\chi_{\rm full} \le -3.5$ in the full sample. Our intent here was to see if a moderately significant excess in $\sigu$ paired with colour would allow us to detect flickering symbiotics deeper into the colour-colour distribution than with colour-colour alone. One of these targets would also have qualified with only the u<16 colour-colour criterion.
    \item We selected 5 targets to have as high as possible $\sigu$, without considering colour, to test if $\sigu$ could be used without colour criteria to successfully detect flickering symbiotics.
    \item Interestingly, while we intentionally chose $\sigu$ as defined in
    \autoref{equation1} 
    to make the flickering symbiotic EF Aql stand out, we discovered that another known symbiotic, V1044 Cen (CD-36 8436), also has $\sigu$ well above 3$\sigma_{\rm eff}$, as shown in \autoref{simbadlabels}. So we added it to our list of spectroscopic (and fast-photometric; see \autoref{Flickering}) targets.
\end{enumerate}

 Otherwise, we excluded all known symbiotics, as well as AGN or galaxies identified as such in SIMBAD, from our spectroscopic target lists. We ignored other SIMBAD labels, on the premise that they could be wrong---and that if they were right, having a spectrum could be useful for identifying other contaminants.

\section{New observations and data reduction}\label{Observations}

To explore the parameter spaces which we constructed in \autoref{theskymapperparameterspace} from archival SkyMapper photometry, we performed our own observations in a layered observing strategy. We conducted an optical spectroscopy survey of the 234 targets selected from SkyMapper photometry, followed by more concentrated follow-up observations of select targets from that set: optical fast photometry of 11 targets, shallow {\it Swift} X-ray observations of 9 targets, and deep {\it Chandra} X-ray observations of 2 targets.

\subsection{Optical spectroscopy}

We obtained optical spectra of the 234 objects selected in \autoref{targeting}. Of these targets, 233 were observed at the South African Astronomical Observatory (SAAO) with the 1.9-meter telescope, using SpUpNIC (Spectrograph Upgrade: Newly Improved Cassegrain; \citealt{Crause2019}). We used the G7 grating, which has a blaze at 5000\AA, a range of 5550\AA, and a plate scale of 2.7 \AA/pixel. The spectral resolution was typically 11 to 12~\AA. 

The majority of spectra, including for all but one of the targets in which we detected emission lines, were obtained in the dome on 2019 June 19 -- July 2 by a single observer (ABL), with about five nights of good weather in varying non-photometric conditions. These observations were supplemented by a smaller set obtained remotely from the Cape Town control room on 2020 May 21--23, 27 and June 24--25, 30 by multiple observers on shared time. We manually inserted a BG38 filter into the arc beam at the start of the 2019 run; we believe that the same was done by staff at the start of the 2020 run. We used a CuAr arc lamp exposure after each spectrum, while the telescope was still pointed at the source, to track the shifts in the wavelength solution caused by frequent fast slews. We took at least four 90-second exposures per night of the standard star CD-32 9927 in the 2019 run, and a few exposures of the same standard on some nights of the 2020 run. We prioritized a time-economical observing pattern instead of consistently observing at the zenith, and it was not possible to adjust the slit angle to match the parallactic angle.

We employed test exposures in an attempt to achieve a final signal-to-noise ratio (SNR) of around 100 per pixel in the pseudo-continuum around H$\alpha$. In some cases it was impossible to achieve that without saturating emission lines or the red end of the continuum, in which case we usually took an additional shorter exposure. Any saturated features in spectra presented in this work are noted when they appear. The final exposure times ranged from 10 to 1800 seconds. Individual exposure durations, timestamps, and airmasses are included in the spectral atlas at the end of Chapter 3 in \citet{Lucy2021}.

The SpUpNIC spectra were reduced following standard procedures in the Image Reduction and Analysis Facility (IRAF; \citealt{IRAF}). We subtracted the bias scaled by the overscan region, and flat-fielded with dome flats from the 2019 run. We observed some curious features in the dome flats, but assessed that they have no impact on our science case; additional details are available in \citet{Lucy2021}. SAAO spectra headers are slightly non-standard, so we created header values for local sidereal time and epoch with the {\it asthedit} task, and manually ran the {\it setjd} and {\it setairmass} tasks. Aperture extractions were performed with automatic resizing of the aperture limits to 5\% of peak flux, with step=30, nsum=30, clean=no, and weights=no. Each aperture trace was inspected, and background regions manually adjusted for each target to be as free of stars as possible. 

Dispersion corrections to the SpUpNIC spectra were performed in {\it doslit} with the lamp arc obtained immediately after each science spectrum, and with the default IRAF CuAr line list slightly shortened to account for the low spectral resolution. The wavelength solution was typically well fit by a three-component cubic spline. The wavelength solution is an extrapolation at wavelengths shorter 3850\AA, where we were not able to identify any arc lines. Science spectra with Balmer emission lines shortward of 3850\AA\, (e.g., Mira variables) indicate that the wavelength solution was typically excellent throughout the Balmer series.

We performed relative flux calibration of the SpUpNIC spectra with the standard star CD-32 9927 and the built-in IRAF model for that star ({\it ctionewcal cd32}), using standard-star exposures on the same night when possible, or the closest available night when not possible for some 2020 nights. We corrected for atmospheric extinction with the CTIO extinction file bundled into IRAF. 

A single spectrum obtained 2018 March 19 for a target falling in the SkyMapper colour-colour selection region was available from a test run we conducted at Complejo Astronómico El Leoncito (CASLEO) with the REOSC spectrograph (R$\sim$700). We attempted relative flux calibration of the CASLEO spectrum using a single spectrum of the standard star HR 4468 and a built-in IRAF model for that star ({\it bstdscal hr4468}), and corrected for atmospheric extinction with the CTIO extinction file bundled into IRAF. Flat-fielding was not performed in this case. We have low confidence in the relative flux calibration for this target (IRAS 15220-6952). Only one other target in our selection region (the symbiotic Hen 3-1768, which we first reported in \citealt{Lucy2018b}) was also observed at CASLEO, but we obtained a second spectrum of that source with SpUpNIC.

\subsection{Optical fast photometry}\label{photreduction}

We conducted 68 hours of time-series optical photometry with the Las Cumbres Observatory (LCO; \citealt{lcogt}) in search of optical flickering in 11 targets meeting our spectroscopic criteria for SkyMapper-discovered symbiotics (SkySy) or symbiotic candidates (SkySyC),\footnote{We describe our classification scheme for these internal designations in \protect\autoref{Definitions}. Internal designations are mapped to SIMBAD-recognized names in \protect\autoref{tableskysy}.} and the known symbiotic V1044 Cen. We used the Sinistro cameras on the global network of 1-meter telescopes to create 3--5 hour light curves, with a time resolution ranging from 37 seconds to 2 minutes, for the 11 targets identified in \autoref{lcogttable}. These data are publicly available, as proposal LCO2020B-009 in semester 2020B, from the LCO Science Archive. Sinistro images are ideal for our purposes: with their large, 26.5 by 26.5 arcmin field of view, 0.389 arcsec pixel scale, and relatively short overhead time (28 seconds per image), they provide the abundance of check stars, spatial resolution, and high cadence that allowed us to determine whether stochastic flickering was real or the result of systematics.

\begin{table*}
\begin{threeparttable}
\begin{tabular}{llccccccc}
\hline
Internal target name & Filter & Start time & Obs.\tnote{*} & Exp.\tnote{$\dagger$} & Length & Comparison star\tnote{+} & Primary check star\tnote{+} \\
 &  & UT & & (s) & (hr) & & \\
\hline
SkySy 1-2 & B & 2020-08-10 17:15:28 & CPT & 32 & 3.8 & TYC 8294-2276-1 & 230.2085 -45.3253 \\
" & U & 2020-08-10 18:55:22 & " & 92 & 1.0 & " & " \\
SkySy 1-4 & B & 2020-07-25 17:34:30 & " & 57 & 4.9 & 246.9642 -29.2877 & 246.9389 -29.2620 \\
SkySy 1-6 & B & 2020-09-08 17:18:07 & " & 80 & 3.4 & 256.8155 -56.9065 & 256.7938 -56.8977 \\
" & " & 2020-09-09 17:47:44 & " & 80 & 3.1 & " & " \\
SkySy 1-10 & B & 2020-07-24 20:30:33 & " & 32 & 4.8 & 279.4726 -20.7615 & 279.3948 -20.9112 \\
SkySy 1-11 & B & 2020-07-27 20:43:00 & " & 32 & 3.3 & 286.5007 -21.1505 & 286.5049 -21.1303 \\
V1044 Cen & B & 2021-01-18 05:45:45 & LSC & 9 & 2.9 & TYC 7275-1680-1 & TYC 7275-1145-1 \\
" & U & 2021-01-18 06:03:45 & " & 32 & 2.8 & " & " \\
" & B & 2021-01-29 23:15:34 & CPT & 9 & 3.2 & " & " \\
\\
SkySyC 1-1 & B & 2021-01-23 05:40:31 & LSC & 32 & 3.0 & 201.3881 -58.9061 & 201.4832 -58.9689 \\
SkySyC 1-5 & B & 2020-07-19 17:25:12 & CPT & 32 & 4.8 & 264.2552 -18.1159 & 264.2662 -18.0856 \\
SkySyC 1-6 & B & 2020-07-24 20:24:45 & " & 32 & 2.9 & TYC 6857-560-1 & 276.7064 -24.2478 \\
SkySyC 1-8 & B & 2020-09-03 17:45:37 & " & 32 & 4.8 & TYC 6860-673-1 & 284.0394 -23.4420 \\
" & U & 2020-09-03 18:44:23 & " & 112 & 1.0 & "& " \\
SkySyC 1-10 & B & 2020-08-07 18:55:36 & " & 57 & 4.9 & 286.3946 -6.0462 & 286.4091 -6.0470 \\
\hline
\end{tabular}
\begin{tablenotes}
\caption{ Metadata for the LCO light curves. \label{lcogttable}}
\item[*]  CPT is the SAAO site in the LCO network, and LSC is the Cerro Tololo Interamerican Observatory site in the LCO network. Both sites have three 1-meter telescopes each, in separate domes, although some telescopes were inoperable at the time of this work. Simultaneous U-band observations were all collected from a neighboring dome at the same site as their corresponding B-band observation.
\item[$\dagger$]  The exposure times are tabulated here. The cadence can be calculated by adding the tabulated values to an overhead of about 28 seconds per exposure.
\item[+]  Comparison and primary check stars are given as SIMBAD names where available, otherwise degrees R.A. decl. in J2000.\\~\\~\\~\\~\\~\\~
\end{tablenotes}
\end{threeparttable}
\end{table*}

\begin{table*}
\begin{threeparttable}
\begin{tabular}{lccccccr}
\hline
Internal target name & Total XRT exp. time & First start date & Last start date & ObsID count & First ObsID & Last ObsID & XRT detection? \\
& ks & UT & UT & & & \\
\hline
SkySy 1-2 & 18.0 & 2019-09-21 & 2020-05-18 & 6 & 00011559001 & 00011559006 & yes \\
SkySy 1-4\tnote{*} & 20.1 & 2019-10-05 & 2020-05-19 & 6 & 00012021001 & 00012021007 & no \\
SkySy 1-6 & 1.0 & 2019-10-16 & 2019-10-21 & 2 & 00012050001 & 00012050002 & no \\
SkySy 1-11 & 9.6 & 2020-06-30 & 2020-07-10 & 5 & 00013581001 & 00013581005 & no \\
\\
SkySyC 1-1 & 10.3 & 2020-03-31 & 2020-06-10 & 6 & 00013303002 & 00013303007 & yes?  \\
SkySyC 1-4 & 0.9 & 2020-01-08 & - & 1 & 00013047001 & - & no \\
SkySyC 1-8 & 6.0 & 2020-08-15 & 2020-08-23 & 2 & 00013639001 & 00013639002 & no \\
SkySyC 1-9 & 1.0 & 2020-08-07 & - & 1 & 00013629001 & - & no \\
SkySyC 1-10 & 0.8 & 2020-02-28 & 2020-03-04 & 2 & 00012171002 & 00012171003 & no \\
\hline
\end{tabular}
\begin{tablenotes}
\caption{ Metadata for the {\it Swift} XRT observations. \label{xrtobstable}}
\item[*]  SkySy 1-4 was incidentally in the footprint of a GRB observation in 2012 with an additional 21 ks of XRT exposure time (GRB120327a). But the count rate of the source, later detected in our {\it Chandra} follow-up, was below the background count rate in {\it Swift}.\\~\\~
\end{tablenotes}
\end{threeparttable}
\end{table*}

With these observations, we created B-band LCO light curves for seven targets that exhibited $\sigma_{\Delta u}$ of at least $2\sigma_{\rm eff}$ within a SkyMapper filter sequence (SkySy 1-2, 1-4, 1-6, 1-11; SkySyC 1-8, 1-10; and V1044 Cen) and four additional targets for which such variability data was not available in SkyMapper (SkySy 1-10; SkySyC 1-1, 1-5, and 1-6). For SkySy 1-2, SkySyC 1-8, and one of our V1044 Cen observations, we were also able to obtain some U-band data simultaneous to the B-band data. Observational metadata for our light curves are tabulated in \autoref{lcogttable}. Queue-scheduled observations with extremely poor data quality due to bad weather have been omitted from this table and our reported results.

We utilized the reductions performed automatically by the \textsc{BANZAI} pipeline maintained by the LCO \citep{BANZAI}. After image processing, the \textsc{BANZAI} pipeline performs source extraction using \textsc{SEP} \citep{SEP}, with outputs including the integrated flux for each source in the image from an adaptively scaled Kron ellipse aperture (equivalent to Source Extractor's \textsc{flux\_auto}) and a bitmap data quality flag for each source in the image.

\subsubsection{Validation checks}\label{photvalidation}

~

For each target, we performed the following steps to obtain light curves for the target, a primary check star, and at least 40 secondary check stars: 

\begin{enumerate}[align=left,leftmargin=*,rightmargin=0ex,labelsep=0ex]\itemsep0.3em

\item Using VizieR \citep{simbad}, we identified a bright comparison star with similar optical colours (and, where possible, a red spectral type), and a primary check star with similar flux and similar optical colours. These are listed in \autoref{lcogttable}.

\item In the first image for each light curve, we determined the coordinates of around 50 or so secondary check stars with \textsc{SEP} fluxes closest to the target's \textsc{SEP} flux, so that at least 40 secondary check stars would be left after quality cuts.

\item We created light curves of the \textsc{SEP} flux for the target and each of the check stars using a cross-matching\footnote{We subsequently checked our results with much smaller cross-matching radii, yielding similar results.} radius of 2 arcsec to capture the \textsc{SEP} catalog index for a given source in each image, dividing each measurement by the flux in that image of the bright comparison star chosen in the previous step.
In this process, we discarded any images in which the target was lost by the \textsc{SEP} source detection algorithm. If a check star was lost in one image where the target was not lost, if more than 10\% of a check star's light curve was flagged by the pipeline as having data quality issues (anything other than flag=0), or if a check star ever came within 100 pixels of the edges of the image, we discarded that check star.

\item We normalized the resultant light curves (of flux divided by the comparison star's flux) to each light curve's median.

\item For each target, along with its comparison star, primary check star, and secondary check stars, we inspected and removed any data points flagged with data quality issues by \textsc{SEP}. In each light curve, we also clipped up to one (two) manually-determined outlying non-flagged suspicious measurement per source, isolated from neighboring measurements by at least 5\% (50\%) of the light curve's median flux.

\item In addition to the light curves, we output the time-dependent curves of the diagnostics output by \textsc{SEP} within the \textsc{BANZAI} pipeline for manual inspection, including for each source the sky and image coordinates, fixed-aperture extractions of 1 and 6 arcsec radius, the subtracted background, the full-width-half-max (FWHM) and major and minor axes, the Kron radius, and the ellipticity and its angular orientation.

\end{enumerate}

\subsection{X-ray spectroscopy}

We obtained shallow X-ray observations of nine targets with {\it Swift}-XRT as Targets of Opportunity, tabulated in \autoref{xrtobstable}. We also used these observations to look for UV variability in {\it Swift} UVOT; we defer a detailed discussion of the UV observations to future papers. We checked for X-ray detections of our targets using the online {\it Swift} XRT Product Builder \citep{Evans2009,Evans2020}. If a source was detected (SkySy 1-2 and SkySyC 1-1), we extracted a spectrum from the co-addition of all available exposures using the online tool. 

Following a 2$\sigma$ X-ray detection of SkySy 1-2 in {\it Swift}-XRT and a preliminary 4$\sigma$ detection of UVW2 rapid variability in SkySy 1-4 in {\it Swift} UVOT, we were awarded Director's Discretionary Time (DDT; proposal 21308721) on the {\it Chandra} X-ray Observatory for two targets: 10 ks on SkySy 1-2 and 25 ks on SkySy 1-4. These targets were observed in Cycle 21 on the S3 chip on ACIS-S without a grating, in Timed Exposure mode with Very Faint telemetry. The on-source time for SkySy 1-2 was 9.72 ks starting on UT 2020 August 19 6:11:40, and the on-source time for SkySy 1-4 was 23.06 ks starting on UT 2020 September 10 16:43:25. We extracted the spectra following standard procedures with {\it specextract} in \textsc{CIAO} \citep{CIAO} v. 4.12 (CALDB v. 4.9.1) with standard 2.5 arcsec apertures and carefully chosen background regions.

\section{Survey results and classification of new sources}\label{Results}

We discovered 12 new symbiotic stars and 10 new symbiotic star candidates, all previously unknown, based on the results of our optical spectroscopic survey of 234 objects together with follow-up optical fast photometry of 11 targets and X-ray spectroscopy of 9 targets (2 with deep {\it Chandra} exposures). We also mapped the locations of other astronomical object types in SkyMapper parameter space, including apparently isolated cool giants of various types (S stars, carbon stars, K giants, M giants, and Miras), post-AGB and other warm or hot stars, and superpositions of an M or K giant with a warm or hot star. Our full set of optical spectra and their individual sources' locations in SkyMapper parameter space are available in the spectral atlas at the end of Chapter 3 in \citet{Lucy2021}. In this paper, we report the highlights.

The names and properties of all the newly discovered symbiotic and symbiotic candidate sources (and V1044 Cen) are tabulated in \autoref{tableskysy}. Pseudo-equivalent widths for H$\alpha$ emission lines were measured relative to the local pseudo-continuum. The presence or absence of other emission lines associated with symbiotic binarity (\ion{He}{i}, [\ion{O}{iii}], \ion{He}{ii}, \ion{O}{vi}) is denoted with boolean flags.

The optical spectra of symbiotics and symbiotic candidates are shown in \autoref{spectra_sy} alongside their locations in SkyMapper parameter space. Where possible, standard star template fits were performed by inspection using empirical stellar UVILIB templates from the TRDS Pickles Atlas version of the \citet{Pickles1998} Stellar Spectral Flux Library, with manual adjustment of a \citet{Fitzpatrick2007} interstellar extinction correction using the \citet{extinction} \textsc{extinction} code, while programatically auto-adjusting the Pickles template scaling to match in a pre-defined region after each reddening change. In some cases, we produced a better fit by adding a flat-spectrum constant value to the Pickles template, a constant value of flux as a function of wavelength; we generally only had to do this for symbiotic stars with strong and obvious hot component spectra, and the component is plotted when present. Our best-guess spectral type from spectral template fitting for each symbiotic and symbiotic candidate's cool giant component is also tabulated in \autoref{tableskysy}.

We also tested the molecular absorption spectral indices [TiO]$_{1}$, [TiO]$_{2}$, and [VO] following the spectral typing method of \citet{Kenyon1987indices} applied to our de-reddened spectra, modulo a typo in their equation for [VO]. In every case where we had been able to achieve a good fit with a cool stellar template alone, the computed indices are consistent with our best-fit spectral template to within 1.0 spectral types (0.5 in the vast majority of cases, e.g. M5.0 versus M5.5). In the other cases---the ones with optical flux from an obvious nebula contribution---the spectral types from computed indices are typically inconsistent with both each other and our best-fit stellar template, as expected. In these cases, the spectral types from computed indices always approach in the direction of our best-fit stellar template with indices of increasing wavelength. This is what would happen if our stellar template fitting was yielding accurate results, since the computed indices are expected to become more accurate at longer wavelengths as the relative flux contribution of the nebula decreases redward.

For example, we are encouraged by the case of the known symbiotic V1044 Cen, in which we had to add a strong flat-spectrum component to our fit. Our fitted stellar template is M6 III. The computed indices [TiO]$_{1}$, [TiO]$_{2}$, and [VO] yield spectral types of M3.5, M5, and M5, respectively, ascending with increasing wavelength. And in the infrared, where the spectral typing of symbiotics is more trustworthy, \citet{Murset1999} obtained M5.5 III, nearly identical to our result.

\subsection{SkySy and SkySyC classification criteria}\label{Definitions}

We adopt the internal designations ``SkyMapper symbiotics'' (SkySy) for the fully-validated set of new symbiotics, and ``SkyMapper symbiotic candidates'' (SkySyC) for the candidate set, for convenience and consistency with \citet{Lucy2021}. The decision tree in \autoref{flowchart} summarizes our classification scheme. To be categorized as at least a symbiotic star {\it candidate} (SkySyC) in \autoref{tableskysy}, an object had to meet {\it all} of the following three criteria:

\begin{enumerate}[align=left,leftmargin=*,rightmargin=0ex,labelsep=0ex]\itemsep0.3em
    \item The object exhibited spectroscopic evidence for a cool K--M I--III star (including dusty Mira spectra), in addition to meeting all the luminosity and IR colour criteria that underlie our photometric sample selection function.
    \item The object exhibited H$\alpha$ in emission\footnote{With a slight exception made for SkySyC 1-8, our weakest candidate, where it is unclear whether H$\alpha$ is in emission or not.} with an integrated flux above the continuum\footnote{Integrated flux in the line minus the local pseudo-continuum, {\it not} pseudo-equivalent width.} that after correcting for our best-estimate interstellar extinction\footnote{The extinction correction to fit a Pickles template also accounts for small uncertainties in Balmer ratios due to the relative flux calibration, differential refraction, aperture extraction, and background subtraction. The exceptions are dusty Mira spectra like SkySyC 1-4, where we have not been able to estimate the interstellar extinction.} was greater than the integrated flux in each other detectable higher-order Balmer line.
    \item Either the object exhibited He I emission, {\it or} there was no evidence that the cool K--M I--III star is an S star or a carbon star, such as ZrO molecular bands. If an object containing an S or carbon star met additional criteria to qualify as a confirmed SkySy (through emission from higher-ionization lines, or follow-up evidence for flickering/X-rays/etc. in data obtained on the basis of its SkyMapper properties), it could be a SkySy, but it could not be a SkySyC without He I.
\end{enumerate}

The idea behind the latter two criteria relates to the phenomenology of Balmer emission produced by shocks in the atmospheres of isolated red giants. This Balmer emission appears most strongly \citep{Castelaz2000}, but not exclusively \citep{Yao2017}, in late Mira variables and other strong pulsators. Recall that Balmer emission in any sort of standard diffuse nebula, including symbiotic stars, exhibits a {\it decrement}, in which the flux in H$\alpha$ > H$\beta$ > H$\gamma$ > H$\delta$ and so on. In contrast, the Balmer emission produced in the atmospheres of oxygen-rich (O-rich) Miras exhibits a Balmer {\it increment} through H$\delta$, in which the flux in H$\alpha$ < H$\beta$ < H$\gamma$ < H$\delta$. The increment in O-rich Miras may be attributable to TiO molecular bands absorbing the H$\alpha$, H$\beta$, and H$\gamma$ lines, although an alternative idea involving radiative transfer effects in the Balmer lines themselves has also been entertained (see \citealt{Castelaz2000}, and references therein). In S and Carbon stars, however, the effect vanishes (perhaps due to the weakening of the TiO bands), and a Balmer decrement is observed \citep{Castelaz2000,Yao2017}. Thus, a Balmer decrement is a promising hint of possible binary interaction in an O-rich star, but not in an S or carbon star.

We caution, however, that there has been little reason for cool giant experts to investigate carefully whether it is uniformly impossible for an isolated O-rich M giant to exhibit a Balmer decrement in emission. Indeed, the issue seems to have been very recently complicated by \citet{Yao2017}, in which M1--M3 giants exhibit a Balmer decrement at least between some of the relevant Balmer orders. There is also a risk of circular reasoning if every O-rich M giant with a Balmer decrement is immediately classified as a binary candidate. For these reasons, where Balmer decrements are our only evidence for binary interaction, we classify the object as only a candidate symbiotic (SkySyC), pending other evidence of binary interaction or future advances in our understanding of red giant Balmer emission.

\begin{figure*}
	\includegraphics[width=\textwidth]{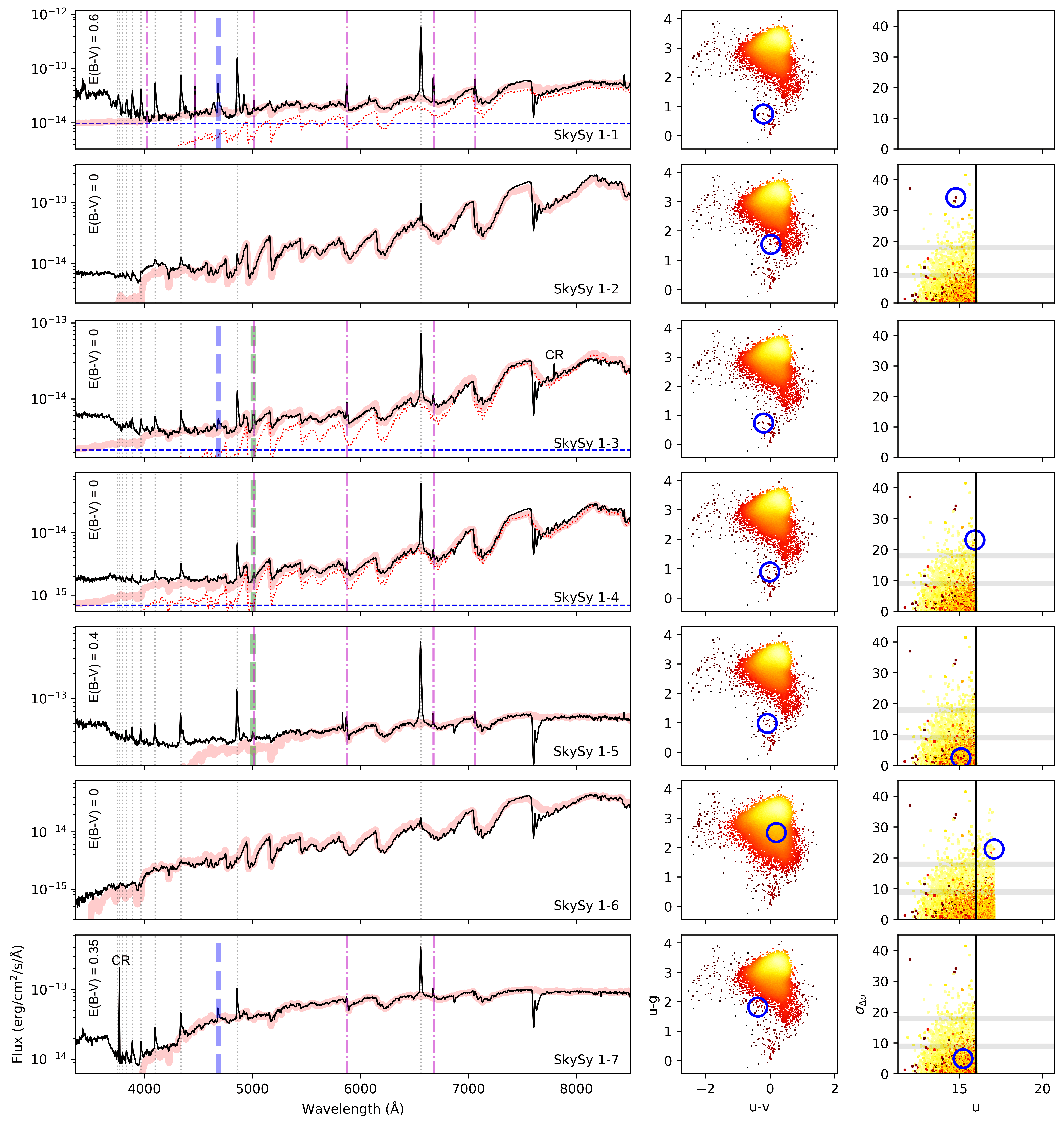}
    \caption{(Three pages.) SpUpNIC spectra of SkySy, SkySyC, and V1044 Cen, and their positions in SkyMapper parameter space. Line plots include the target spectrum (medium black line) and our spectral fit (thick pale red line). When no other line plots are included, the  thick pale red line represents only a \citet{Pickles1998} template. When an additional flat-spectrum component was necessary to achieve a good fit, the thick pale red line represents the sum of the flat-spectrum component (horizontal blue dashed line) and a \citet{Pickles1998} template (thin dotted red line). Cosmic ray artifacts in the target spectra are annotated with ``CR''. The upper left of each plot with a good spectral fit is labeled with the applied E(B-V) dereddening; these values should be treated with caution, because the slit was not aligned to the parallactic. The entire Balmer series is denoted with vertical dotted grey lines when any member of the Balmer series is in emission. \ion{He}{i} emission is denoted with vertical dash-dotted magenta lines where it appears (in order of how commonly they appear: 5876\AA, 6678\AA, 5015\AA, 7065\AA, 4026\AA, 4471\AA, and 4922\AA). [\ion{O}{iii}] emission is denoted with vertical thick dashed green lines (typically 5007\AA, occasionally 4363\AA). \ion{He}{ii} emission is denoted with vertical thick dashed blue lines (4686\AA). Raman O VI 6825\AA\ and/or 7082\AA\ is denoted with vertical thick dashed red lines. The blue open circles in the right panels denote the position of the object in SkyMapper parameter space. Each colour-colour diagram includes only those objects with u brighter than the target or u brighter than 16, whichever is more inclusive. Each $\sigu$ plot is left blank for targets for which a $\sigu$ metric was not available.}
    \label{spectra_sy}
\end{figure*}

\begin{figure*}\ContinuedFloat
	\includegraphics[width=\textwidth]{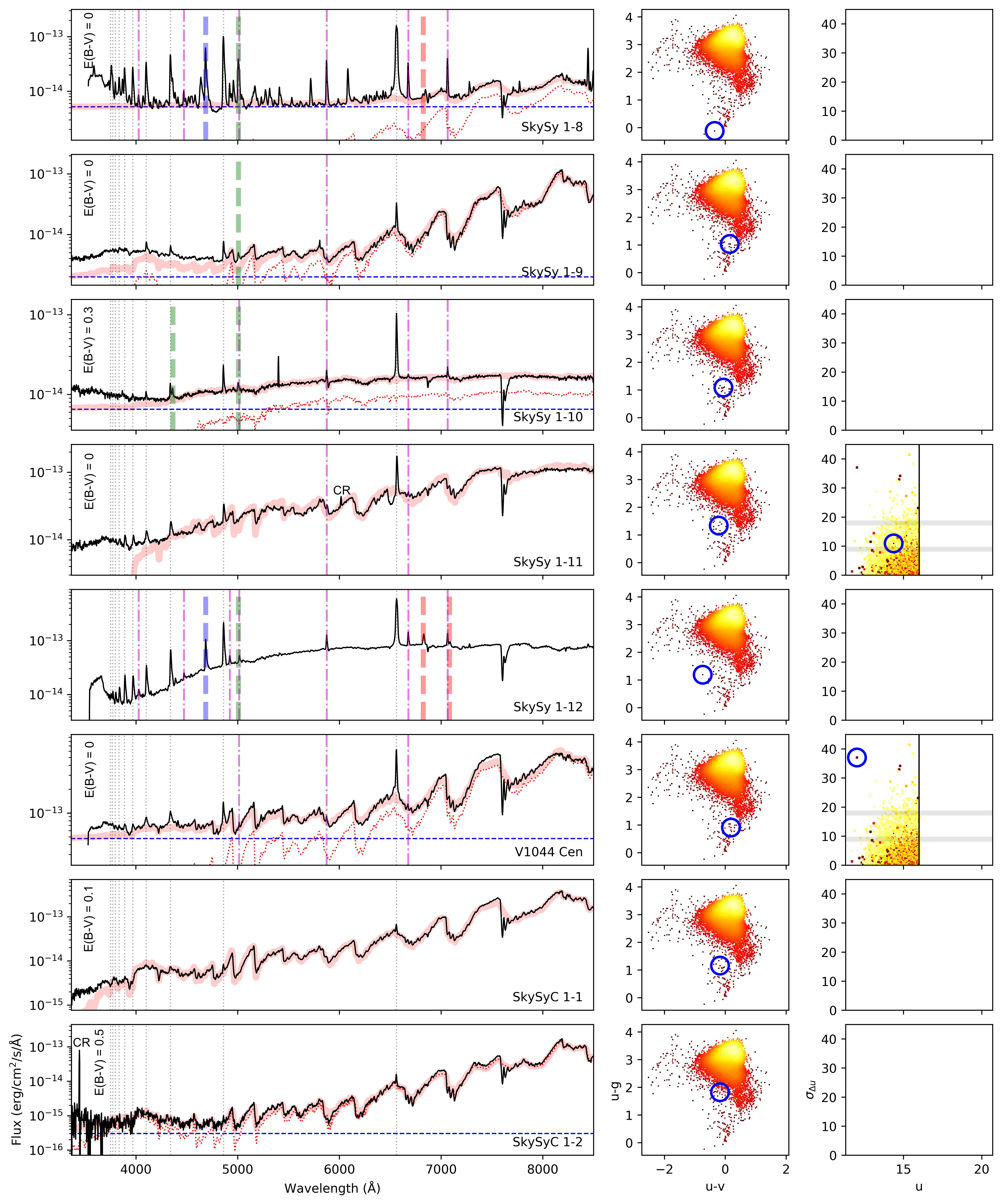}
    \caption{Continued (page 2 of 3).}
\end{figure*}

\begin{figure*}\ContinuedFloat
	\includegraphics[width=\textwidth]{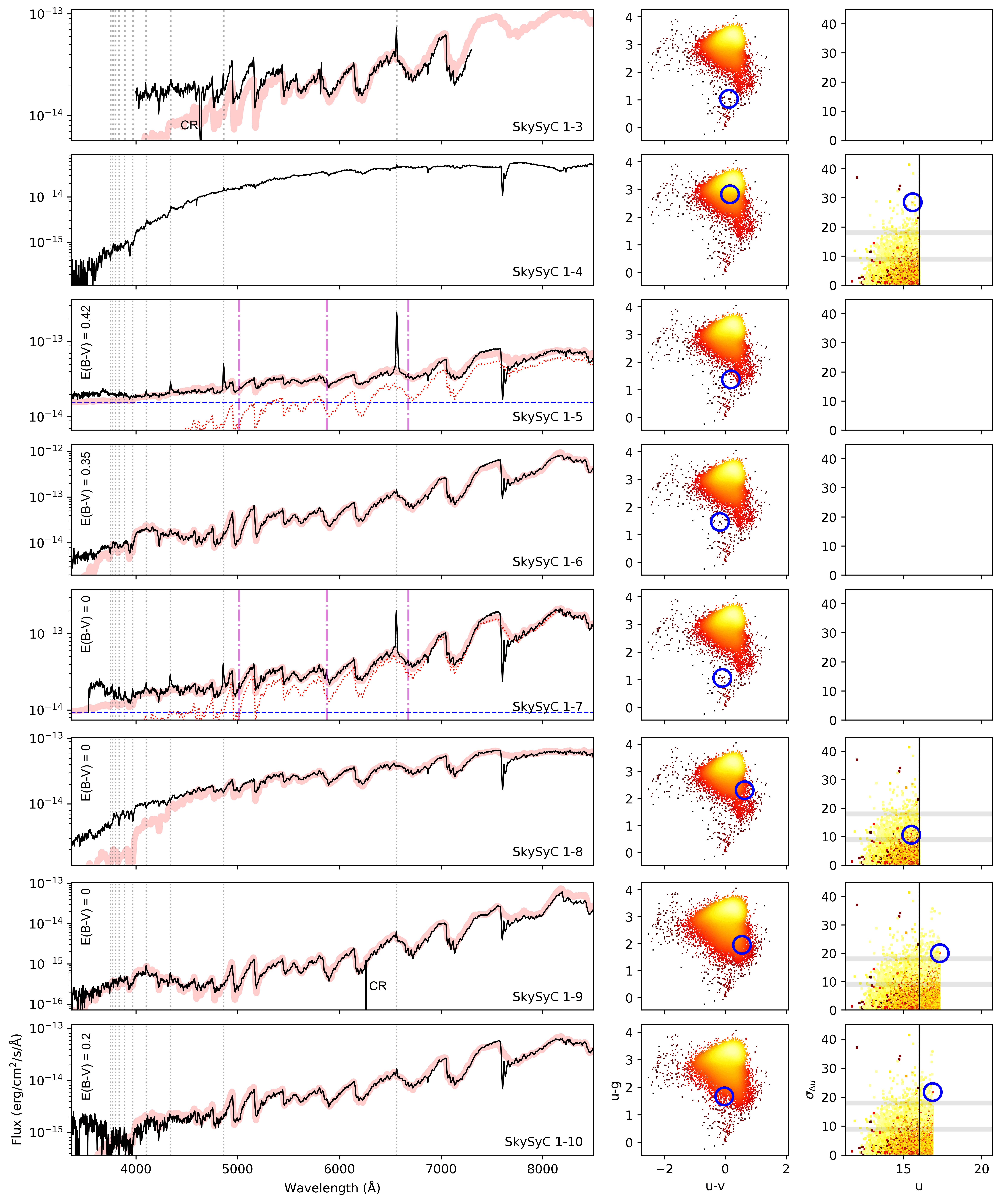}
    \caption{Continued (page 3 of 3).}
\end{figure*}

\begin{landscape}
\begin{table}
\begin{threeparttable}
\begin{tabular}{llllrlllrcccccc}
\hline
Internal name & SIMBAD name & RA Dec & Select & D & M$_J$ & u & Giant & H$\alpha$ |pEW| & \ion{He}{i} & [\ion{O}{iii}] & \ion{He}{ii} & \ion{O}{vi} & Flickering & X-rays \\
& & J2000 & & pc & & & & \AA & & & & & & \\
\hline
{\it SkySy} & & & & & & & & & & & & & & \\
SkySy 1-1 & WRAY 16-145 & 14:08:47.0 -57:04:36.6 & C & 5460 & -4.6 & 15.3 & M4 III & 220 & \cellcolor{green!20} Y & N & \cellcolor{green!20}Y & N & - & - \\
SkySy 1-2 & IRAS 15175-4508 & 15:20:58.3 -45:19:44.4 & Cf & 879 & -3.2 & 14.8 & M6 III & 11 (7) & N & N & N & N & \cellcolor{green!20} Y & \cellcolor{green!20} Y \\
SkySy 1-3 & NSV 20400 & 15:47:50.8 -32:43:59.3 & C & 4350 & -4.8 & 14.0 & S & 70 & \cellcolor{green!20} Y & \cellcolor{green!20} Y & \cellcolor{green!20} Y & N & - & - \\
SkySy 1-4 & Haro 1-10 & 16:27:49.8 -29:16:45.5 & Cf & 3885 & -3.4 & 15.9 & M6 III & 100 & \cellcolor{green!20} Y & \cellcolor{green!20} Y & N & N & \cellcolor{green!20} Y & \cellcolor{green!20} Y \\
SkySy 1-5 & UCAC4 307-090112 & 16:38:07.8 -28:42:07.6 & C & 8820 & -4.9 & 15.1 & M0.5 III & 100 & \cellcolor{green!20} Y & \cellcolor{green!20} Y & N & N & - & - \\
SkySy 1-6 & 2MASS J17071056-5653174 & 17:07:10.6 -56:53:17.5 & CF & 5066 & -4.3 & 17.1 & M5 III & 6 & N & N & N & N & \cellcolor{green!20} Y & - \\
SkySy 1-7 & WRAY 15-1790 & 17:50:47.5 -39:01:17.7 & C & 8019 & -5.3 & 15.2 & M0 III & 40 & \cellcolor{green!20} Y & N & \cellcolor{green!20} Y & N & - & -  \\
SkySy 1-8 & 2MASS J18022627-4135544 & 18:02:26.3 -41:35:54.4 & C & 9295 & -5.3 & 14.3 & M6 III & 750 & \cellcolor{green!20} Y & \cellcolor{green!20} Y & \cellcolor{green!20} Y & \cellcolor{green!20} Y & - & - \\
SkySy 1-9 & V589 CrA & 18:25:40.0 -42:14:22.8 & C & 9908 & -7.2 & 15.7 & M7 III (Mira) & 19 & \cellcolor{green!20} Y & \cellcolor{green!20} Y & N & N & - & - \\
SkySy 1-10 & V1918 Sgr & 18:37:34.8 -20:53:25.0 & C & 11199 & -4.6 & 15.7 & K4 I & 60 & \cellcolor{green!20} Y & \cellcolor{green!20} Y & N & N & N & - \\
SkySy 1-11 & CSS 1102 & 19:05:59.1 -21:09:25.0 & Cf & 7468 & -6.0 & 14.4 & S & 40 & \cellcolor{green!20} Y & N & N & N & \cellcolor{green!20} Y & - \\
SkySy 1-12$^{*}$ & Hen 3-1768 & 19:59:48.4 -82:52:37.4 & C & 7067 & -5.4 & 13.0 & K? & >120 & \cellcolor{green!20} Y & \cellcolor{green!20} Y & \cellcolor{green!20} Y & \cellcolor{green!20} Y & - & - \\
V1044 Cen & V1044 Cen & 13:16:01.4 -37:00:10.8 & Cf$^{\dagger}$ & 1795 & -4.7 & 12.0 & M6 III & $\approx$35 & \cellcolor{green!20} Y & N & N & N & \cellcolor{green!20} Y & - \\
\\
\hline
{\it SkySyC} & & & & & & & & & \\
SkySyC 1-1 & IRAS 13224-5839 & 13:25:41.7 -58:55:05.9 & C & 585 & -2.4 & 15.8 & M6.5 III & 6 (2) & N & ? & N & N & N? & Y? \\
SkySyC 1-2 & IRAS 15057-5252 & 15:09:24.0 -53:03:28.0 & C & 7434 & -7.6 & 15.4 & M8 III (Mira?) & 8 & N & N & N & N & - & - \\
SkySyC 1-3 & IRAS 15220-6952 & 15:26:57.3 -70:03:10.7 & C & 1929 & -4.8 & 12.6 & M4.5? III$^{\ddagger}$ & 10 & N & N & N & N & - & - \\
SkySyC 1-4 & KU Aps & 17:21:51.0 -81:00:36.5 & CF & 9135 & -5.2 & 15.6 & Mira & 1 & N & N & N & N & - & - \\
SkySyC 1-5 & SS 316 & 17:36:59.4 -18:06:26.6 & C & 4876 & -4.5 & 15.4 & M4 III & 60 & \cellcolor{green!20} Y & N & N & N & ? & - \\
SkySyC 1-6 & IRAS 18237-2417 & 18:26:46.9 -24:15:47.2 & C & 3238 & -6.6 & 15.6 & M6.5 III & 2 & N & N & N & N & N & - \\
SkySyC 1-7 & 2MASS J18344959-3824568  & 18:34:49.6 -38:24:56.9 & C & 1936 & -3.9 & 14.1 & M5.5 III & 20 & \cellcolor{green!20} Y & N & N & N & - & - \\
SkySyC 1-8 & ATO J284.0167-23.4539 & 18:56:04.0 -23:27:13.9 & CF & 5822 & -4.6 & 15.5 & M3 III & <1 & N & N & N & N & N & - \\
SkySyC 1-9 & 2MASS J19045180-1728398 & 19:04:51.8 -17:28:39.9 & CF & 8710 & -6.5 & 17.3 & M7.5 III (Mira?) & 3 & N & N & N & N & - & - \\
SkySyC 1-10 & GSC 05140-03255 & 19:05:48.0 -6:05:49.6 & CF & 3839 & -4.6 & 16.9 & M6 III & 4 & N & N & N & N & N & - \\
\end{tabular}
\begin{tablenotes}
\caption{ SkyMapper symbiotics (SkySy) and SkyMapper symbiotic candidates (SkySyC) discovered through this work---and the previously known symbiotic V1044 Cen, in which flickering was discovered through this work. Columns are described below.\\
$*$  We first reported SkySy 1-12 (Hen 3-1768) in \citet{Lucy2018b}.\\
$\dagger$  V1044 Cen is a previously-known symbiotic. If it had not already been discovered, this is the selection mechanism by which we would have identified it. \\
$\ddagger$  We have low confidence in the flux calibration and spectral typing for SkySyC 1-3, our only target not observed with SAAO SpUpNIC. However, we are confident that it exhibits a Balmer decrement. \label{tableskysy}}
\item[Internal name]  Internal SkySy or SkySyC designation.
\item[SIMBAD name]  A SIMBAD-recognized name for external reference.
\item[RA Dec]  The SkyMapper coordinates in J2000.
\item[Select]  Flag for SkyMapper selection mechanism employed, including targets selected by colour-colour alone (C), targets selected by both colour-colour and $\sigu$ (CF), and targets selected by both colour-colour and $\sigu$ but for which colour-colour alone would have been sufficient to select (Cf).
\item[D]  \citet{Bailer2018} Bayesian inference of the distance using {\it Gaia} DR2.
\item[M$_{\rm J}$]  Absolute J magnitude calculated using 2MASS J magnitude and the distance D. Not corrected for extinction.
\item[u]  SkyMapper u\_psf magnitude from the dr2.master catalog.
\item[Giant]  Estimated spectral type of the donor star. S refers to s-process-enhanced stars. 
\item[H$\alpha$ |pEW|]  Absolute value of the H$\alpha$ pseudo-equivalent width, relative to the surrounding pseudo-continuum. H$\alpha$ was in emission in every case.
\item[\ion{He}{i}, {[}\ion{O}{iii}{]}, \ion{He}{ii}, Raman \ion{O}{vi}]  Flags for the detection of emission lines from these ions: ``Y'' for detected, ``N'' for not detected, ``?'' for ambiguous case that may warrant future follow-up observations.
\item[Optical flick.]  Flag for optical flickering in hours-long B-band light curves (\autoref{Flickering}): ``Y'' for detected, ``N'' for not detected, ``N?'' or ``?'' for ambiguous cases discussed in the main text, ``-'' for not observed.
\item[X-rays]  Flag for X-ray detections (which always include hard-components with photon energies above 2.4 keV; \autoref{X-rays}): ``Y'' for detected, ``Y?'' for ambiguous case discussed in the main text, otherwise ``-''.
\end{tablenotes}
\end{threeparttable}
\end{table}
\end{landscape}

\begin{figure*}
	\includegraphics[width=\textwidth]{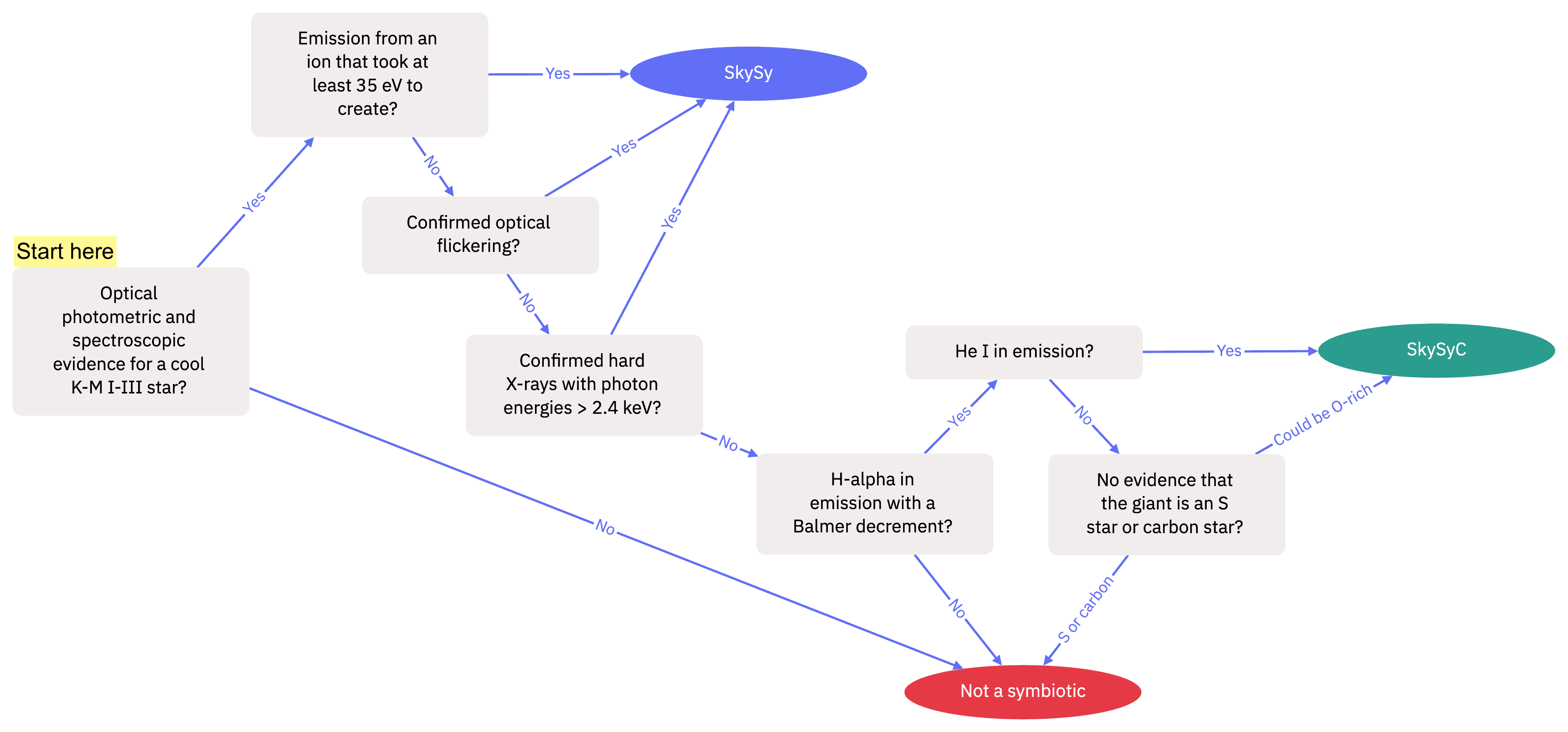}
    \caption{Decision tree summarizing our observational criteria for classifying an object as a fully-validated symbiotic (SkySy), a candidate symbiotic (SkySyC), or a contaminant. Read from left to right.}
    \label{flowchart}
\end{figure*}

We exercise similar caution when it comes to our detected optical \ion{He}{i} emission lines, coming from a state with an excitation energy of 23 eV (which in the nebular case is reached through recombination, requiring a 24.6 eV photon to first ionize He to He$^{+1}$). That is much higher than the excitation of Balmer emitters, but it is significantly lower than the 35 eV needed to ionize an O$^{+1}$ ion to O$^{+2}$ and thereby allow for collisional excitation to the upper level of the [\ion{O}{iii}] emission line, which is sometimes treated as effectively the lowest-ionization species sufficient to classify an object as a symbiotic star (e.g., \citealt{Mikolajewska2014}). Again, there is not much motivation for cool giant experts to conclusively demonstrate that shocks in an isolated giant categorically cannot produce \ion{He}{i} emission, so out of an abundance of caution, we do not treat \ion{He}{i} as sufficient evidence to classify an object as a confirmed symbiotic (SkySy). We do, however, consider \ion{He}{i} to be sufficient to classify an S or carbon star as a candidate (SkySyC), because we are not aware of any evidence that S/C stars emit \ion{He}{i} on their own.

Instead, to be categorized as a {\it fully-validated symbiotic star} (SkySy) in \autoref{tableskysy}, an object had to exhibit spectroscopic evidence for a cool K--M I--III star and meet {\it at least one} of the following criteria, incorporating both spectral information from this section and other information from \autoref{Flickering} and \autoref{X-rays}:

\begin{enumerate}[align=left,leftmargin=*,rightmargin=0ex,labelsep=0ex]\itemsep0.3em
\item The object exhibited emission from an ion that took at least 35 eV to create, such as [\ion{O}{iii}] and/or \ion{He}{ii} emission.
\item The object exhibited optical flickering in hours-long continuous light curves, subject to extensive validation checks (\autoref{Flickering}).
\item The object exhibited unambiguous X-ray emission with photon energies above 2.4 keV (\autoref{X-rays}).
\end{enumerate}

\subsection{Results from follow-up optical fast photometry}\label{Flickering}

We detected B-band variability on time-scales of minutes---apparent flickering (\autoref{introflickering})---in our LCO light curves with very high confidence in SkySy 1-2, 1-4, 1-6, 1-11, and with moderate confidence in V1044 Cen. All these also exhibited $\sigu$ above 2$\sigma_{\rm eff}$ in our SkyMapper parameter space. The light curves of the flickerers (and SkySyC 1-5, discussed below) are shown in the left panels of \autoref{lightcurves1}. We also plot each primary check star's light curve, and the bright comparison star's FWHM, a measure of the seeing.

\begin{figure*}
\centering
\includegraphics[width=5.0in]{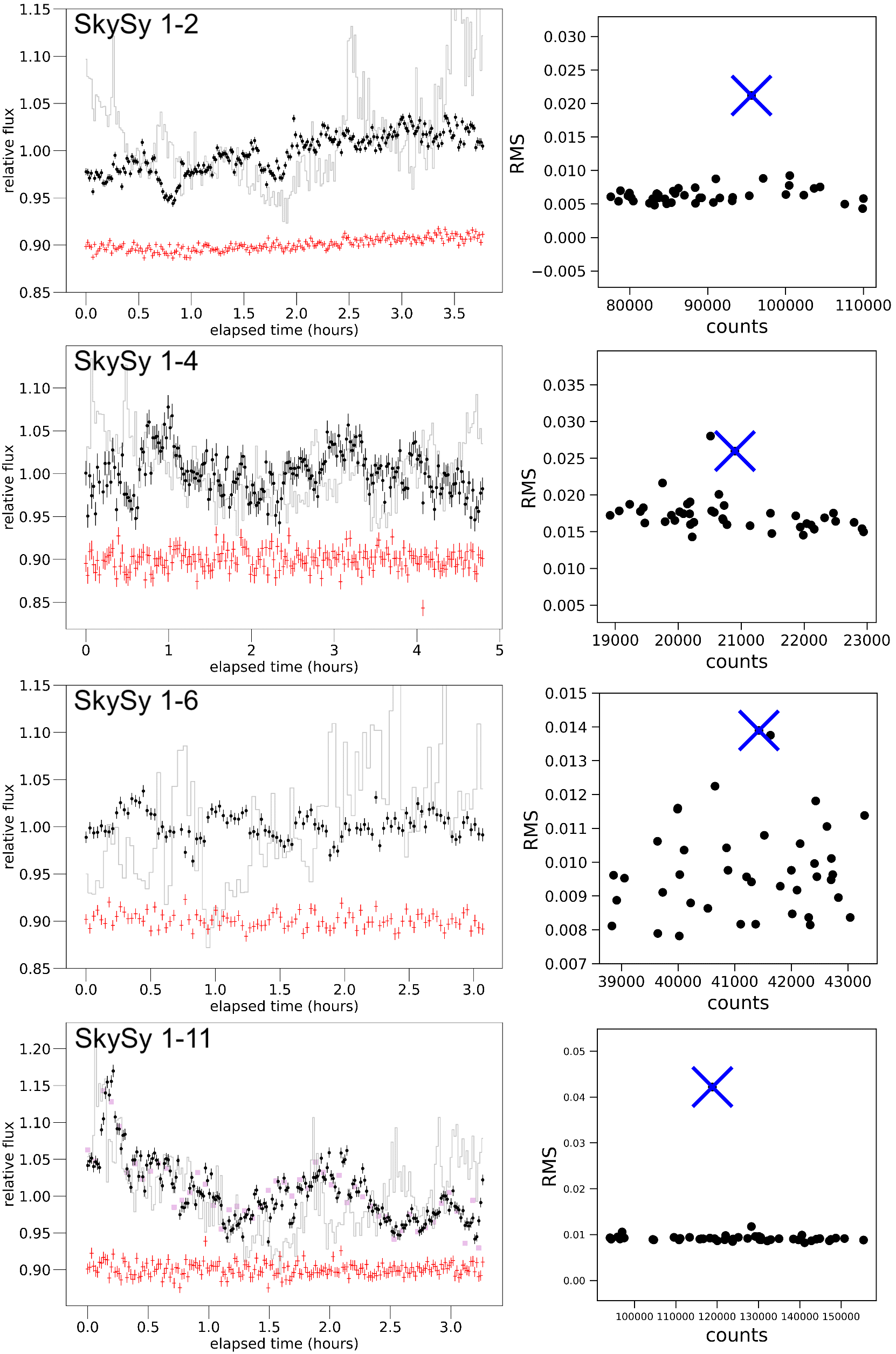}
\caption{(Two pages.) B-band LCO light curves show flickering in (from top to bottom rows) SkySy 1-2, SkySy 1-4, SkySy 1-6, and SkySy 1-11. On the left side, the light curve of the target is shown in flux relative to the median flux (black circles with error bars), the seeing is tracked relative to the median seeing (pale grey line), and the light curve of a check star with similar colour and flux is shown offset by 0.1 on the vertical axis (red crosses with error bars). For SkySy 1-11 (bottom row), we also plot the identical results of an extremely large 16 arcsec-radius aperture extraction of the target (light purple squares). On the right side, we show that the light curve RMS for the target star (blue cross) is higher than almost all of the 40 or so check stars (black circles) closest in flux to each target.}\label{lightcurves1}
\end{figure*}
\begin{figure*}\ContinuedFloat
\centering
\includegraphics[width=5.0in]{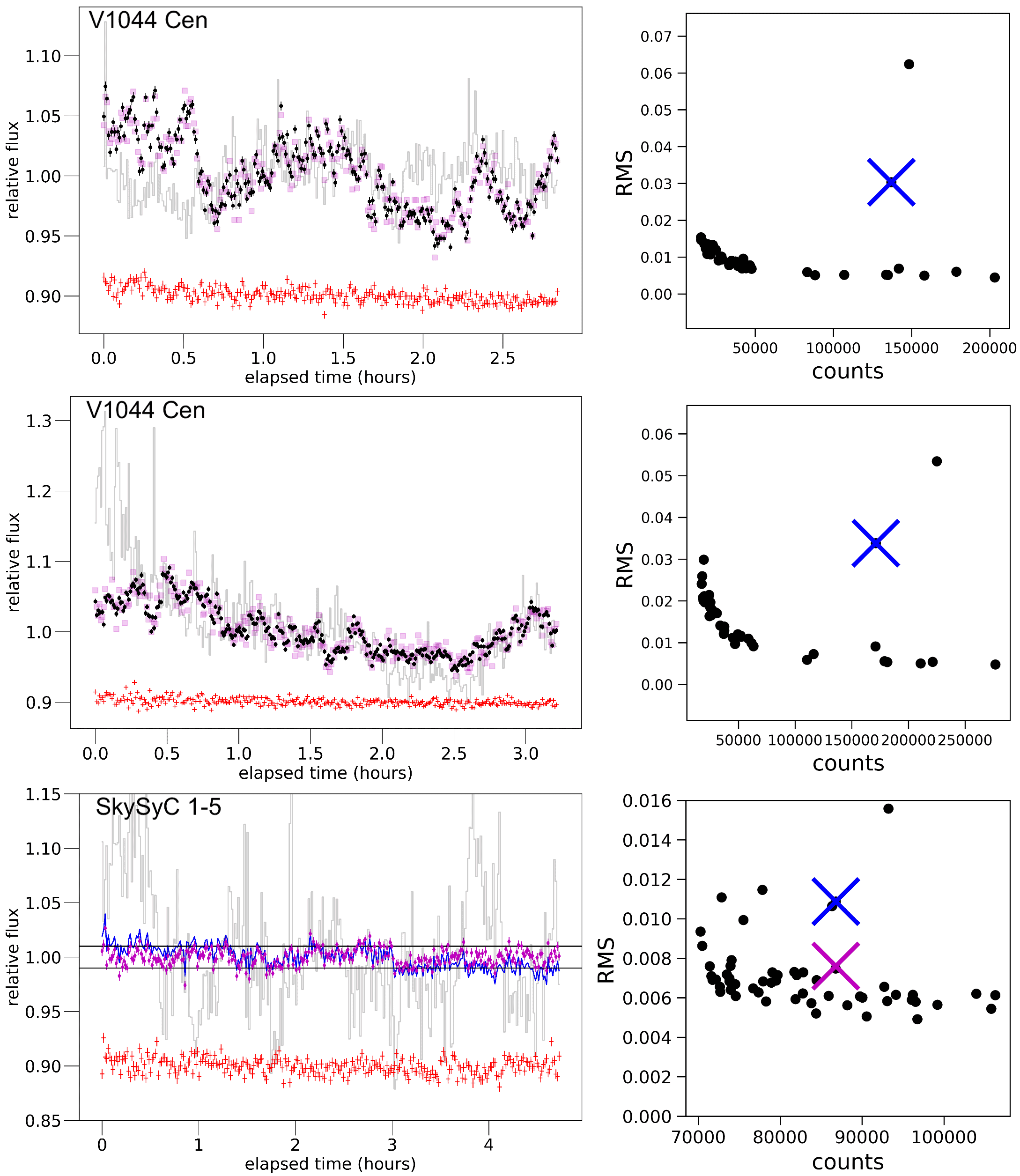}
\caption{Continued (page 2 of 2). B-band LCO light curves show likely flickering in V1044 Cen (top and middle rows), and a dubious hint of flickering in SkySyC 1-5 (bottom row). For SkySyC 1-5 (bottom row), we plot both the original light curve (blue line) and a light curve de-trended by a linear fit (purple circles with error bars), bracketed by horizontal lines at $\pm$1\% relative flux; in the lower-right panel, the corresponding light curve RMS values for SkySyC 1-5 are plotted as blue and purple crosses, respectively.}
\end{figure*}

To validate the pipeline results for SkySy 1-2, 1-4, 1-6, 1-11, and V1044 Cen, we performed fixed-aperture photometry of the target, comparison star, and primary check star in \textsc{AstroImageJ} \citep{astroimagej}, and examined the resultant light curves and output diagnostics. We tested aperture radii of 10, 20, and 40 pixels (4, 8, 16 arcsec). Our results remained identical in these tests, with the exception of the 16 arcsec radius extraction of the dim SkySy 1-4 (which became dominated by noise, but the 8 arcsec radius extraction shows clear flickering) and the 16 arcsec radius extraction of SkySyC 1-5 (which likewise became dominated by noise, but the 8 arcsec radius extraction looked the same as the normal reduction). In cases where relationships between the flux and the seeing were of any possible concern, we overplotted the 16 arcsec radius \textsc{AstroImageJ} extractions.

In the right panels of \autoref{lightcurves1}, inspired by the RMS plot in \mbox{\citet{Zamanov2017}}, we plot the unweighted RMS of the calibrated and normalized light curve of each target flickerer and 40 or so similarly-bright check stars from the same field (see \autoref{photvalidation}). The plotted units of the RMS are in fraction of median flux. The expected statistical behavior in these plots is for the check stars to have small RMS values at high count rates, rising at low count rates. The RMS of the flickering targets SkySy 1-2, 1-4, 1-6, 1-11, and V1044 Cen was consistently higher than all 40 check stars in each field, with minor exceptions: 

\begin{enumerate}[align=left,leftmargin=*,rightmargin=0ex,labelsep=0ex]\itemsep0.3em

\item In SkySy 1-4 (\autoref{lightcurves1}, second row), there is one  check star with higher RMS than the target. Unlike the target star, the anomalous check star's variability vanishes with a fixed-aperture \textsc{AstroImageJ} extraction even at 4 arcsec radius, while SkySy 1-4 remains flickering at 8 arcsec radius. The anomalous check star also exhibits a correlation between flux and seeing, and two very close blended neighbors with centroids within about 5 arcsec (with about 14\% and 19\% of the check star's flux).

\item Likewise, for SkySy 1-6 (\autoref{lightcurves1}, third row), the variability in the check star with RMS just below that of the target vanishes with a fixed-aperture \textsc{AstroImageJ} extraction of 4 arcsec, while the flickering in \mbox{SkySy 1-6} remains with a 16 arcsec extraction. An additional night (not shown) on SkySy 1-6 had poorer atmospheric conditions, with 6 check stars exhibiting correlations between flux and seeing. Even then, \textsc{AstroImageJ} fixed-aperture photometry removed the variability of the check stars without removing the variability of the target.

\item The check star with higher RMS than V1044 Cen (\autoref{lightcurves1}, top two rows on the second page) exhibits a smooth section of a sine wave and is classified in SIMBAD as an eclipsing binary named TYC 7275-1968-1.

\end{enumerate}

We also manually examined the \textsc{SEP}- and \textsc{AstroImageJ}-output diagnostics  
and did not find anything unusual about any of the target star's diagnostics, or any worrying correlations between the diagnostics and the flickering target's flux, with a singular exception: The flux of V1044 Cen was mildly correlated with the seeing, especially in the second night. The Pearson R (Spearman R) correlation coefficients between the flux and the \textsc{AstroImageJ} X-width\footnote{The FWHM in the horizontal image direction. It was slightly more correlated than the FWHM output by either program, so we report correlations for the X-width here to be conservative.} of the comparison star were 0.33 (0.35) in the first light curve and 0.67 (0.75) in the second. There is an extremely dim, close neighbor with a flux of 1\%  of the target, though an extraction of the neighbor's light curve showed no issues or correlations with the target's variability. Additionally, the aforementioned eclipsing binary is of about equal brightness and is located 22.9 arcsec away, exhibiting a smooth single-crest section of a sine wave in the first night, and a smooth single-trough section of a sine wave in the second night. There was also a slight downtick in the background (throughout the image, not just at the target) where the second light curve sloped up in the last $\sim$15\% of the target light curve. These issues limit our confidence somewhat in our claim of flickering for V1044 Cen. 

However, as shown in \autoref{lightcurves1}, V1044 Cen's light curves when using an exceedingly large extraction aperture radius of 16 arcsec in \textsc{AstroImageJ} are completely identical to the standard \textsc{SEP} reduction. We also tested a fixed-position aperture extraction (i.e., without re-centroiding in each image), with no change in the light curve. An exceptionally small aperture of 1.6 arcsec radius also showed flickering, as did variable-width apertures with radii on the order of the FWHM. Even when seeing was at its worst, the radial profile of V1044 Cen and the eclipsing-binary neighbor each had a half-width-at-half-maximum of less than 2 arcsec, with 99\% of the flux confined within a radius of 6 arcsec. And we did not notice any other issues or unexpected correlations with the other diagnostics output by \textsc{SEP} or \textsc{AstroImageJ}, including the background. Finally, a 20 arcsec-radius fixed-coordinate extraction centered between and encompassing both the target and its eclipsing binary neighbor, roughly mimicking what an instrument with lower spatial resolution like TESS might observe, showed the target's flickering features superimposed on its neighbor's sine wave. 

It is unclear what sort of seeing effect, in this context, could simulate flickering throughout this suite of tests. None of the sharp features in V1044 Cen's light curves are mirrored in the smooth, less than half-period light curves of its eclipsing-binary neighbor, and some of its sharps features are not reflected in the seeing. We conclude that V1044 Cen is likely flickering, but that follow-up observations in stable seeing would be advisable.

\subsubsection{Dubious hints and non-detections}

We detected a dubious hint of low-amplitude LCO flickering in SkySyC 1-5 (plotted in the bottom row of \autoref{lightcurves1} on the second page) with no confidence. SkySyC 1-5 did not have a $\sigu$ measurement in SkyMapper. In its LCO light curve,  it exhibited much higher RMS than almost all its check stars, but its variability was restricted to about $\pm$1\%. Furthermore, de-trending its light curve with a linear fit reduced the RMS to only 1.2 times the median RMS of the check stars. Some of the check stars with higher RMS can be accounted for with unambiguous flux/seeing correlations---but not all. Accordingly, we do not have any confidence in claiming a flickering detection for SkySyC 1-5. On the other hand, the RMS excess and light curve structure remains the same in \textsc{AstroImageJ} fixed-aperture extractions of 4 and 8 arcsec. Precise observations from space or in very stable seeing may be necessary to determine conclusively whether this target exhibits flickering or other short-timescale variability.

SkySyC 1-1 (not shown, and without a $\sigu$ measurement in SkyMapper) exhibited excess RMS relative to most check stars in the pipeline LCO photometry, but it suffered from severe seeing issues, with every substantial feature in the light curve reflected in the seeing. These issues may connect to two neighbors with fluxes around 20\% of the target flux and centroids within about 5 arcsec, or to optical spikes from a bright neighbor to the west. Three check stars in this field also suffered from similarly strong flux/seeing correlations leading to comparably high excess RMS. And unlike our flickerers, variability in excess of the noise almost entirely vanished with a 4 arcsec fixed-aperture extraction in \textsc{AstroImageJ}.

The remainder of our LCO targets, including SkySyC 1-8 and 1-10 (which exhibited $\sigu$ above 2$\sigma_{\rm eff}$ in SkyMapper) and SkySy 1-10 and SkySyC 1-6 (which did not have $\sigu$ measurements in SkyMapper) did not exhibit flickering in our LCO light curves. Specifically, the RMS of their light curves was less than the median RMS of their 40 check stars. We note in particular we have not reproduced our previously claimed flickering detection in American Association of Variable Star Observers (AAVSO) photometry of SkySyC 1-10 reported in \citet{Lucy2019}.

\subsection{Results from follow-up X-ray spectroscopy}\label{X-rays}

We detected at least two (likely three) targets in X-rays, yielding {\it Chandra} and {\it Swift} XRT spectra for SkySy 1-2, a {\it Chandra} spectrum for SkySy 1-4, and a low-count {\it Swift} XRT spectrum of SkySyC 1-1. These X-ray data are presented in \autoref{xrayspectrach3}. Our main results are that there is a hard X-ray component in SkySy 1-2 and SkySy 1-4, and a tentative detection of hard X-rays from SkySyC 1-1 as well. Following the schema in \citealt{Luna2013}, we define hard X-rays to be those with photon energies above 2.4 keV.

\subsubsection{SkySy 1-2}

{\it Chandra} detected about 55 counts from SkySy 1-2 in 9720 seconds at the expected target position, with significantly detected photons out to 6.3 keV. Following the models used by \citet{Luna2013} and \citet{Patterson1985b} for X-ray emission from boundary layers around accreting white dwarfs, we fit the {\it Chandra} spectrum with collisionally-ionized, optically-thin plasma emission (\textsc{APEC} in \textsc{Sherpa}, a package in \textsc{CIAO}; \citealt{Sherpa}) multiplied by photoelectric absorption with Wisconsin cross-sections (\textsc{WABS} in \textsc{Sherpa}). The temperature of the emitting plasma is degenerate with the fit normalization and the absorption column, so we can only calculate a minimum, but for a single-temperature fit the plasma temperature was kT$\geq$2 keV at the 90\% confidence level. At this 2 keV lower-limit temperature, we used \textsc{Sherpa} \textsc{sample\_flux} to estimate the observed luminosity (not corrected for absorption) at $1.2^{+0.9}_{-0.7} \, \times10^{31} \mathrm{erg \, s^{-1}} \, \left(\frac{d}{879 \, \mathrm{pc}}\right)^2$ with uncertainties corresponding to the 90\% confidence level.

\begin{figure}
\centerline{\includegraphics[width=\columnwidth]{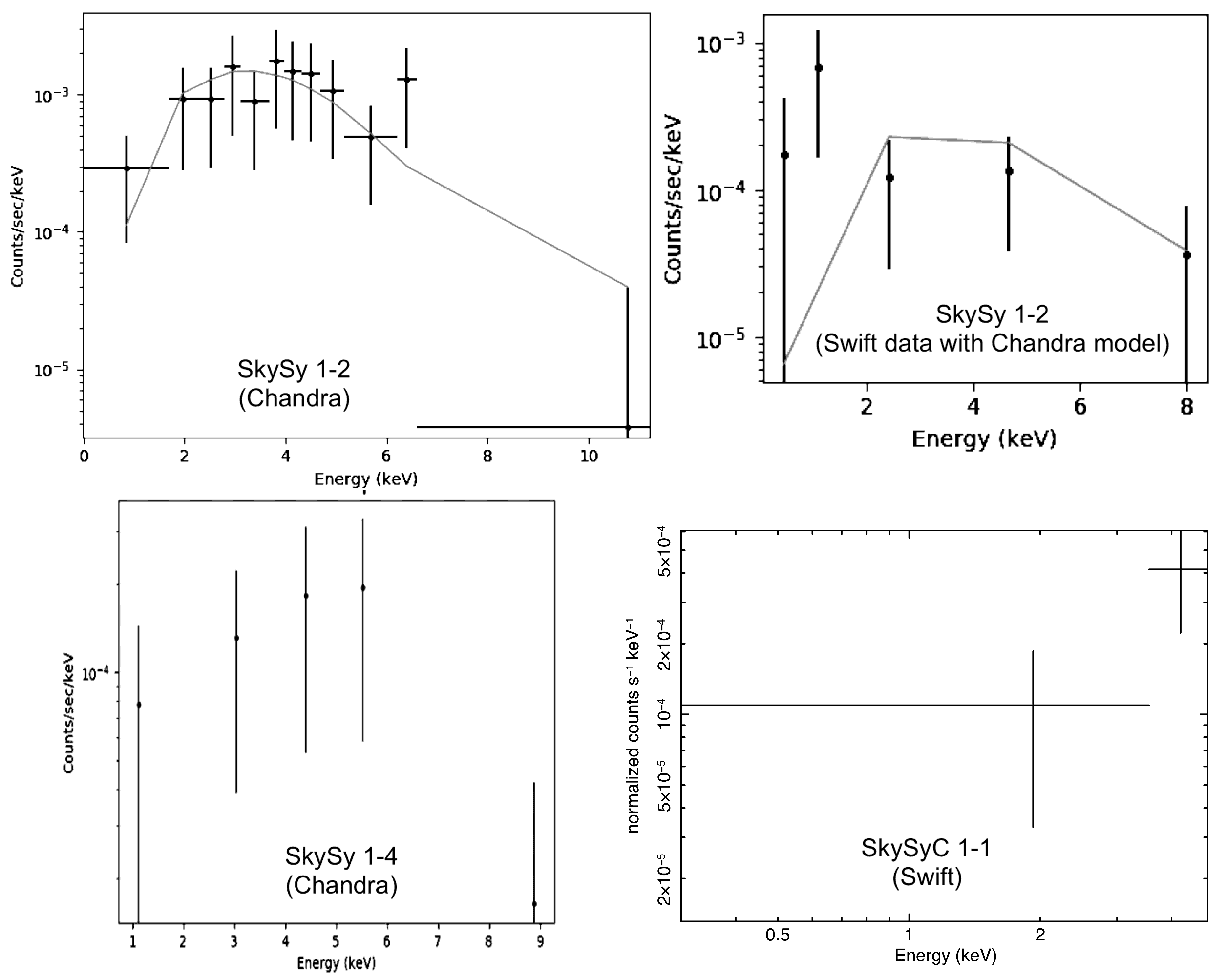}}
\caption{{\it Top left}: single-temperature fit to {\it Chandra} spectrum of SkySy 1-2 with kT=9.4 keV, nH=4. {\it Upper right}: SkySy 1-2's single temperature {\it Chandra} fit on its combined {\it Swift} XRT spectrum. {\it Lower-left}: {\it Chandra} spectrum of SkySy 1-4; it has a lower count-rate and a fit was not attempted, but it resembles SkySy 1-2. {\it Lower-right}: {\it Swift} XRT spectrum of SkySyC 1-1, showing a hint of hard X-rays.} \label{xrayspectrach3}
\end{figure}

The {\it Swift} XRT spectrum of the same target SkySy 1-2 has lower SNR, but it shows a hint of an additional softer component peaking at photon energies around 1 keV, absent from the {\it Chandra} spectrum. Contemporaneous V magnitudes from the AAVSO are about 12.6, so optical loading was likely not an issue. The single-temperature {\it Chandra} model missed this soft component completely (upper right panel of \autoref{xrayspectrach3}). We cannot, however, confidently attribute this component's absence from the {\it Chandra} spectrum to variability; although {\it Chandra} generally has a far larger collecting area than {\it Swift} XRT, {\it Swift} is more sensitive than {\it Chandra} below about 0.8 keV.

\subsubsection{SkySy 1-4}

SkySy 1-4 was also detected in {\it Chandra}, at the expected target position, with only 20 counts in 23060 seconds. We have not attempted a fit for its spectrum, but it almost identically resembles a lower-count version of the SkySy 1-2 spectrum, so it has a similar hard component but with less rigorous constraints on temperature. The spectrum is dominated by photons well above 2.4 keV. Estimating solely by comparing the count rate and {\it Gaia} distance of SkySy 1-4 to that of SkySy 1-2, SkySy 1-4 may be about three times as luminous as SkySy 1-2.

\subsubsection{SkySyC 1-1}

SkySyC 1-1 was marginally detected in {\it Swift}-XRT with a significance of 4.6$\sigma$, with an average of 1.9$^{+0.7}_{-0.6}\times10^{-3}$ counts s$^{-1}$ at $\alpha$=13h 25m 42.6s, $\delta$=-58$^{\circ}$ 54$^{\prime}$ 59.7$^{\prime\prime}$, well within the XRT error circle for the optical position. The extracted spectrum has more counts above 2.4 keV than below.

\subsection{Locations of different object types in SkyMapper parameter space}

\begin{figure*}
\centerline{\includegraphics[width=\textwidth]{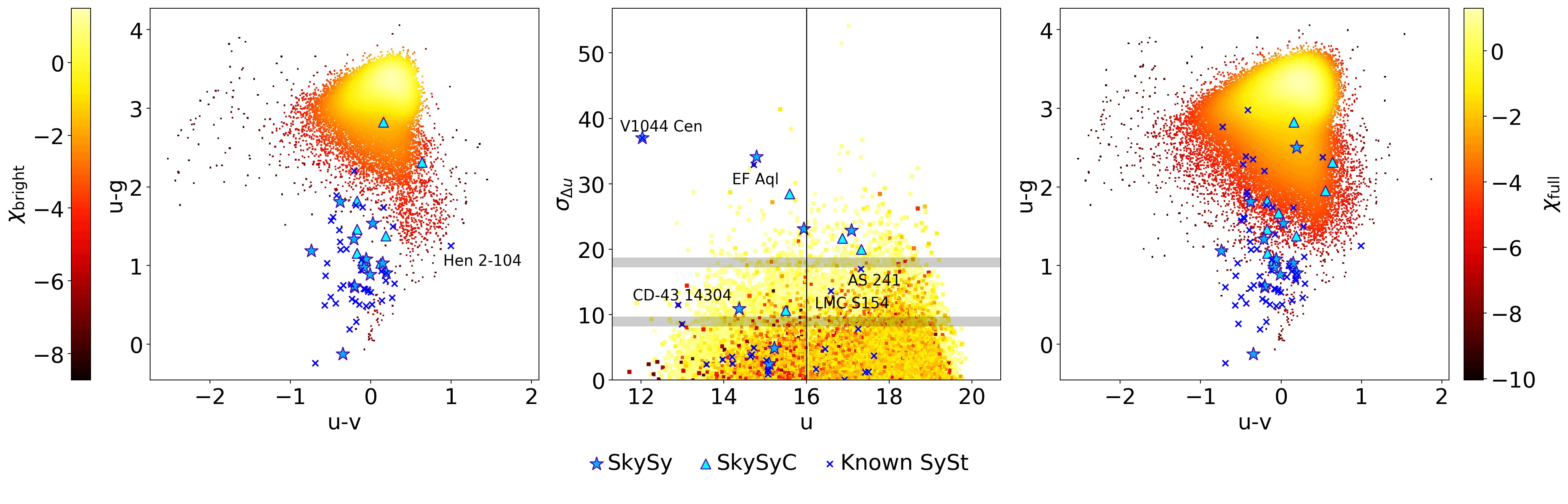}}
\caption{The SkyMapper symbiotics (SkySy; star symbols) and symbiotic candidates (SkySyC; triangles), alongside the sample of previously known symbiotics (small crosses), overlaid on the \autoref{parameterspace} parameter space.} \label{symbioticsinspace}
\end{figure*}

\begin{figure*}
\centerline{\includegraphics[width=\textwidth]{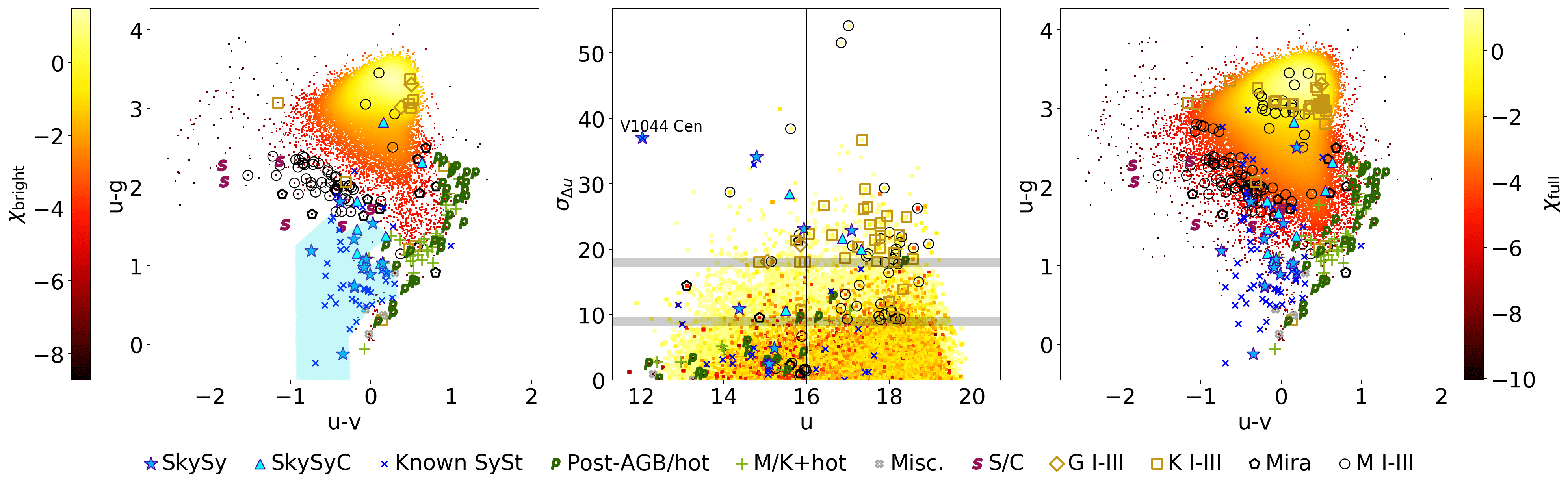}}
\caption{Object type classification based on our optical spectroscopy on the 234 objects targeted in \autoref{targets}, overlaid on \autoref{parameterspace}. These include newly reported SkyMapper symbiotics (SkySy; star symbols), newly reported SkyMapper symbiotic candidates (SkySyC; triangles), post-AGBs and other warm/hot stars (P symbols), superpositions of M-type TiO bands with a warm/hot stellar spectrum (plus signs), S stars and carbon stars (S symbols), G0--8 I--III stars (diamonds), K0--7 I--III stars (squares), M0--8 I-III stars (pentagons; excluding Miras), Miras/suspected Miras (open circles), and a small miscellaneous group (open crosses; including an AGN, a Wolf-Rayet, a PN, and an unclassified PN-like object). The \citet{Merc2019} sample of previously known symbiotics is also included (small crosses). The mostly-symbiotic zone in the bright subsample's colour-colour plot is illustrated, with arbitrary boundaries, as a blue region in the left panel.}\label{allresults}
\end{figure*}

\begin{figure*}
	\includegraphics[width=6.0in]{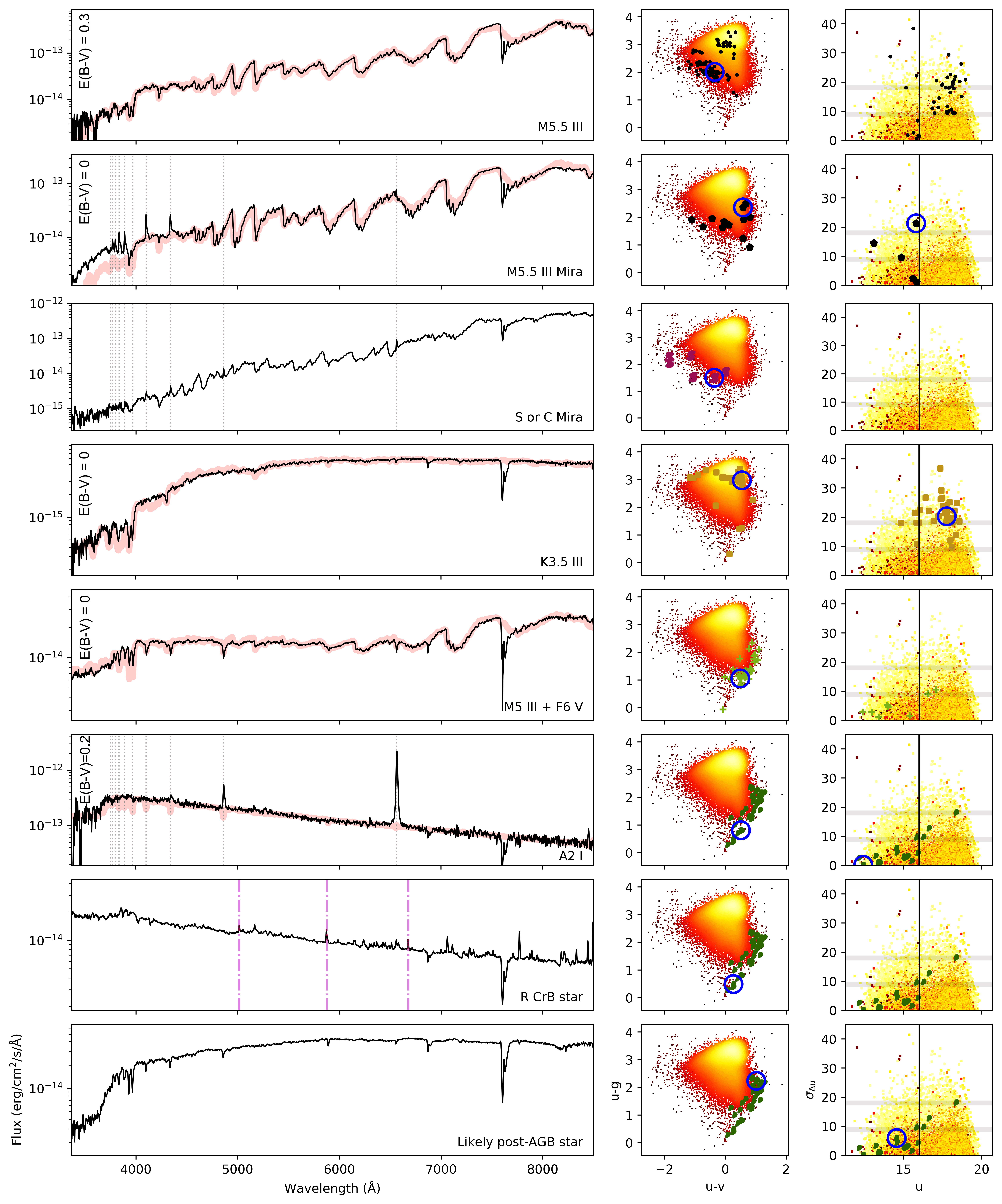}
    \caption{Spectra of emblematic examples of non-symbiotic targets, with overplotted Pickles templates of the corresponding spectral type where applicable (thick pale red lines), alongside the distribution of their target category in SkyMapper parameter space. Distributions are for {\it observed spectroscopy targets only}. We caution against over-interpreting target selection effects as representing meaningful physics. In particular, targets were only drawn from the colour-colour stellar locus if they had high $\sigu$, and isolated giants are concentrated in the stellar locus, so it is unsurprising that isolated giants have high $\sigu$ in these plots. Rows from top to bottom: 1. IRAS 13412-6515 representing M I--III stars (non-Miras only). The distribution in the SkyMapper panels reflects our selection criteria for optical spectroscopy. 2. V1339 Cen representing O-rich Miras. Distribution dominated by our selection criteria for optical spectroscopy convolved with mild peripherality in colour-colour space. Note the Balmer increment. 3. WRAY 18-250, a known S or carbon Mira, representing S and carbon stars. Very blue in u-v. Note the Balmer decrement. 4. {\it Gaia} DR3 5016002492939071744 representing K giants. Distribution dominated by our selection criteria for optical spectroscopy. 5. {\it Gaia} DR3 5340083854481489920 and/or 5340083854494748672 representing M/K star + warm/hot star superpositions. u-v strongly affected by the warmer star's spectrum. 6. CPD-75 116, a known hot supergiant, in the hot/post-AGB stars group. Members of this group well-fit by Pickles templates tend towards bluer u-g values within the group. 7. MV Sgr, a known R CrB post-AGB member of the hot/post-AGB stars group. 8. {\it Gaia} DR3 4202289873211624704, which we propose is a yellow post-AGB star, with a steep drop-off at short wavelengths due to strong Balmer jump absorption of the sort described in \citet{Bond1997}, yielding redder u-v values. Members of the hot/post-AGB stars group with steep drop-offs like this tend towards redder u-g and u-v values within the group.}
    \label{spectra_contaminants}
\end{figure*}

Rather than simply optimizing the discovery rate of symbiotics, our optical spectroscopic survey aimed to perform a census of the outlying regions of SkyMapper parameter space. In \autoref{symbioticsinspace} we plot the distribution of SkySy, SkySyC, and previously-known symbiotics in SkyMapper parameter space. In \autoref{allresults}, we present a summary figure illustrating the positions of all our 234 targets and previously-known symbiotics in SkyMapper parameter space, split into different categories of object classification.  In \autoref{spectra_contaminants}, we plot spectra of emblematic examples of non-symbiotic targets alongside their broader categories' distribution in parameter space. The object type categories include:

\begin{enumerate}[align=left,leftmargin=*,rightmargin=0ex,labelsep=0ex]\itemsep0.3em

    \item SkyMapper symbiotics (SkySy) and SkyMapper symbiotic candidates (SkySyC), subject to the criteria described in \autoref{Definitions}. The ones found through colour-colour alone are extremely well-localized to a mostly-symbiotic zone in the lower-left of the u<16 colour-colour plot, to the point where one could draw a somewhat wiggly zone in SkyMapper u-g/u-v space in which 100\% of the objects are symbiotics with the exception of easily-filtered AGN. This region, the precise boundaries of which should not be taken seriously, is drawn for reference in \autoref{allresults}. As expected, it appears that the hot component, whether burning or accreting-only, can cause symbiotics to have a flatter slope in the short-wavelength spectrum probed by SkyMapper u, v, and g. The remaining SkySy and SkySyC were found deeper in the colour-colour distribution with the help of $\sigu$.
    
    \item Post-AGB stars or warm/hot stars (concatenated into one group). These are objects resembling spectral type A or F, and which in the majority of cases bear a strong resemblance to our spectra of a few known post-AGB stars that we ended up accidentally targeting. The most distinctive feature is what appears to be extremely strong Balmer jump absorption. This is expected from yellow post-AGB stars; the Balmer jump is very sensitive to surface gravity in the A--F temperature range, and post-AGB stars are very low-mass and therefore have very low surface gravities. This idea is discussed by \citealt{Bond1997}; see their Table 1, showing surface gravities for yellow post-AGBs of log(g) in the range 0.4 to 1.25. Post-AGBs have high luminosity and probably still have plenty of circumstellar dust leftover from the AGB phase, explaining how they met our IR criteria. The extremely strong Balmer jump absorption explains why our post-AGBs have very red u-v colours and were selected as outliers; in fact, \citet{Bond1997} proposed using a Gunn u-band filter, which is quite similar to SkyMapper's Str{\"o}mgren u, to find yellow post-AGBs. We find that SkyMapper's u and violet v filters, which are intentionally designed to bracket the Balmer discontinuity, effectively implement \citet{Bond1997}'s idea with an all-sky survey, and are very effective tools for finding yellow post-AGBs. Interestingly, we did not find any new symbiotic stars in this post-AGB and superpositions region of SkyMapper parameter space, even though Hen 2-104 (the Southern Crab) is located in this region.
    
    \item Superpositions of M or K I--III stars with a warm or hot star, often resembling an A or F star on the short-wavelength end, with TiO bands on the long-wavelength end. At least a few of these are almost certainly spatial superpositions of unrelated stars. Others may be genuine binaries comprised of an M giant and a luminous A or F star (cf. \citealt{Neugent2018}).\footnote{Two of these superpositions have interesting archival light curves that may be worth tentatively considering as symbiotic candidates. A source at 10:43:25.6 -53:05:47, with a composite spectrum of M6 III and an A or F-like star, exhibits hints of stochastic variability around 0.15 mag peak-to-peak on timescales of a month in ASAS-SN light curve data. HD 123356, which has been conceptualized in the literature as a possible "Vega-type" object with a circumstellar disk and/or a possible superimposed late secondary \citep{hd1,hd2}, has an ASAS-SN light curve with possible low-amplitude oscillations on timescales of months superimposed on a secular decline of 0.2 mag over a few years.}
    
    \item Miscellaneous objects, including two planetary nebulae (PNe), one massive Wolf-Rayet WC7/8 star, and one AGN. These are all located around the ``known AGN'' region of SkyMapper parameter space from \autoref{simbadlabels}. Recall that we intentionally did not make cuts to remove AGN for fear of cutting EF Aql-alikes, and that planetary nebulae can be both very red \citep{Corradi2008} and very luminous. The new AGN, serendipitously, turned out to be of interest for its extreme luminosity \citep{Onken2022}.
    
    \item S and carbon stars. These are located near the known SIMBAD S and carbon stars labelled in \autoref{simbadlabels}, but with slightly bluer u-g colour. It is not clear why S and carbon stars overall are so often blue outliers in u-v; higher SNR on the short-wavelength end of the spectrum would be helpful towards answering that question.
    
    \item Various separate categories of G, K, M, Mira I-III stars. There is a high density of M giants in the low-density colour-colour outskirts just above and to the left of the mostly-symbiotic zone in the left panel. They comprise a wide range of temperatures, and it is unclear what made these M giants stand out from the main distribution in the bright subsample. Miras are more widely distributed, including in the vicinity of the lower-right tail of the colour-colour distribution. Most stars in the highest-density concentration at the top of the colour-colour distribution are isolated cool giants of various kinds. We will discuss the regions in which our $\sigu$ metric worked to identify symbiotics, and where it did not work to identify symbiotics, in the next section.
    
\end{enumerate}

\section{Discussion}\label{Discussion}

\subsection{Accretion discs detected}

\begin{figure*}
\centerline{\includegraphics[width=\textwidth]{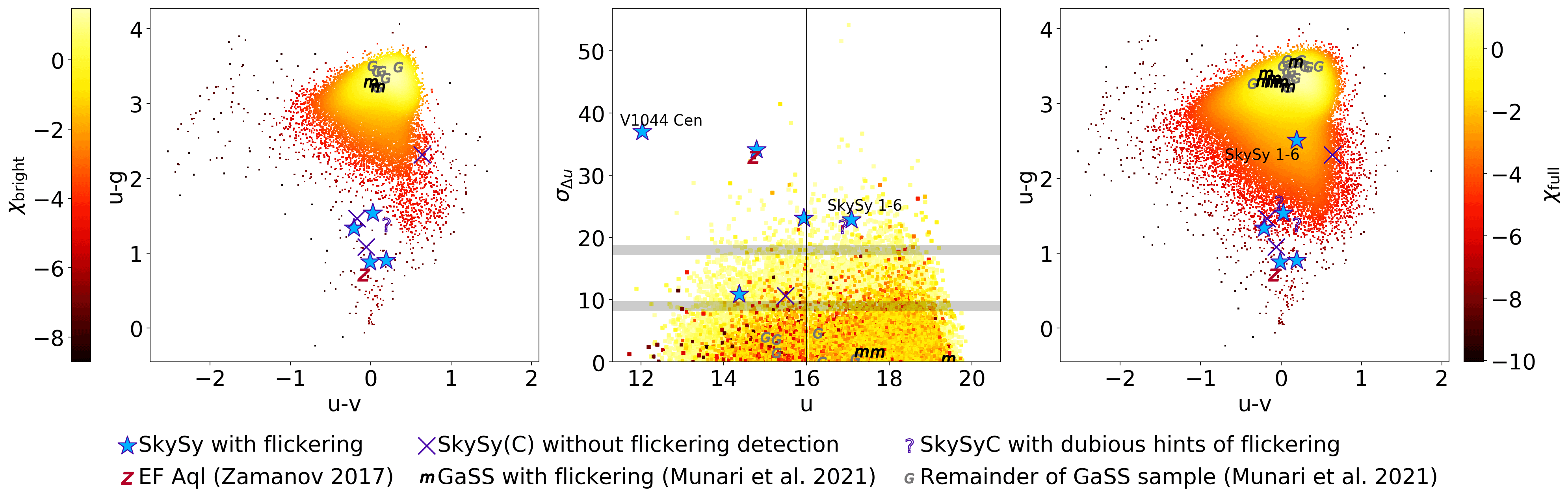}}
\caption{The intersection of our luminous red objects sample with flickering symbiotics, overlaid on \protect\autoref{parameterspace}. Our results include SkySy with confirmed detected flickering (plus the previously-known symbiotic V1044 Cen, in which we have discovered likely flickering for the first time), SkySy and SkySyC that were also observed with an hours-long continuous optical light curve but in which flickering was not detected, and SkySyC with dubious hints of flickering in either LCO or AAVSO data (SkySyC 1-5 and SkySyC 1-10). The previously-known flickering symbiotic EF Aql from \citet{Zamanov2017} is also in the luminous red objects sample. Of particular note is our confirmed flickerer SkySy 1-6, which demonstrates that neither colour-colour deviations alone nor variability alone are sufficient to find all the flickering symbiotic stars. We have also plotted the intersection of our luminous red objects sample with the \citet{Munari2021} GaSS symbiotics, including those with reported optical flickering (m symbols) and the remainder (G symbols). Where data are available, there was no evidence for flickering or colour deviations in the GaSS objects in our sample, indicating that we are probing different regimes of parameter space from \citet{Munari2021}.} \label{flickerers}
\end{figure*}

We have discovered optical accretion disc flickering in five symbiotic stars: the newly discovered SkySy 1-2, 1-4, 1-6, and 1-11, and likely the previously known symbiotic V1044 Cen, which had not previously been shown to flicker. We are aware of no other possible explanation for stochastic variability with these amplitudes observed on time-scales of minutes in a cool giant system. Cool giant pulsations are too slow and monotonic, and the observed peak-to-peak variability of 5--15\% in 
at most about 30 minutes is substantially greater than the expected thermal-timescale WD shell-burning variability of 
at most about 1\% change per 10 minutes (\autoref{introflickering}). The peak-to-peak time-scales of around 10--30 minutes in the optical flickering suggest a compact-object accretor, because the dynamical and viscous time-scales at an accretion disc inner radius set by the size of a main sequence accretor would be expected to be orders of magnitude longer \citep[e.g.,][]{Sokoloski2010}.

Of these five optically flickering symbiotics, all five exhibit Balmer emission with a decrement, three exhibit emission He I emission, one exhibits [\ion{O}{iii}] emission, and none exhibit higher ionization states (\autoref{tableskysy}). The H$\alpha$ pseudo-equivalent widths are low among these flickering symbiotic stars, ranging from 6\AA ~to 100\AA. These emission line properties are consistent with the idea that flickering is most often detectable in accreting-only symbiotic stars (acc-SySt in the nomenclature of \citealt{Munari2021}) without WD shell burning.

Besides flickering, another direct accretion signature is hard X-rays photons with energies above 2.4 keV: $\delta$-type according to the taxonomy proposed by \citet{Luna2013}. This idea is based on the \citet{Patterson1985b} model for hard X-ray emission from the optically-thin shocked boundary layer, or the optically-thin halo of the optically-thick boundary layer, between the quickly-rotating accretion disc and the slowly-rotating WD. Hard X-rays have not been detected in symbiotics with the tell-tale signatures of shell burning, never being present alongside nuclear-powered\footnote{By specifying ``nuclear-powered'', we exclude only the lower-luminosity very soft X-ray component in T CrB during its accretion high state, which likely originated in the accretion disc boundary layer as it became optically thick \citep{Luna2018}---quite unlike classical supersoft components in symbiotics, which originate in shell burning.} supersoft X-ray emission \citep{Luna2013} or high-ionization Raman \ion{O}{vi} emission. \citet{Luna2013} proposed that extreme ultraviolet (EUV) photons from shell burning cool the accretion disc boundary layer in burning symbiotics via Compton cooling \citep{Weast1981}, suppressing the production of hard X-rays by hot plasma. Perhaps for this reason, hard X-rays have only been detected in about 25 symbiotics, and about 10 of those are likely neutron star accretors \citep[e.g.,][and references therein]{Luna2013,Mukai2016,Merc2019,Lima2024}. The light from WD shell burning is also reprocessed into the UV and optical wavebands by the surrounding wind nebula, outshining the flickering accretion disc and explaining the correlation observed between UV/optical flickering and $\delta$-type X-ray emission \citep[e.g.,][]{Sokoloski2001, Luna2013}.

Supporting an accretion disc origin for our flickering targets, and the absence of shell burning, we detected X-rays with photon energies above 2.4 keV in both targets for which we obtained {\it Chandra} X-ray exposures: SkySy 1-2 and SkySy 1-4.\footnote{It should be noted that, while there had previously been a claimed X-ray detection for our flickering V1044 Cen as well, this X-ray source is in fact associated with the same neighboring eclipsing binary that we mentioned in our \autoref{Flickering} \citep{Lima2024}.} The highest-count spectrum (SkySy 1-2) requires an emitting plasma temperature higher than around 1$\times10^{7}$ K. The X-ray spectra closely resemble the $\delta$ or $\beta/\delta$ types in the \citet{Luna2013} schema for symbiotic star X-ray classification. The X-ray {\it Chandra}-band observed luminosity for SkySy 1-2 was at least 5 $\times10^{30}$ erg s$^{-1}$ with 90\% confidence, and the low-count spectrum of SkySy 1-4 nevertheless had a higher count-rate than SkySy 1-2 would at the former's {\it Gaia} distance. This luminosity is towards the low end of the typical range for symbiotic stars \citep{Luna2013}, but red giants probably\footnote{We rely here on the current community consensus ascribing X-rays from M giants in the $10^{30}$ -- $10^{31}$ erg s$^{-1}$ range to accretion onto a companion \citep{Sahai2015, Schmitt2024}. The long-standing standard paradigm is that giants redder than the ``X-ray dividing line'' do not have exposed hot coronae (e.g., \citealt{Schmitt2024}, and references therein). Supporting this consensus, coronal X-ray luminosities in warmer stars are usually far lower than $10^{30}$ -- $10^{31}$ erg s$^{-1}$ \citep{Schmitt2024} except in the case of rapid rotation. And coronal temperatures in warmer stars with X-ray-emitting coronae are typically less than about 1$\times10^{7}$ K (1 keV), with values up to a few $\times10^{7}$ K possible for some stellar types. Sensitive XMM exposures on two unusually magnetic AGB stars have yielded no detections, with upper limits of $10^{30}$ erg s$^{-1}$ if the temperature is assumed to be at least $10^{7}$ K \mbox{\citep{Kastner2004}}.} cannot produce X-ray emission with this temperature and luminosity on their own \citep{Sahai2015,Schmitt2024}. The $\delta$-type appearance of the X-ray spectra hint at a WD accretor, but their low SNR cannot rule out an absorbed power law from a neutron star accretor.

\subsection{Finding symbiotics with detectable accretion disc flickering}\label{findingflickerers}

In the process of discovering these flickering symbiotics, we have shown that variability between the three u-band exposures within a 20-minute SkyMapper Main Survey {\it uvgruvizuv} filter sequence (of which there are up to two per object) can efficiently detect accretion disc flickering. The locations in parameter space of the new flickering symbiotics are shown in \autoref{flickerers}, along with other samples to be discussed. Among our emission-line cool giants observed with follow-up LCO light curves, 5 out of 7 with SkyMapper $\sigu$>2$\sigma_{\rm eff}$ exhibited LCO B-band flickering. And of course, as motivated our search design in the first place, the only previously-known flickering symbiotic in our luminous red objects sample---EF Aql \citep{Zamanov2017}---also exhibited SkyMapper $\sigu$>2$\sigma_{\rm eff}$. In contrast, 0--1 out of 4 emission line cool giants without a SkyMapper $\sigu$ measurement exhibit B-band flickering in LCO.

Four out of the 5 new optical flickerers would have been selected as a symbiotic by the u-g/u-v colour-colour diagram alone, but we would not have known to look for flickering in them without a $\sigu$ measurement. One of those four, V1044 Cen, has been classified as a symbiotic star since 1968 \citep{Carlson1968}, but was never reported to be flickering. Extensive searches in the known sample of symbiotics often have not revealed optical flickering; for example, \citet{Sokoloski2001} obtained B and/or U band light curves of 35 symbiotic stars, and only detected unambiguous flickering in five that had already been known to flicker beforehand. It is apparent that a selection mechanism to detect optical flickering in symbiotics is motivated, and that SkyMapper's filter sequences are well suited to the task.

However, we have found that there is an important caveat to the utility of the SkyMapper {\it uvgruvizuv} filter sequences in finding flickering symbiotics, that must guide future search strategies: $\sigu$ does not work without additional selection criteria. As shown in \autoref{allresults}, the vast majority of targets even with $\sigu$ above $3\sigma_{\rm eff}$ are cool giants without any hint of emission lines in our optical spectra. This includes targets selected by $\sigu$ alone (the green triangles in \autoref{targets}, none of which exhibit emission lines) and targets on the edges of the densest part of the colour-colour diagram (the line of normal giants at u-g$\approx$3 in the right panel of \autoref{allresults}). Targets with $\sigu$ between $2\sigma_{\rm eff}$ and $3\sigma_{\rm eff}$ and close to the edge of the luminous red objects distribution in colour-colour space (but not in the mostly-symbiotic zone of colour-colour space; rather, see the arc of black squares just interior to the arc of magenta circles in the right panel of \autoref{targets}) also show no signs of being symbiotic. Admittedly, it is conceivable that all these null results are flickering symbiotic stars during a low-accretion state without emission lines, because it would have been prohibitively time-consuming to observe them with LCO, {\it Chandra}, or long-term monitoring, and we do not know the duty cycle of symbiotic activity (e.g., \citealt{Pujol2023}). But they are probably just isolated giants.

\begin{figure*}
\centerline{\includegraphics[width=\textwidth]{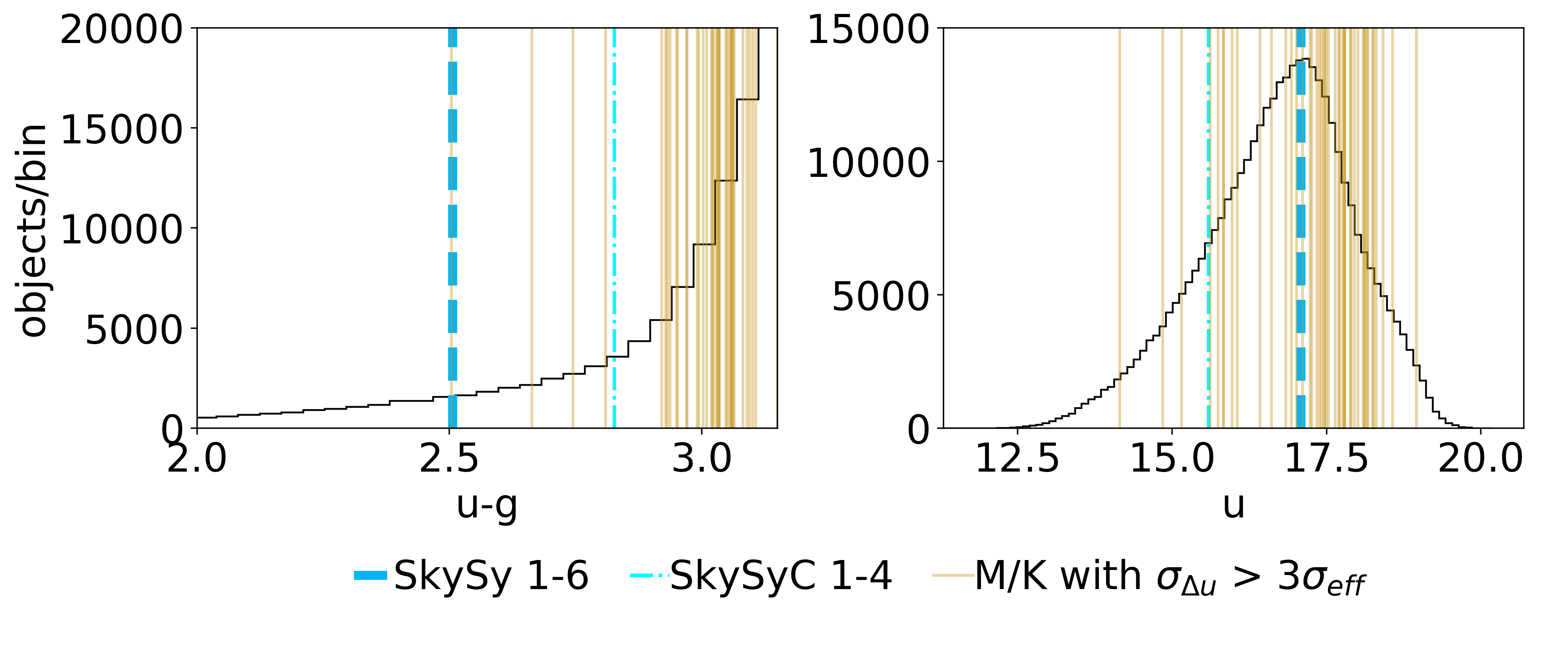}}
\caption{The distribution in u-g and in u of spectroscopically normal M and K giants that had $\sigu > 3\sigma_{\rm eff}$ in the middle panel of \autoref{allresults} (vertical gold lines), the full sample of luminous red objects (black histogram), and the locations of SkySy 1-6 (vertical dashed thick dark-blue line) and SkySyC 1-4 (vertical dash-dot light-blue line) as a function of SkyMapper u-g colour (left) and SkyMapper u magnitude (right). SkyMapper variability detections in symbiotic candidates are more likely to signify real flickering when the u-g colour is bluer. Our results imply that u flickering is usually only detectable when there is sufficient accretion disc light to produce an excess in u.} \label{distribution}
\end{figure*}

The solution, and the method with which we found flickering symbiotics, is to incorporate both $\sigu$ {\it and} colour criteria. Certainly, anything within the mostly-symbiotic zone of the colour-colour diagram in the left panel of \autoref{allresults} and high $\sigu$ is likely to be a flickering symbiotic star, which is how we found 4 of our new flickerers. But we were also able to find the flickering symbiotic SkySy 1-6 (labelled in the right panel of \autoref{flickerers}), which is located very deep into the distribution of luminous red objects in the colour-colour diagram.

The parameter-space feature that allowed us to discover the hidden SkySy 1-6 is that there is an abrupt cutoff around u-g$\approx$2.9, redward of which $\sigu$>$3\sigma_{\rm eff}$ is generally spurious, but blueward of which a target with $\sigu$>$3\sigma_{\rm eff}$ has at least a 1/3 chance of being a flickering symbiotic. We show this in \autoref{distribution}, in which we plot the locations of SkySy 1-6, SkySyC 1-4, and spectroscopically normal M and K giants with $\sigu$>$3\sigma_{\rm eff}$ superimposed on a histogram signifying the distribution of the full sample of luminous red objects. It is not the case that a constant fraction of $\sigu$>$3\sigma_{\rm eff}$ objects are flickering symbiotics independent of u-g value; if it were, based on the results blueward of u-g$\approx$2.9 (i.e., u-g < 2.9), about a third of the spectroscopically normal giants clustered around u-g=3 in the left panel of \autoref{distribution} would be symbiotic. Nor can the u-g$\approx$2.9 cutoff be explained by a higher rate of spurious $\sigu$ detections at dim u-band magnitudes; the concentration of spectroscopically normal giants at u>17.1 suggests this might conceivably play a role, but if it were the sole factor, SkySy 1-6 would not be at the u=17.1 distribution peak, dimmer than 14 spectroscopically normal giants in the right panel. And while the confluence of two rare phenomena (like blue u-g and $\sigu$ variability) would be unlikely to occur spuriously by chance even if the physical processes giving rise to colour and variability were physically independent of each other, blue u-g is not actually particularly rare in the distribution relative to the location of the spectroscopically normal giants, with about 1600 objects/bin at the u-g of SkySy 1-6 and about 6000--15000 objects/bin at the u-g of the normal giants cluster.

Rather, the fact that bluer u-g colour makes $\sigu$ variability more likely to be real flickering is probably 
attributable to an intuitive physical connection: that u flickering is usually only detectable when there is sufficient accretion disc light to produce an excess in u. This is consistent with our argument that we are detecting real flickering produced by a mechanism operating in the u band (the accretion disc) that is distinct from the mechanism operating at redder wavelengths (the cool giant). It also suggests an outline of a strategy for future searches for flickering symbiotics using SkyMapper: restrict future  searches to three overlapping groups, including (1) known symbiotics with $\sigu$ above $2\sigma_{\rm eff}$, (2) luminous red objects with colours inside the mostly-symbiotic zone of our colour-colour diagram and $\sigu$ above $2\sigma_{\rm eff}$, and most excitingly (3) luminous red objects with \mbox{u-g $\lesssim$ 2.9} and $\sigu$ above $3\sigma_{\rm eff}$.

\subsection{Completeness of prior searches}

As shown in \autoref{tableskysy}, the flickering accreting-only symbiotics SkySy 1-6 and (depending on the date of observation) \mbox{SkySy 1-2} could have been missed by any narrow-band emission line photometry survey requiring a minimum H$\alpha$ pseudo-equivalent width of 10\AA, and the flickering accreting-only symbiotics SkySy 1-11 and V1044~Cen could have been missed by any such survey requiring a minimum H$\alpha$ pseudo-equivalent width of 50\AA~(such as the symbiotic searches by \citealt{Corradi2008} and \citealt{Rodriguez2014}). Most symbiotic stars have historically been found through objective prism plate surveys, as discussed in \citet{Munari2021}. The threshold for the detection of H$\alpha$ amid TiO bands in 20th-century objective prism spectra is to our knowledge not well constrained, but it is probably at least 10\AA. This suggests that our SkyMapper selection methods (even without, but especially with, the incorporation of $\sigu$) are less biased against the selection of accreting-only and optically flickering symbiotics than the traditional methods of both narrow-band emission line photometry and objective prism plate surveys---though our colour-colour method has proven effective in finding burning symbiotics too.

Quantitatively, we have sufficient data to address two overlapping categories of symbiotics to determine what fraction were missing from past surveys: missing symbiotics that can be found in the mostly-symbiotic zone of the u<16 u-g/u-v colour-colour diagram, and missing flickering symbiotics.

\subsubsection{Missing symbiotics in the mostly-symbiotic zone}

\begin{figure}
\includegraphics[width=\columnwidth]{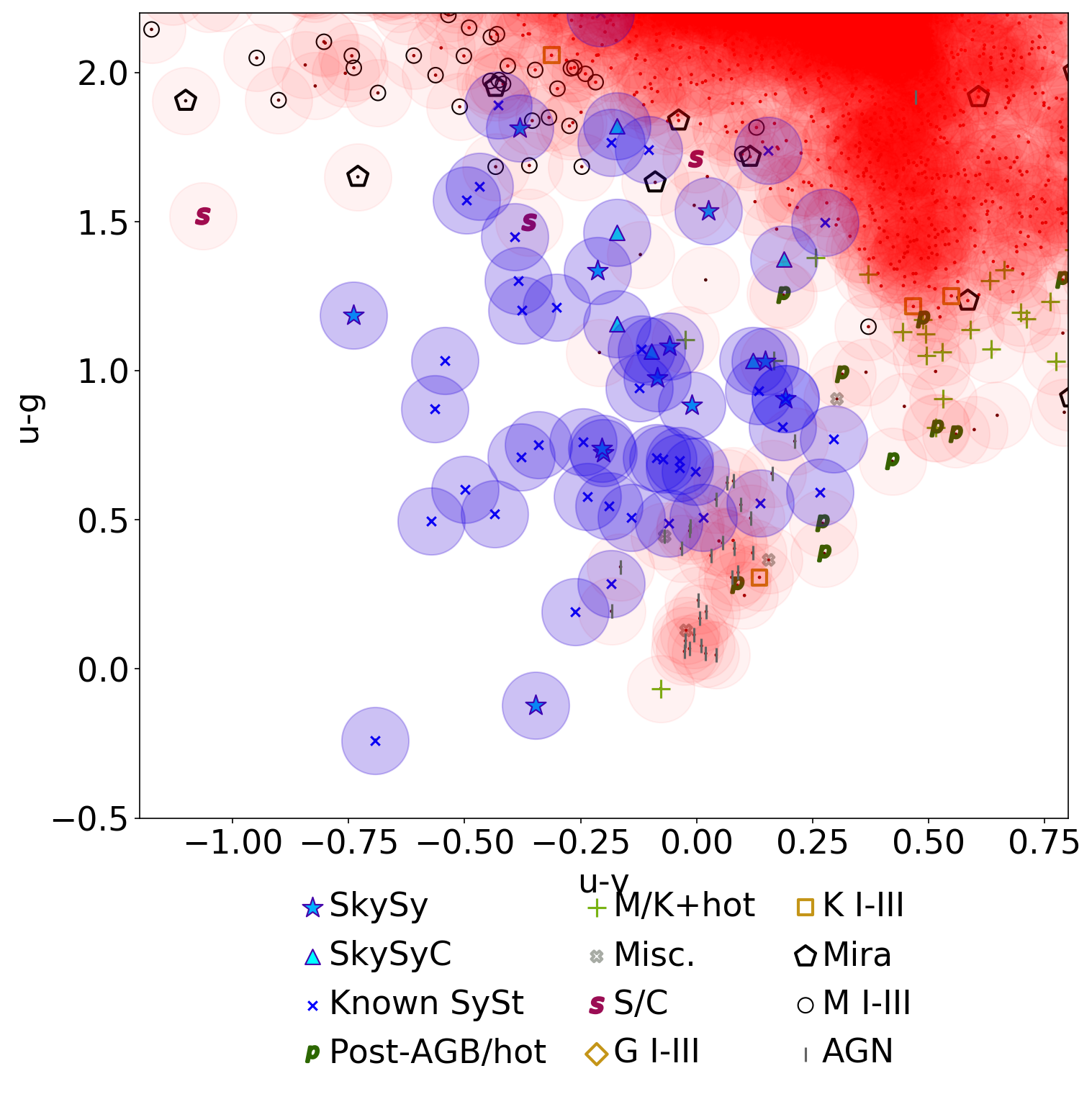}
\caption{Zoomed-in portion of the left panel of \autoref{allresults} showing the types of objects we identified in the mostly-symbiotic zone of the colour-colour diagram in our bright subsample. A large transparent red circle has been added to each object (whether we obtained follow-up optical spectroscopy or not) to help illustrate the source density. To illustrate the predominance of symbiotics in the mostly-symbiotic zone, red circles are replaced by blue circles for SkySy, SkySyC, and previously-known symbiotics. The mostly-symbiotic zone is bounded by M giants to the upper left, post-AGBs/hot stars/superpositions to the right, a mix of AGN and other contaminants to the lower-right, and a rapid rise to high source density to the upper right.} \label{zoom}
\end{figure}

For symbiotics with strong u excess in u-g and u-v, we have almost fully explored the mostly-symbiotic zone of the u<16 bright subsample of our luminous red objects (\autoref{zoom}) modulo a very few targets with inconvenient sky coordinates. The mostly-symbiotics zone has clear boundaries, with M giants to the upper left, post-AGBs/hot stars/superpositions to the right, a mix of AGN (easily filtered using AGN catalogs, IR colours, or {\it Gaia} proper motions) and other contaminants to the lower right, and a rapid rise to high source density to the upper right that would be too time-consuming to explore. Thirty-nine out of the 42 previously-known symbiotics (as catalogued and validated by \citealt{Merc2019}) intersecting with our luminous red objects sample and with u<16 are in this zone, leaving out just one barely outside the boundary, one much deeper in the distribution, and the Southern Crab. We have found 11 SkySy and 6 SkySyC in this zone. 

So the completeness of the old sample of symbiotics with strong u-g and u-v excess and u<16 was about 70--80\% depending on how many SySyC are real symbiotics. We can generalize this completeness level to the full set of 274 \citet{Merc2019} symbiotics, corresponding to a roughly estimated 205 \citet{Merc2019} symbiotics with u<16.\footnote{Based on the ratio of targets with u<16 to targets with u>16 among those that intersect with our luminous red objects sample.} We conclude that there were about 51 to 88 symbiotics with u<16 (of which 11 to 17 have been reported in this work) missing from the \citet{Merc2019} sample prior to our first observation run, that can be easily found with a near-zero contamination rate through the u-g/u-v colour-colour diagram when {\it uvg} colours are available for the whole sky. This can be accomplished by future data releases of SkyMapper in the south, and the future Multi-channel Photometric Survey Telescope (MEPHISTO; \citealt{Mephisto}) in the north, which will use a {\it uvgriz} filter set very similar to SkyMapper. If improved depth in the final data releases of SkyMapper and MEPHISTO resolves the distinction between the u<16 and u>16 colour-colour diagram through improved SNR, one could then expect to find a net 68 to 117 symbiotics with any u through this method.

\subsubsection{Symbiotic/Post-AGB transitional objects}

Intriguingly, the previously-known symbiotic Hen 2-104 (the Southern Crab) appears to have a completely unique location for a symbiotic in our colour-colour diagram, with a large positive u-v colour, perhaps implying strong Balmer jump absorption. It is surrounded in our colour-colour diagram by post-AGB candidates and M giant/hot star superpositions. It has been argued that the donor in Hen 2-104 is at the end of its AGB phase, with an unusually massive nebula, constituting a planetary nebula precursor ``in which the nebular shaping and excitation is done by the WD companion, which `anticipates' the post-AGB evolution of the Mira'' \citep{Santander2008}. Its unique location amid post-AGB stars in our parameter space may support this idea of Hen 2-104 as a transitional object. There is one unclassified source near Hen 2-104 in SkyMapper colour-colour space, at ICRS coordinates $\alpha$=08h 22m 58.9s, $\delta$=-32$^{\circ}$ 50$^{\prime}$ 00.9$^{\prime\prime}$, which we have not observed.

\subsubsection{Missing optically-flickering symbiotics}

With respect to optically-flickering symbiotics, there was only one optically-flickering\footnote{We concentrate on optical flickering in this work, but the situation is not so different in the ultraviolet -- of the four southern-hemisphere possible UV flickerers in \citet{Luna2013}, only BD-21 3873 and Haro 1-2 intersect with our luminous red objects sample. Their SkyMapper colours resemble those of other symbiotics. BD-21 3873 is borderline in SkyMapper variability space at u=12.5 and $\sigu$ just below the 2$\sigma_{\rm eff}$ threshold; in \citet{Luna2013}, it exhibited UV variability in one sequence but not in its other sequence. Haro 1-2 does not have a $\sigu$ measurement here.} symbiotic intersecting with our luminous red objects sample before our search, and our search increased that count to six. This suggests that the completeness of past samples of known optical flickerers was at most around 17\% before our first observing run. Thirty known symbiotics and new SkySy (excluding SkySyC) have a $\sigu$ measurement in our sample, and out of these, six have now been shown to flicker in both $\sigu$ and targeted light curves. This suggests that at least around 20\% of the true population of symbiotics should be expected to have optical flickering detectable through SkyMapper $\sigu$. We note that the previously-known symbiotics LMC S154 and AS 241 are strong candidates for flickering in light of these results and their position in the middle panel of \autoref{symbioticsinspace}, especially given LMC S154's status as a symbiotic recurrent nova \citep{Ilk2019}. CD-43 14304 is a burning symbiotic with Raman \ion{O}{vi} emission and $\alpha$-type supersoft X-rays \citep{Schmid1998,Muerset1997}, so its apparent $\sigu$ excess in \autoref{symbioticsinspace} may make it a good candidate for detecting coherent oscillations like those in Z~And \citep{Sokoloski1999}. Finally, there are 4 SkySyC in \autoref{tableskysy} with excess $\sigu$ that may warrant further followup, two of which were not observed with LCO.

\begin{figure*}
\centerline{\includegraphics[width=\textwidth]{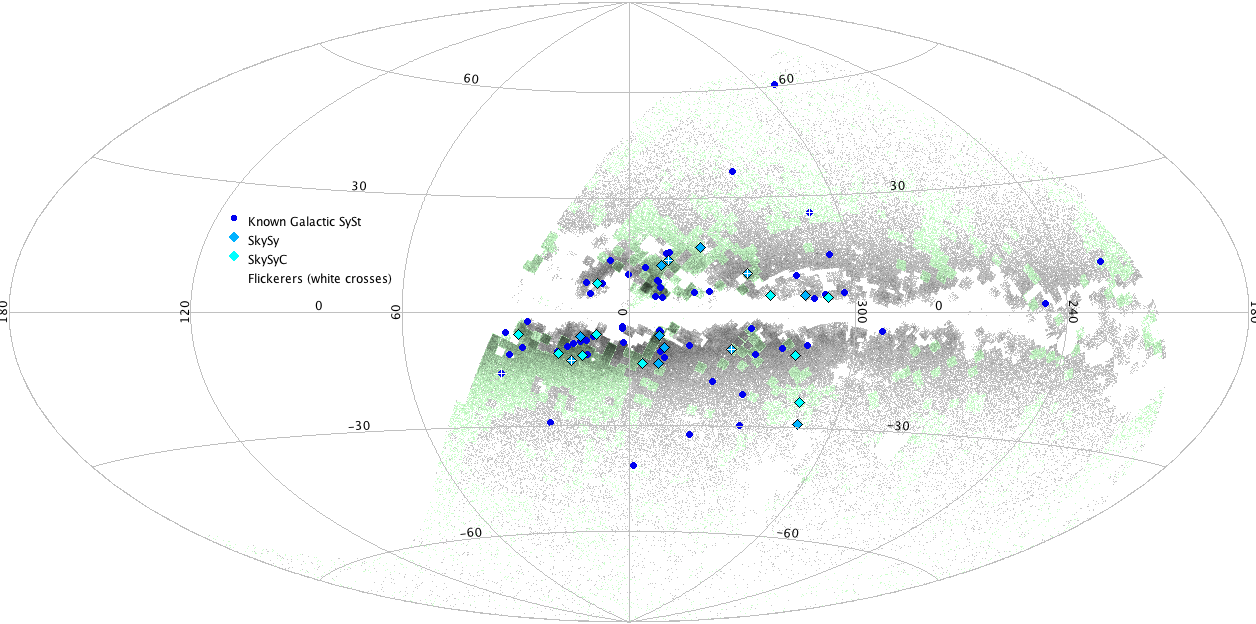}}
\caption{Density-shaded scatter plot in galactic coordinates of our luminous red objects sample (grey); an overlay of the sample for which $\sigu$ information was available (light green); and the members of the sample which are SkySy (blue diamonds), SkySyC (light blue diamonds), or previously-known symbiotics intersecting with our luminous red objects sample (dark blue circles). Optically-flickering symbiotics are overplotted with white crosses.} \label{mapresults}
\end{figure*}

As shown in \autoref{mapresults}, we appear to be searching in roughly the same parts of the sky as prior surveys, so the differences between our search and prior symbiotic star surveys are methodological, not spatial. The exceptions are SkySy 1-12 and SkySyC 1-4, which while typical in Galactic coordinates, have more southern declinations than any previously-known symbiotic (\citealt{Merc2019}). We also note that flickering symbiotics exhibit a hint of being slightly preferentially at higher galactic latitudes in this figure, but they are also slightly preferentially closer in {\it Gaia} distances, so these do not translate into greater heights above the Galactic plane. As shown in \autoref{mapresults}, the lack of SkySy and SkySyC towards the Galactic anti-center reflects both observing restrictions and the smaller number of stars overall towards the anti-center, and we do not draw any new conclusions about the overall spatial distribution of symbiotic stars. The distribution of both previously-known symbiotics and new SkySy/SkySyC stars over galactic latitude is just a low-resolution version of the distribution of luminous red objects in our full sample.

We note that symbiotic candidates selected through SkyMapper colours alone cannot evidently be classified as burning versus accreting-only without some other observable. Burning symbiotics are themselves just barely outside of the locus of luminous red objects in u-g/u-v colour space. Accreting-only symbiotics probably have an even tougher time distinguishing themselves from the locus of red giants, and those that manage it are those which resemble the optical colours of burning symbiotics. Objects inside the colour-colour locus of luminous red objects, however, are more likely to be accreting-only; the trick is detecting them, through a combination of SkyMapper variability and subtle colour effects as discussed in \autoref{findingflickerers}, or through other observables such as faint Balmer lines \citep{Munari2021}, X-rays \citep{Mukai2016}, or infrared properties \citep{Akras2019}.

The question of whether there is a hidden population of accreting-only symbiotics that is {\it larger} than the population of burning symbiotics remains unanswered. One issue that prevents us from assessing completeness in a more comprehensive way is that the duty cycle with which 2--3 exposures in a 20 min SkyMapper filter sequence capture flickering is still unconstrained. What we do know now, however, is that there is a significant population of optically-flickering symbiotics hidden both within and beyond the known catalogs of symbiotic stars.

\subsection{Comparison to recent surveys}\label{comparison}

Until this point, we have for the sake of historical clarity focused on comparing our results to the state of affairs that preceded our first observing run. Several innovations in symbiotic discovery have occurred since then. Here we endeavor to put our results in conversation with the recent surveys most relevant to our results: \citet{Munari2021}, \citet{Akras2019}, \citet{Akras2023}, and \citet{Merc2024}.

\subsubsection{GALAH flickering symbiotics with non-pulsating donors (Munari et al. 2021)}

While we were analyzing our observations, \citet{Munari2021} published the initial results of an independent search for accreting-only symbiotic stars without nuclear burning, using the Galactic Archaeology with HERMES (GALAH) survey. They present evidence that they have detected 33 acc-SySt after selecting for M giants without coherent radial pulsations\footnote{The idea is that M giant Balmer emission is stronger from strongly pulsating giants (see our \autoref{introflickering}), so by avoiding strong pulsators, \citet{Munari2021} increase the likelihood that any Balmer emission originates from an accreting companion rather than the giant.} and with an H$\alpha$ peak at least 50\% above a spectral template in high-resolution GALAH spectra, including 12 acc-SySt for which they reported low-level optical flickering.

Interestingly, our work and \citet{Munari2021} appear to be probing completely different regimes in SkyMapper parameter space. We include in \autoref{flickerers} the intersection of our luminous red objects sample with the \citet{Munari2021} GALAH Symbiotic Star (GaSS) flickerers and (separately) the full GaSS sample. The GaSS sample, including the low-level flickerers, are located in the densest part of the colour-colour diagram, appearing as completely normal giants. There is no hint of any $\sigu$ excess in the reported flickerers or in the full sample. We also inspected objects which do not intersect with our luminous red objects sample but which do have a SkyMapper u-g colour; there is only one GaSS with a SkyMapper DR2 u-g less than 3.1, GaSS 1-2 at u-g=1.8 in the SkyMapper dr2.master catalog.\footnote{There is no v-band measurement, and the u and g measurements are separated from each other by four days: at least two reasons for its exclusion from our luminous red objects.} The lack of SkySy and SkySyC in the GaSS sample is at least in part simply because none of them were observed in GALAH DR3 \citep{galah}, but there might be another reason too, as follows.

The GaSS sample objects are clustered around early M spectral types, ranging from M0 to M4 in the flickering subsample. In contrast, four of our new flickering symbiotics contain giants in a narrow range from M5--6. Our fifth, SkySy 1-11, contains an S star otherwise reminiscent of M4. The earlier spectral types in the GaSS sample are an expected consequence of Munari et al.'s exclusion of M giants with strong radial pulsations.

It is important to note that while some of our SkySy have extremely faint Balmer lines compared to most known symbiotics, the GaSS Balmer lines are even fainter than that. Their GALAH spectra have a spectral resolving power of 28,000, compared to our 600; and they measure emission line height above a spectral template with absorption lines, while we can only look at the neighboring pseudo-continuum. Most of the GaSS H$\alpha$ line profiles have a P-Cygni-like or otherwise complexly-absorbed profile in a bandpass smaller than our spectral resolution, which would cause them to be non-detections, or even weaker detections than SkySy 1-6, if smoothed to our lower spectral resolution. This includes all of the GaSS objects with sufficient SkyMapper data to appear in the middle panel of our \autoref{flickerers}.

Our preliminary impression is that the GaSS flickerers may be genuine flickering symbiotics, but with lower accretion disc contributions to the optical and consequently lower-amplitude flickering than our flickering SkySy. This would be consistent with both the relatively low-amplitude of variability in the GaSS light curves and the weaker Balmer emission in the GaSS spectra. In contrast to the earlier spectral type giants in the GaSS sample, we suspect that the later spectral type giants in the SkySy are transferring mass at a higher rate, consistent with the higher mass loss rates of late giants and leading to stronger u-band excesses and stronger Balmer lines in the SkySy flickerers.

\subsubsection{IR selection techniques (Akras et al. 2019)}

\citet{Akras2019} used machine learning to reliably distinguish between symbiotic stars and other objects (including isolated M giants) based solely on IR colours. The physical mechanism determining why known symbiotics have different IR colours than isolated giants is unclear, given that the giant dominates in the IR. Possible explanations include dust heated by the light of the symbiotic accretor, heating of the giant by the light of the accretor, IR emission from the cool edges of the accretion disc, or some property of symbiotic giants that correlates with higher mass transfer rates. In \autoref{tableskysyakras}, we run the 2MASS and ALLWISE colours of our SkySy/SkySyC through the decision trees illustrated by \citet{Akras2019} in their Figure 7 (a general tree disambiguating between symbiotics and mimics, landing in either a bin dominated by symbiotics or a bin in which symbiotics are rare), their Figure 9 (disambiguating between different IR subtypes of symbiotics, including S, S+IR, D, and D'), and their Figures A1 through A7 (disambiguating between symbiotics and specific types of mimics, assuming that the object is either a symbiotic or the types of mimics addressed by the particular decision tree). 

Interestingly, 10 out of our 12 SkySy (including all 4 flickering SkySy), and 4 out of our 10 SkySyC, meet the criteria of the general decision tree (their Figure 7) to count as a strong symbiotic star candidate, suggesting that there is substantial overlap in the effects of our selection methods despite operating at such disparate wavelengths. Four of our SkySy and SkySyC, however, get classified as M giants instead of symbiotics in their Figure A5---including the flickering SkySy 1-2 and 1-6, by extremely thin margins. This suggests that our flickering, accreting-only symbiotics may be on the boundary of the \citet{Akras2019} selection criteria, presumably because the available training set of known symbiotic stars is dominated by burning symbiotics, and that whatever physical mechanism changes the IR colours of symbiotic stars is operating to a lesser degree in accreting-only symbiotics than in burning symbiotics. Similarly, several other SkySy and SkySyC get classified as Miras in their Figure A2 or K giants in their Figure A5---probably accurately with respect to the nature of the donor, but inaccurately with respect to the presence of binary interaction. And two SkySy, along with six SkySyC, do not meet the criteria of the \citet{Akras2019} Figure 7 general decision tree to count as symbiotic candidates. These are often Miras in Akras Figure A2 and in SIMBAD, reinforcing what \citet{Akras2019} have themselves noted: that IR criteria have difficulty distinguishing symbiotics from Mira variables.

\subsubsection{GALEX colours (Akras et al. 2023)}

\citet{Akras2023} showed that the GALEX FUV-NUV colour can be used to uncover symbiotic stars. Only SkySy 1-12 and SkySyC 1-5 have data available in GUVCAT\_AIS (\href{https://doi.org/10.17909/t9-pyxy-kg53}{doi:10.17909/t9-pyxy-kg53}; \citealt{bianchi2017}), so we cannot say much about our search mechanism's compatibility with the GALEX UV criterion. SkySy 1-12 meets the FUV-NUV < 1 criterion, while SkySyC 1-5 does not.

\subsubsection{TESS flickering (Merc et al. 2024)}

\citet{Merc2024} used TESS to systematically search for optical flickering in a large sample of symbiotics, and found 20 flickerers of which 13 were not previously known to be flickering. Four of those 13 new flickerers are available in our sample of luminous red objects. However, none of these four has sufficient SkyMapper DR2 data to obtain a $\sigu$ measurement, so we cannot say anything about SkyMapper's ability to pick up their variability. We note also that \citet{Merc2024} discovered optical flickering in V420 Hya, which was eliminated from our sample of luminous red objects on at least the basis of saturated 2MASS photometry and, according to the SkyMapper DR2 pipeline, another source within 6 arcsec.

\citet{Merc2024} were able to use TESS to confirm the presence of optical flickering from SkySy 1-2 (whose flickering was also detected by SkyMapper in this work), and detect optical flickering for the first time in SkySy 1-3 (which did not have enough data for a $\sigu$ measurement in SkyMapper DR2). The latter classification means that the only two confirmed symbiotics in our sample of luminous red objects with S star donors (per the latest version of \citealt{Merc2019}), SkySy 1-3 and SkySy 1-11, both flicker.

\begin{landscape}
\begin{table}
\begin{threeparttable}
\begin{tabular}{llllll}
\hline
Internal name & Giant & Akras 7 & Akras 9 & Akras A1--A7 & Small $\Delta$ \\
\hline
{\it SkySy} & & & & & \\
SkySy 1-1 & M4 III & Symbiotic & S & \cellcolor{red!10} M giant in A5; Symbiotic in A1--A4, A6--A7 & $\Delta$(K-W3)=0.025 (A5) \\
SkySy 1-2 & M6 III & Symbiotic & S & \cellcolor{red!10} M giant in A5; Symbiotic in A1--A4, A6--A7 & $\Delta$(K-W3)=0.05 (A5) \\
SkySy 1-3 & S & Symbiotic & S & Symbiotic & - \\
SkySy 1-4 & M6 III & Symbiotic & S & Symbiotic & - \\
SkySy 1-5 & M0.5 III & Symbiotic & S & Symbiotic & - \\
SkySy 1-6 & M5 III & Symbiotic & S & \cellcolor{red!10} M giant in A5; Symbiotic in A1--A4, A6--A7 & $\Delta$(K-W3)=0.016 (A5) \\
SkySy 1-7 & M0 III & Symbiotic & S & Symbiotic & -  \\
SkySy 1-8 & M6 III & \cellcolor{red!10} Mimic & S+IR & \cellcolor{red!10} Mira in A2; YSO in A3; Symbiotic in A1, A4--A7 & - \\
SkySy 1-9 & M7 III (Mira) & \cellcolor{red!10} Mimic & S+IR & \cellcolor{red!10} Mira in A2; YSO in A3; Symbiotic in A1, A4--A7 & - \\
SkySy 1-10 & K4 I & Symbiotic & S & \cellcolor{red!10} K giant in A5; Symbiotic in A1--A4, A6--A7 & -\\
SkySy 1-11 & S & Symbiotic & S & Symbiotic & - \\
SkySy 1-12 & K? & Symbiotic & S & Symbiotic & - \\
V1044 Cen & M6 III & Symbiotic & S & Symbiotic & - \\
\\
\hline
{\it SkySyC} & & & & & \\
SkySyC 1-1 & M6.5 III & \cellcolor{red!10} Mimic & S & \cellcolor{red!10} Mira in A2; Symbiotic in A1, A3--A7 & - \\
SkySyC 1-2 & M8 III (Mira?) & \cellcolor{red!10} Mimic & S & \cellcolor{red!10} Mira in A2; YSO in A3; Symbiotic in A1, A4--A7 & - \\
SkySyC 1-3 & M4 III & Symbiotic & S & Symbiotic & - \\
SkySyC 1-4 & Mira & \cellcolor{red!10} Mimic & S & \cellcolor{red!10} Mira in A2; WR in A4; WTT in A6; Symbiotic in A1, A3, A5, A7 & - \\
SkySyC 1-5 & M4 III & \cellcolor{red!10} Mimic & S & \cellcolor{red!10} Mira in A2; Symbiotic in A1, A3--A7 & - \\
SkySyC 1-6 & M6.5 III & \cellcolor{red!10} Mimic & S & \cellcolor{red!10} Mira in A2; Symbiotic in A1, A3--A7 & - \\
SkySyC 1-7 & M5.5 III & Symbiotic & S & Symbiotic & - \\
SkySyC 1-8 & M3 III & Symbiotic & S & \cellcolor{red!10} M giant in A5; Symbiotic in A1--A4, A6--A7 & - \\
SkySyC 1-9 & M7.5 (Mira?) & \cellcolor{red!10} Mimic & S & \cellcolor{red!10} Mira in A2; Symbiotic in A1, A3--A7 & - \\
SkySyC 1-10 & M6 III & Symbiotic & S & Symbiotic & - \\
\end{tabular}
\begin{tablenotes}
\caption{  Flags for the decision trees that \citet{Akras2019} used to identify symbiotic stars with IR colour criteria. Columns are described below. Instances in which our selection methodology led to different results than \citet{Akras2019} are highlighted in light red.\label{tableskysyakras}}
\item[Internal name]  SkySy or SkySyC identification number.
\item[Giant]  Estimated spectral type of the donor star from \autoref{tableskysy}.
\item[Akras 7]  Classification of the object using figure 7 from \citet{Akras2019}, a general tree for disambiguating between symbiotics and mimics.
\item[Akras 9]  Classification of the object based on whether it lands in a bin dominated by stellar-type (S), stellar-type with IR excess (S+IR), dusty type (D), or dusty yellow type (D') in figure 9 from from \citet{Akras2019}, a tree for disambiguating between different kinds of symbiotics.
\item[Akras A1--A7]  Classification of the object based on figures A1 through A7 from \citet{Akras2019}, trees for disambiguating between symbiotics and specific types of mimics, assuming that the object is either a symbiotic or the types of mimics addressed by the particular decision tree. 
\item[Small $\Delta$] If a criterion was missed by a narrow margin, we note the difference between the colour of the object and the colour that would be required to qualify as a symbiotic in that tree. The \citet{Akras2019} figure to which the tabulated $\Delta$ applies is noted in parentheses.
\end{tablenotes}
\end{threeparttable}
\end{table}
\end{landscape}

\subsubsection{Final thoughts}

The synchronicity between our results and the \citet{Akras2019} results, and the stark differences between our results and the \mbox{\citet{Munari2021}} results, are both surprising outcomes, considering that our interest in accreting-only symbiotics more closely aligns with \citet{Munari2021} than \citet{Akras2019}. With SkyMapper $\sigu$ and SkyMapper colours, large-scale spectroscopic surveys like GALAH, GALEX UV colours, 2MASS/ALLWISE IR colours, and TESS light curves, a wide variety of different strategies to find accreting-only symbiotic stars are emerging, with exciting prospects for reconciling and combining different selection criteria.

\section{Conclusions}\label{Conclusions}

We conducted a multi-step, multi-wavelength search for symbiotic stars, aiming to identify their properties in a parameter space defined by photometry from the SkyMapper Southern Sky Survey. We were particularly interested in developing search techniques that are less biased against accreting-only symbiotics than preceding surveys. First, we built a sample of luminous red objects and calculated the SkyMapper parameters most likely to separate symbiotics from isolated cool giants. Second, we obtained optical spectra of 234 luminous red objects with outlying SkyMapper parameters. Third and finally, we obtained follow-up observations of select targets, including optical light curves and X-ray spectra. We have learned the following:

\begin{enumerate}[align=left,leftmargin=*,rightmargin=0ex,labelsep=0.5ex]\itemsep0.3em

    \item We discovered 12 symbiotic stars (SkySy) and an additional 10 symbiotic star candidates (SkySyC), spanning a wide range of subtypes from accreting-only to burning, with various kinds of donors including M III stars, S stars, and a K supergiant (\autoref{tableskysy}).
    
    \item We discovered fully-validated optical accretion-disc flickering in hours-long B-band light curves of four accreting-only symbiotics, and likely optical accretion-disc flickering in the previously known symbiotic V1044 Cen (CD-36 8436). We observed two of the flickering SkySy with {\it Chandra}, and detected hard $\delta$-type X-ray emission in both. We can constrain, at least for SkySy 1-2, the X-ray luminosity and emitting plasma temperature to be above what would be plausible for an isolated red giant under the standard paradigm of coronal X-ray emission from giant stars. Our results support prior work asserting a correlation between optical flickering, hard X-rays, weak optical emission lines, and the absence of higher-ionization optical emission lines---all signatures of an accreting-only symbiotic.

    \item The SkyMapper colours u-g and u-v are, together, effective at distinguishing most known symbiotic stars (including burning symbiotics and some accreting-only symbiotics) from other luminous red objects. The hot components appear to cause symbiotics to have flatter slopes in the short-wavelength optical spectral region probed by SkyMapper u, v, and g. There is a zone in the u-g/u-v colour-colour diagram that is virtually 100\% symbiotic stars. This zone is bordered by distinct regions dominated by (clockwise from the left in, e.g., the left-most panel of \autoref{allresults}) isolated cool M I--III and S stars, an unexplored abrupt rise to high source density, post-AGB stars and superpositions of M I--III stars with warm/hot stars, and a group of AGN that are mostly already cataloged in SIMBAD.

    \item The significance of rapid u-band variability between the two to three exposures within a single SkyMapper Main Survey {\it uvgruvizuv} filter sequence, as measured by $\sigu$ (\autoref{equation1}) normalized to an effective $\sigma_{\rm eff}$ to account for systematics, can be used in conjunction with a u-g$\lesssim$2.9 colour cut to detect symbiotic stars with optical accretion disc flickering. For known symbiotics and objects in the mostly-symbiotic zone of the u-g/u-v colour-colour diagram, the appropriate threshold above which to search is around 2$\sigma_{\rm eff}$. For objects outside the mostly-symbiotic zone but with u-g$\lesssim$2.9, the appropriate threshold is around 3$\sigma_{\rm eff}$. $\sigu$ detections in objects redder than u-g$\approx$2.9 are typically spurious, probably because real flickering is not detectable in u without a contribution from the accretion disc producing excess flux in u.
    
    \item Two of the flickering SkySy have an H$\alpha$ pseudo-equivalent width of less than 10\AA\, in at least one observation, and a third flickering SkySy has an H$\alpha$ pseudo-equivalent width  of less than 50\AA, below the detection threshold of at least some symbiotic star surveys that use narrow-band emission line photometry. Even without the involvement of $\sigu$ data, SkyMapper colours may be less biased against the selection of accreting-only symbiotics than narrow-band emission and objective prism surveys, though they are also effective at selecting burning symbiotics. With $\sigu$, SkyMapper is even more powerful, enabling the detection of more difficult-to-find symbiotics like SkySy 1-6.
    
    \item With all-sky {\it uvg} colours becoming available through later DRs of SkyMapper and MEPHISTO, we expect between about 51 and 117 symbiotics missed by previous surveys (of which 11 to 17 have been reported in this work) to be discoverable through the u-g/u-v colour-colour diagram alone, with a near-zero contamination rate.
    
    \item We expect at least about 20\% of the true population of symbiotics to have optical flickering detectable through SkyMapper $\sigu$. There is a significant population of optically-flickering symbiotics hidden both within and beyond the known catalogs of symbiotic stars.

    \item For those GALAH Symbiotic Stars (GaSS) from \citet{Munari2021} that are in our sample of luminous red objects and for which data are available, the GaSS are located in the densest part of the u-g/u-v colour-colour diagram and exhibit no evidence for excess $\sigu$, suggesting that we are probing very different regimes (\autoref{flickerers}). The flickerers in the GaSS sample have donors with earlier spectral types than our flickering SkySy, which can be attributed at least in part to the intentional exclusion of strongly pulsating giants in \citet{Munari2021}. The absence of u flux excess and $\sigu$ excess in the GaSS suggests that these earlier spectral types are transferring mass at a lower rate, resulting in dimmer accretion discs with less influence on the optical continuum, and yielding weaker Balmer emission lines.
    
    \item Ten out of 12 SkySy (including all 4 flickering SkySy), and 4 out of 10 SkySyC, meet the criteria for IR colour selection of symbiotic stars presented by \citet{Akras2019} in their Figure 7. This suggests that there is substantial overlap in the effects of our selection methods, even though we selected outliers only on the basis of uvg colours and variability while \citet{Akras2019} selected outliers only on the basis of IR colours. However, not all of our SkySy are correctly categorized by their other decision trees; most interestingly, two flickering SkySy miss getting correctly classified by the Akras symbiotic vs. M giant tree in their Figure A5 by an extremely thin margin---suggesting that the physical cause of abnormal IR colours in symbiotics is operating less strongly in some accreting-only symbiotics.
        
    \item The symbiotic star Hen 2-104 (the Southern Crab) has a large positive u-v colour, extremely unusual for a symbiotic star. It is located next to post-AGBs with strong Balmer jump absorption in the colour-colour diagram. SkyMapper u-v (especially when paired with u-g) is effective at distinguishing yellow post-AGB stars with strong Balmer jump absorption from most other luminous red objects.
    
    \item The only two confirmed symbiotics with S star donors in our sample of luminous red objects, SkySy 1-3 and SkySy 1-11, have been found to be optical flickerers: SkySy 1-11 in this work, and SkySy 1-3 in \citet{Merc2024}. Many known S stars (and carbon stars) are extreme outliers in SkyMapper u-v with large negative values; however, these two symbiotics with S star donors do not have large negative u-v.
    
    \item A single IR colour-cut of (J-K$_{\rm s}$)$_{0}$>0.85 works well to select samples of cool giants, paired with an M$_{\rm J}$<0 luminosity cut enabled by {\it Gaia}. However, some AGN and a number of post-AGB stars are included. Most selected AGN at the SkyMapper survey depth are already known and labelled in SIMBAD, but most selected post-AGB stars are not.
    
    \item When working with samples containing highly variable objects, such as Mira variables or even some semi-regular variables, SkyMapper colours built from the SkyMapper {\it dr2.master} catalog are not physically meaningful. This is because the SkyMapper catalog pipeline clips outlying measurements in its construction of the master catalog. To solve this problem, we presented an effective technique for reconstructing an average of nightly snapshot colours from the full {\it dr2.photometry} table of individual measurements, without which symbiotic stars and many isolated Miras would share the same space in the u-g/u-v diagram.

\end{enumerate}

\newpage

\section*{Acknowledgments}

We acknowledge the Gamilaroi people as the traditional owners of the land on which the SkyMapper Telescope stands.

We thank Mike Shara and David Zurek for extensive and vital assistance in obtaining and preparing SAAO observing runs. We thank Patrick Woudt, Paul Groot, Laura Rogers, John Southworth, and Lisa Crause for important logistical help. We thank the referee for their prompt and very useful report.

ABL thanks the skilled cooks at the SAAO hostel. We thank Sarah Tuttle, Adam Miller, Lucianne Walkowicz, and Dimitri Pourbaix for useful conversations. We thank the technical and cleaning staff of the SAAO and the Las Cumbres Observatory network.

We acknowledge contributed spectral observations from ARAS observers \citep{ARAS}, including Terry Bohlsen, Forrest Sims, Fran Campos, Christophe Boussin, James R. Foster, Colin Eldridge, A. Garcia, Pascal Le Du, and Berrand Guegan. We acknowledge with thanks the variable star observations from the AAVSO International Database contributed by observers worldwide and used in this research. This work makes use of observations from the Las Cumbres Observatory global telescope network. This paper uses observations made at the South African Astronomical Observatory (SAAO). This research has made use of data obtained from the {\it Chandra} Data Archive provided by the {\it Chandra} X-ray Center (CXC). We acknowledge the use of public data from the {\it Swift} data archive. This work made use of data supplied by the UK Swift Science Data Centre at the University of Leicester. This research has made use of the Astrophysics Data System, funded by NASA under Cooperative Agreement 80NSSC21M00561. 

ABL was supported by the STScI Fellowship from the Space Telescope Science Institute, which is operated by the Association of Universities for Research in Astronomy, Inc. under NASA contract NAS 5–26555. ABL was also supported by the NSF GRFP under grant DGE-1644869. ABL and JLS were supported by {\it Chandra} award DD6-17080X, NASA award DD0-21118X, and NSF awards AST-1616646 and AST-1816100. G.J.M.L. is a member of the CIC-CONICET (Argentina).

The national facility capability for SkyMapper has been funded through ARC LIEF grant LE130100104 from the Australian Research Council, awarded to the University of Sydney, the Australian National University, Swinburne University of Technology, the University of Queensland, the University of Western Australia, the University of Melbourne, Curtin University of Technology, Monash University and the Australian Astronomical Observatory. SkyMapper is owned and operated by The Australian National University's Research School of Astronomy and Astrophysics. The survey data were processed and provided by the SkyMapper Team at ANU. The SkyMapper node of the All-Sky Virtual Observatory (ASVO) is hosted at the National Computational Infrastructure (NCI). Development and support of the SkyMapper node of the ASVO has been funded in part by Astronomy Australia Limited (AAL) and the Australian Government through the Commonwealth's Education Investment Fund (EIF) and National Collaborative Research Infrastructure Strategy (NCRIS), particularly the National eResearch Collaboration Tools and Resources (NeCTAR) and the Australian National Data Service Projects (ANDS).

This research has made use of \protect{\sc{SAOImageDS9}}, developed by Smithsonian Astrophysical Observatory, and of \protect{\sc{Astropy}} \citep{astropy1,astropy2} and \protect{\sc{JSkyCalc}}. \autoref{flowchart} was created with \protect{\sc{flowchart.fun}}.\footnote{\url{https://github.com/tone-row/flowchart-fun}}

\section*{Data Availability}
 
SkyMapper DR2 data are publicly available through various ADQL-compatible endpoints.\footnote{\url{https://skymapper.anu.edu.au/how-to-access/}} {\it Chandra} (\href{https://doi.org/10.25574/23178}{doi:10.25574/23178} and \href{https://doi.org/10.25574/23179}{doi:10.25574/23179}) and {\it Swift} (see ObsIds in \autoref{xrtobstable}) data are publicly available through the HEASARC. Las Cumbres Observatory data (proposal LCO2020B-009) are publicly available through the LCO Science Archive.\footnote{\url{https://archive.lco.global/}} Optical spectra are available upon request from the authors, who will endeavor to eventually archive them in VizieR if time allows.



\bibliographystyle{mnras}
\bibliography{references} 




\appendix

\vspace{-3 mm}

\section{Snapshot colour reconstruction}\label{Snapshots}

\begin{figure}
{\includegraphics[width=\columnwidth]{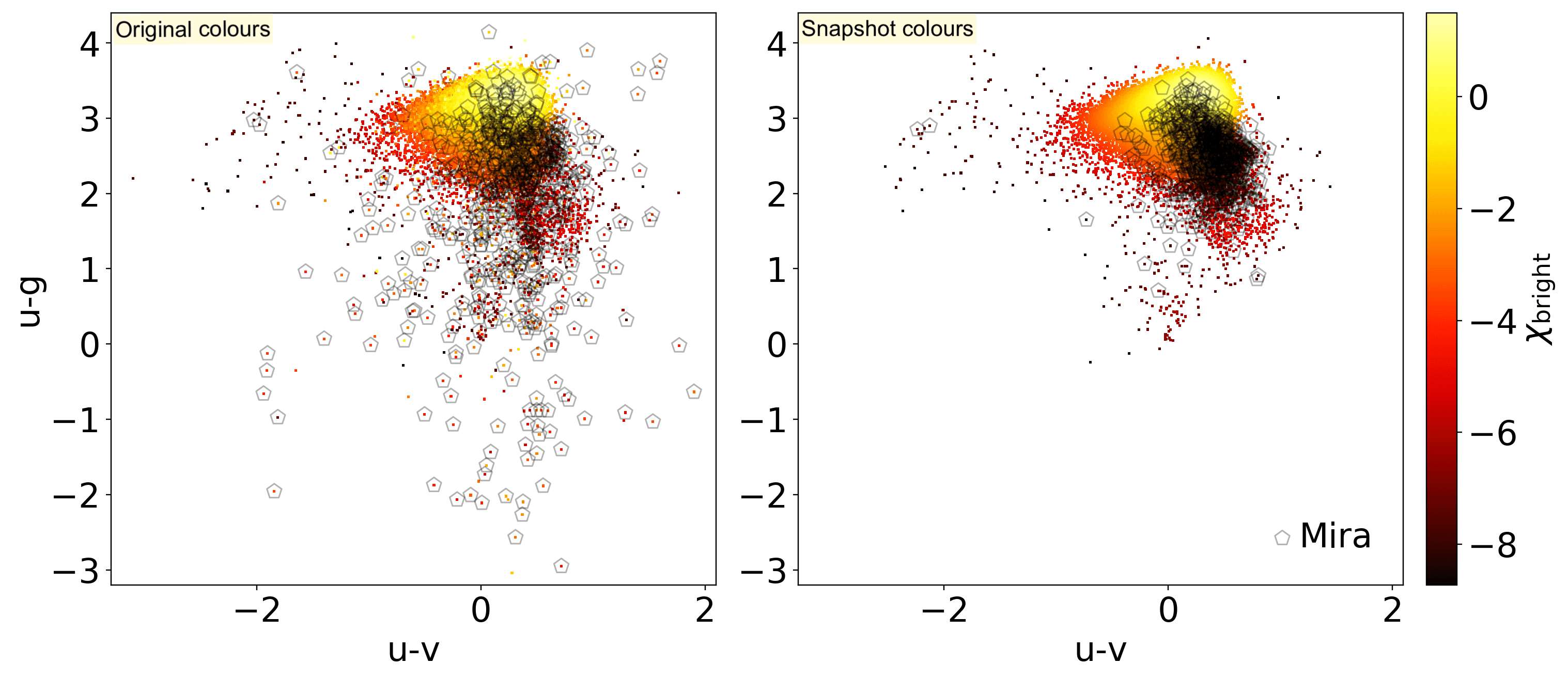}}
\caption{The left panel shows the broad distribution of colours for Mira variables in our luminous red objects sample, if we build the colour-colour diagram from the SkyMapper {\it dr2.master} catalog. In contrast, the right panel shows the more physical, densely packed distribution of colours for the same Miras in our colour-colour diagram if we reconstruct colours from nightly averaging of SkyMapper's {\it dr2.photometry} table of individual-exposure measurements. The colourbar scale for the underlying sample is described in \autoref{parameterspace}. The point centered in a hollow hexagon is the same object (i.e., Mira variable) represented by the hollow symbol.} \label{colorfix}
\end{figure}

In the course of our data exploration, we found that it was inadvisable, for our sample, to calculate u-g and u-v colours from the SkyMapper {\it dr2.master} catalog magnitudes for those filters; instead, it was necessary for us to reconstruct colours directly from {\it dr2.photometry}. The problem is illustrated in \autoref{colorfix}. In the left panel, we show that Mira variables---AGB stars with optical pulsation amplitudes of up to many magnitudes---are widely scattered in the u-g/u-v colour-colour diagram built from {\it dr2.master}. Although SkyMapper observes snapshot colours in both its Main Survey and Shallow Survey filter sequences, the SkyMapper catalog construction pipeline discards some measurements if it identifies them as outliers or if they have data quality issues \citep{skymapperdr1,skymapperdr2}, and those clipped measurements do not make it into the {\it dr2.master} averages. This measurement clipping can time-imbalance what would have been, for example, the average of two Main Survey snapshot colours, or the average of a Shallow Survey snapshot colour and a Main Survey snapshot colour. For Miras, and a wide variety of other red giant pulsators and highly variable stars, a colour calculated from the difference of band measurements taken on different months is physically and practically meaningless.

However, the information necessary to reconstruct an average of true snapshot colours is still contained within the full table of SkyMapper measurements, {\it dr2.photometry}. We first retrieved the full table of individual {\it uvg} PSF magnitude measurements for each object in our luminous red objects sample, including both Main and Shallow Survey measurements.\footnote{The SkyMapper TAP endpoint does not accept table uploads. For details on our ad-hoc procedural workaround, see section 2.1.3.2 in \citet{Lucy2021}.} Then, to replicate the SkyMapper DR2 filtering strategy up until the outlier-clipping stage, we required that {\it used\_in\_clipped} be a boolean value (True or False, but not null). This reduced the number of {\it uvgriz} measurements from 11.4 to 8.7 million. Rejecting null values of {\it used\_in\_clipped} rejects measurements that did not make it to the clipping stage of the SkyMapper pipeline, and amounts to rejecting source extractor flags$\geq$4, nimaflags$\geq$5, and incidents in which multiple Source Extractor detections of the same object were identified in a single observation. Unlike the SkyMapper DR2 catalog pipeline, we accepted {\it used\_in\_clipped}=False, because outlier clipping is not appropriate for variable objects with so few data points. The result was a concatenated table of 8.7 million individual {\it uvgriz} band measurements and their observational metadata for our full sample of luminous red objects.

We then computed as follows, for each object, a weighted average of nightly u-v and u-g colours. First, we split our concatenated photometry table by filter band. The uncertainty for each SkyMapper PSF magnitude measurement was taken to be the SkyMapper {\it e\_mag\_psf} uncertainty in {\it dr2.photometry} propagated in quadrature with a 0.01 mag systematic uncertainty (from the SkyMapper catalog pipeline procedure, in which a 0.01 mag uncertainty is added to reflect flat field uncertainties; \citealt{skymapperdr1}). We used these uncertainties to compute a variance-weighted\footnote{Weighting was performed in logarithmic magnitudes (geometric means), not linear units.} average measurement and propagated variance for each night in each band, using {\it groupby.apply} in \textsc{pandas} \citep{pandassoftware,pandaspaper} to group on the {\it object\_id} and {\it night\_mjd} columns from the {\it dr2.master} and {\it dr2.images} tables and apply the calculation functions. We then performed an inner join of the resultant u and g (or u and v) tables with \textsc{pandas} {\it merge} on the {\it object\_id} and {\it night\_mjd} columns, and computed a nightly u-g (or u-v) colour. After that, we computed an average of the nightly colours weighted by the propagated variances, using groupby.apply to group on {\it object\_id} and again apply the weighted-average calculation function. Finally, we performed an inner join of the resultant u-g and u-v tables with {\it merge} on the {\it object\_id} column. 

The result was a table of reconstructed u-v and u-g colours for each of our luminous red objects, excluding objects for which a nightly colour could not be computed. 7\% of our working catalog was removed by this exclusion. The remaining 366,721 objects comprise our final catalog of luminous objects, the substrate from which we selected outlying targets. A colour-colour diagram from the reconstructed colours of u<16 objects is shown in the right panel of \autoref{colorfix}. Reconstructing u-v and u-g colours using {\it dr2.photometry} reduced the large scatter of pulsating Miras; in the right panel of \autoref{colorfix}, the scatter is now densely localized to a small region of parameter space, allowing true colour-colour outliers to be seen.

Our procedure means that each colour for an object was obtained from self-consistent coeval photometry, but the different colours for each object are still not necessarily coeval to each other. This was done to maximize the number of objects for which we would obtain data. {\it uvg} observations were usually obtained coevally in all-band sequences, but quality cuts/flags sometimes removed filters from a night's sequence in our analysis. The reader should therefore {\it not} attempt to, say, infer the v-g colour from the u-v and u-g colours presented in this paper.


\bsp	
\label{lastpage}
\end{document}